\documentclass[twocolumn]{aastex62}

\usepackage{soul}

\received{April 20, 2022}
\revised{September 13, 2022}
\accepted{September 22, 2022}

\submitjournal{ApJ}

\shorttitle{Polarization of SNRs}
\shortauthors{Shanahan et al.}

\begin{document}

\title{Polarized Emission From Four Supernova Remnants In The THOR Survey}

\correspondingauthor{Russell Shanahan}
\email{rpshanah@ucalgary.ca}

\author[0000-0002-3186-8369]{Russell Shanahan}
\affil{University of Calgary, 2500 University Dr. NW, Calgary AB, T2N 1N4}

\author[0000-0003-2623-2064]{Jeroen M. Stil}
\affil{University of Calgary, 2500 University Dr. NW, Calgary AB, T2N 1N4}

\author[0000-0001-8800-1793]{Loren Anderson}
\affil{Department of Physics and Astronomy, West Virginia University, Morgantown, WV 26506, USA}
\author[0000-0002-1700-090X]{Henrik Beuther}
\affil{Max Planck Institute for Astronomy, K{\"o}nigstuhl 17, 69117 Heidelberg, Germany}
\author[0000-0002-6622-8396]{Paul Goldsmith}
\affil{Jet Propulsion Laboratory, California Institute of Technology, 4800 Oak Grove Drive, Pasadena, CA 91109, USA}
\author[0000-0001-8224-1956]{J{\"u}rgen Ott}
\affil{National Radio Astronomy Observatory, 1003 Lopezville Road, Socorro, NM 87801, USA}
\author{Michael Rugel}
\affil{Max-Planck-Institut f{\"u}r Radioastronomie, Auf dem H{\"u}gel 69, 53121 Bonn, Germany}
\author[0000-0002-0294-4465]{Juan Soler}
\affil{Max Planck Institute for Astronomy, K{\"o}nigstuhl 17, 69117 Heidelberg, Germany}
\author[0000-0003-4322-8120]{Jonas Syed}
\affil{Max Planck Institute for Astronomy, K{\"o}nigstuhl 17, 69117 Heidelberg, Germany}

\keywords{ISM: supernova remnants --- radio continuum: ISM ---
ISM: magnetic fields --- polarization}

\begin{abstract}

We present polarization and Faraday rotation for the supernova remnants (SNRs) G46.8$-$0.3, G43.3$-$0.2, G41.1$-$0.3, and G39.2$-$0.3 in L-band (1-2 GHz) radio continuum in The HI/OH/Recombination line (THOR) survey.  We detect polarization from G46.8$-$0.3, G43.3$-$0.2 and G39.2$-$0.3 but find upper limits at the 1\% level of Stokes $I$ for G41.1$-$0.3.  For G46.8$-$0.3 and G39.2$-$0.3 the fractional polarization varies on small scales from 1\% to $\sim$6\%.  G43.3$-$0.2 is less polarized with fractional polarization $\lesssim$3\%.  We find upper limits at the 1\% level for the brighter regions in each SNR with no evidence for associated enhanced Faraday depolarization.  We observe significant variation in Faraday depth and fractional polarization on angular scales down to the resolution limit of 16\arcsec.  Approximately 6\% of our polarization detections from G46.8$-$0.3 and G39.2$-$0.3 exhibit two-component Faraday rotation and 14\% of polarization detections in G43.3$-$0.2 are multi-component.  For G39.2$-$0.3 we find a bimodal Faraday depth distribution with a narrow peak and a broad peak for all polarization detections as well as for the subset with two-component Faraday rotation.  We identify the narrow peak with the front side of the SNR and the broad peak with the back side.  Similarly, we interpret the observed Faraday depth distribution of G46.8$-$0.3 as a superposition of the distributions from the front side and the back side.  We interpret our results as evidence for a partially filled shell with small-scale magnetic field structure and internal Faraday rotation. 

\end{abstract}

\section{Introduction} \label{sec:intro}

\indent	The study and understanding of supernova remnants (SNRs) is important for providing insight into chemical enrichment, energy injection into the interstellar medium (ISM) and magnetism within galaxies.  Since the primary source of emission from SNRs is synchrotron radiation of relativistic electrons interacting with a magnetic field, radio observations of these sources are useful for probing the SNR's internal magnetic field and the Galactic magnetic field along the line of sight \citep[see][for reviews]{Chevalier1977, Raymond1984, Reynolds2012, Dubner2015, Han2017}. 

\indent	Since radio emission from SNRs is primarily synchrotron, the radiation is linearly polarized.  Polarization was used to identify SNRs in the crowded Galactic plane by \citet{Dokara2021}.  By observation of the polarized electric field vector, one can determine the orientation of the orthogonally aligned magnetic field in the plane of the sky.  From Faraday rotation the electric field vector of the polarized emission is rotated during the propagation within the SNR and the ISM, thereby compromising the observed polarization.  Thus, the observation of true position angles of the magnetic field lines is best done through observing the Faraday rotation of radio waves. The simplest case is where the polarization angle $\chi$ of a linearly polarized wave changes with wavelength $\lambda$ according to $\delta \chi = \phi \lambda^2$.  The Faraday depth $\phi$ is defined as
\begin{equation}
\label{eqn:FD1}
\phi = 0.81 \int \left( \frac{n_e}{\text{cm}^{-3}} \right) \left( \frac{B_{\parallel}}{\mu\text{G}} \right) \left( \frac{dl}{\text{pc}} \right),
\end{equation}
with $\phi$ in rad m$^{-2}$, $n_e$ is the free electron density, $B_{\parallel}$ is the magnetic field component along the line of sight, and $l$ is the path length from the emitting position to the observer  \citep{Burn1966, Klein2015}.  This integral is evaluated from the source to the observer where positive $\phi$ indicates $B_{\parallel}$ pointing towards the observer.

\indent	It has been proposed that the orientation of the magnetic field within SNRs shows a pattern that depends on their age \citep{Dubner2015}.  Early observations of the young SNR Cas A show a radial magnetic field with respect to the shock front \citep{Mayer1968}, whereas the radio polarization observed from the old SNR Vela reveal a tangential magnetic field in the regions of brighter emission \citep{Milne1968}.    The polarization observation of over 27 SNRs by \citet{Milne1987} confirmed that the magnetic fields in young SNRs is predominantly in the radial direction and older SNRs have a dominant magnetic field orientation that is parallel to the shock front or tangled.  This picture of magnetic field orientations with respect to a SNR's age is confirmed by subsequent observations.  

\indent	The tangentially ordered fields observed in older SNRs are generally conceived as being the result of compression in radiative shocks with large shock compression ratios, however the origin of the radial component is still controversial \citep{Dubner2015}.  \citet{Reynoso2013} find that while the orientation of the magnetic field vectors across the SNR are radial, the majority of the magnetic field vectors lie parallel to the Galactic plane.  Therefore, \citet{Reynoso2013} concluded that the ambient magnetic field must be roughly parallel to the Galactic plane and that this original orientation remains even after the passage of the shock front.  From their investigation of polarization of SN1006, \citet{Reynoso2013} also conclude that the most efficient particle acceleration occurs for shocks where the magnetic field orientation is quasi-parallel to the shock normal.  Pointing out a selection effect due to the distribution of cosmic-ray electrons (CREs), \citet{West2017} find that CREs accelerated by quasi-parallel shocks can have spatial distributions that result in an apparent radial magnetic field from radio synchrotron observations where the true magnetic field is turbulent and disordered. 

\indent	The use of SNRs as a probe to study the Galactic magnetic field provides the ability to study a spatially extended region in the Milky Way as opposed to single lines-of-sight that pulsars and extra-galactic sources provide.  \citet{Kothes2009} suggest that a correlation between the bilateral axis angle of SNRs and the orientation of the Galactic magnetic field has the potential to reveal important information on the large-scale Galactic field.  Expanding on this, \citet{West2015} demonstrated that their models can reproduce observed morphologies as well as predicted magnetic fields (tangential and radial) that are consistent with the observed magnetic fields.  

\indent	The observation of radio emission in SNRs provides constraints on the polarization and propagation in turbulent magnetic fields \citep{Reynolds2008}.  The turbulent magnetic field will reduce the fractional polarization by differential Faraday rotation because of reversals in the line of sight component of the magnetic field \citep{Burn1966, Sokoloff1998}.  Also, the isotropic component of the turbulent magnetic field in the plane of the sky contributes to total intensity but not polarized intensity, further reducing the fractional polarization \citep[for e.g.]{Sokoloff1998}. \citet{Petruk2017} developed a model and simulated images of polarized synchrotron emission from Sedov SNRs.  A consequence of their models is that a change in magnetic field projection along the line of sight will enhance the level of depolarization.  Due to Faraday depolarization effects in radio bands, \citet{Bykov2020} uses observations of polarized X-ray synchrotron radiation to reveal how within an anisotropic magnetic turbulence cascade model the magnetic field near the shock is primarily radial and this pattern will produce polarization that is predominantly parallel to the shock front.  \\

\indent 	In this work we describe linear polarization from the THOR survey \citep{Beuther2016} of four SNRs: G46.8$-$0.3, G43.3$-$0.2, G41.1$-$0.3 and G39.2$-$0.3.  We outline the observations, calibration and imaging in Section \ref{sec:obs}.  The methods used in this study to investigate polarized emission are presented in Section \ref{sec:meth}.  In Section \ref{sec:results} we present the results in investigating polarization and Faraday rotation of the SNRs G46.8-0.3 (Section \ref{sec:G46}), G43.3-0.2 (Section \ref{sec:G43}), G41.1-0.3 (Section \ref{sec:G41}) and G39.2-0.3 (Section \ref{sec:G39}).  We present a discussion of our investigation in Section \ref{sec:Disc} and conclusions are found in Section \ref{sec:conclusions}.  We define the detection criteria used in analyzing polarization in Appendix \ref{appendix:DT}.  \\

\section{Observations, Calibration and Imaging} 
\label{sec:obs}

\indent 	The THOR survey is a large program at the Karl G. Jansky Very Large Array with approximately 215h of observing time that covers the inner Galaxy in the longitude range $14\fdg5 < \ell <67\fdg4$ and latitude $-1\fdg25 < b < 1\fdg25$ in C-array configuration in L-band (1 - 2 GHz) \citep{Beuther2016}.  Since the primary beam size over 1 to 2 GHz changes by a factor of 2, the actual areal coverage of THOR depends on frequency, but the approximate areal coverage of THOR is $\sim132$ square degrees.  The survey includes the $\lambda$21 cm line of atomic hydrogen, OH lines, radio recombination lines and the continuum in 512 channels from 1 to 2 GHz, where each channel has a frequency width of 2 MHz.  The $\lambda$21 cm line and total intensity continuum were combined with archive data from the VLA Galactic Plane Survey \citep[VGPS,][]{Stil2006} and the single dish observations at 1.4 GHz by the Effelsberg continuum survey by \citep{Reich1990} to include spatial scales down to 60\arcsec.  However, only the C-configuration data exist for the other spectral lines and continuum polarization, which samples the continuum at 1.5 GHz on angular scales ranging from $\sim15''$ to $\sim5'$.  

\indent 	The full survey was calibrated in total intensity as well as polarization with the CASA software package.  The calibration of the pilot region of the survey was done using the CASA version 4.1.0 and a modified VLA pipeline version 1.2.0.  The second half of the survey was calibrated with newer versions, CASA 4.2.2 and VLA pipeline 1.3.1, for calibration the difference from the older versions are minimal.  Even though frequency bands with strong Radio Frequency Interference (RFI) and bad antennas were flagged manually prior to calibration, the VLA pipeline applies automated RFI flagging on the calibrators in order to improve the calibration solutions and data quality.  However, in the data analysis of SNRs, further flagging was done manually.

\indent	 The flux, bandpass and polarization calibration were completed for all fields using the quasar 3C286.  For the complex gain calibration and polarization leakage calibration two different calibrators were used: J1822-0938 for the observations between Galactic longitudes $14\fdg5$ and $39\fdg1$, and J1925+2106 for the remaining fields at Galactic longitudes $>39\fdg1$.  After calibrator RFI flagging the flux, bandpass and gain calibration was applied using standard procedures.  As described in \citet{Beuther2016}, no Hanning smoothing was performed during calibration, and the weights were not recalculated because of the effect that the operation can have on bright sources.  The calibration was done iteratively, where after a full calibration quality checks and flags were additionally applied, after which calibration was implemented again.  The details of total intensity calibration and imaging is outlined in detail in \citet{Beuther2016}.

\indent 	The polarization calibration was performed per channel in CASA following standard procedures after bandpass, flux and gain calibration were implemented.  Polarization angle calibration was derived from 3C286.  The phase calibrator (J1822-0938) was used to derive solutions for instrumental polarization as well as cross-hand delay terms for linear polarization.  The polarization calibration applies to the centre of each field, therefore the instrumental polarization increases with distance from the field centre.  Calibration for off-axis polarization is not yet available.  After performing experiments on sources of bright thermal emission, we find that leakage in our mosaics is constrained to fractional polarization $\lesssim 1\%$ centred around $\phi \approx 0$ rad m$^{-2}$ (see Appendix \ref{appendix:DT}).  

\indent	In order to reduce processing time, 8 MHz channel averaged images were made averaging four 2 MHz channels.  In doing so, the time required to produce channel averaged images is reduced by a factor of 4, while reducing the noise to approximately 0.4 mJy beam$^{-1}$ for each 8 MHz channel.  For a single mosaic ($1\fdg25 \times 2\fdg5$), 80 different frequency images over the entire bandpass are made and combined into a single cube.  This process is repeated for each of Stokes $I, Q, U$ separately.  

\indent	The restoring beam was calculated by the default algorithm within CASA, yet each channel image was smoothed to the beam size at the lowest frequency in the respective spectral window.  This was done to allow the ability to analyze the upper frequency band separately at higher resolution as well as to compensate for the effects of the changing resolution from the low to high sides of the frequency band.  The implication is that analysis of the cubes must take into account varying resolution across the band.  In order to do this, the change in resolution is accounted for when extracting fluxes from the image cubes.

\indent	In order to recover large scale emission, the CASA multi-scale cleaning algorithm was implemented.  The setup in CASA allows for the selection of different spatial scales, and after experimenting with many different setups we settled on using four spatial scales: point sources, the synthesized beam, then 2 and 5 times the synthesized beam.  With this setup we could better recover large scale emission than previous CLEAN algorithms.  The images were made using Briggs weighting with a default robust value of 0. 

\newpage

\section{Methods}
\label{sec:meth}

\indent	Faraday rotation causes a rotation of the polarization angle ($\chi$) by an amount that is proportional to wavelength-squared ($\lambda^2$) for a Faraday thin source \footnote{A source that is Faraday thin is well approximated by a Dirac $\delta$-function of $\phi$ \citep{Brentjens2005}.  A point source with no internal Faraday rotation behind a Faraday screen is a good approximation of a Faraday thin source.}.  Differential Faraday rotation occurs when both synchrotron emission and Faraday rotation occur in the same volume.  Differential Faraday rotation over a single observed frequency range can lead to depolarization, thus causing sources of strong Faraday rotation to be unpolarized.  Broadband multifrequency observations allow Faraday Rotation Measure (RM) Synthesis to solve for the unknown Faraday depth, polarized intensity and polarization angle simultaneously \citep{Brentjens2005}.  The polarized signal of the source is typically identified as the maximum absolute value in the Faraday depth spectrum.  

\begin{figure*}[htb!]
\centering
   \centerline{\includegraphics[width=0.8\linewidth, angle=0]{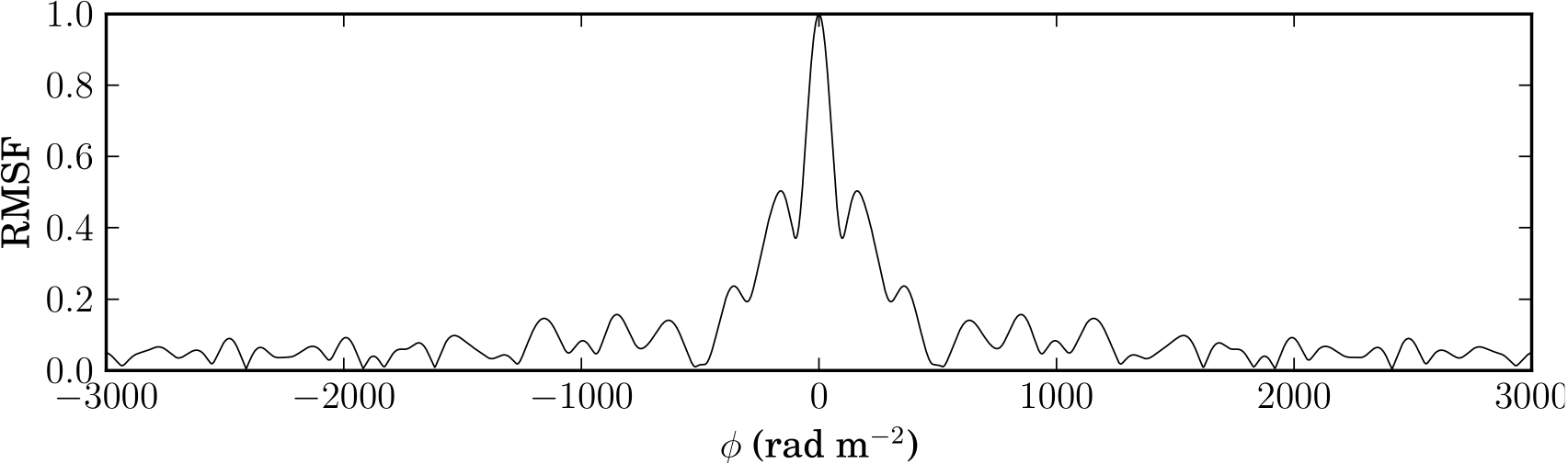}}
   \caption{The Rotation Measure Spread Function (RMSF) after RFI flagging is performed on the polarization image cubes, with a FWHM of 101.7 rad m$^{-2}$.  This RMSF is from the data used in the investigation of SNR G46.8$-$0.3.  The RMSF for the other SNRs in this study have a similar shape with the FWHM falling within the range from 100 rad m$^{-2}$ to 103 rad m$^{-2}$ due to different RFI flagging.} 
   \label{fig:RMSF} 
\end{figure*}

\indent	The analysis of the THOR polarization image cubes is first done by implementing the Faraday RM synthesis algorithm as defined in \citet{Brentjens2005}, where the Faraday depth is defined in Equation \ref{eqn:FD1}.  The complex polarized intensity, as expressed in terms of Stokes $Q$ and $U$ to be $P (\lambda^2) = Q + iU$, is the Fourier transform of the Faraday dispersion function $\tilde{F}(\phi)$, 
\begin{equation}
\tilde{F}(\phi) = \frac{1}{K} \int_{-\infty}^{\infty} P(\lambda^2) W(\lambda^2) \exp[-2 i \phi \lambda^2] d\lambda^2.
\label{eqn:FD}
\end{equation}
Here, $K$ is the integral of the weight function, $W(\lambda^2)$.  $W(\lambda^2) = 1$ when measurements exist and $W(\lambda^2) = 0$ where there are no measurements, including when $\lambda^2 < 0$.  The Faraday dispersion function is the complex polarized surface brightness per unit of Faraday depth \citep{Brentjens2005}. The Fourier transform of the weight function $W(\lambda^2)$ is the Rotation Measure Spread Function (RMSF) which serves as a point-spread function in the Faraday depth regime.  A consequence of RFI flagging in the THOR polarization image cubes is that the Faraday depth resolution $\delta\phi$, the FWHM of the RMSF,  is reduced to $\sim100 \text{ rad m}^{-2}$ as well as raising the sidelobes of the RMSF (as shown in Figure \ref{fig:RMSF}).  Due to different RFI flagging in each image cube, the FWHM of the RMSF changes slightly for each SNR but falls within the range of 100 to 103 rad m$^{-2}$.

\indent	By integrating over frequency, solid angle or the existence of different emission regions along the line of sight it is possible for multiple values of $\phi$ to arise due to the blending of different rates of Faraday rotation.  This effect is referred to as Faraday complexity.  A consequence of this effect is that the fractional polarization will change with wavelength along with a non-linear relation between $\chi$ and $\lambda^2$.  

\indent	Due to Faraday rotation within channels, it is possible for large Faraday rotation to become depolarized within a single frequency channel.  \citet{Brentjens2005} describe this effect in the parameter for maximum observable Faraday depth

\begin{equation}
\parallel \phi_{max} \parallel \approx \frac{\sqrt{3}}{\delta\lambda^2},
\end{equation}
where $\delta\lambda^2$ is the channel width expressed in $\lambda^2$.  This relation is approximate because the channel width in $\lambda^2$ is larger at the low end of the frequency band and smaller at the high frequency side of the band.  At 1.5 GHz with 8 MHz channels we find $\parallel \phi_{max} \parallel = 4.0 \times 10^3 \text{ rad m}^{-2}$, and at 2 GHz $\parallel \phi_{max} \parallel = 9.7 \times 10^3 \text{ rad m}^{-2}$.  Sources where polarization is detected were analyzed using the RM Clean algorithm outlined in \citet{Heald2009} as part of the RMtools \footnote{https://github.com/CIRADA-tools/RM} package \citep{Purcell2020}.  RM Clean is a one-dimensional analog of the CLEAN algorithm in aperture synthesis, where the RMSF acts as the dirty beam.  A difference between these algorithms is that RM Clean acts on a complex function.  The tool for QU fitting \citep{Law2011} was also used but is not a main focus of the results presented in this work.  


\section{Results}
\label{sec:results}

\subsection{SNR G46.8$-$0.3}
\label{sec:G46}

\indent 	G46.8$-$0.3 (HC30) was first identified near the HII region G046.495$-$0.241 using 1.7 and 2.7 GHz observations \citep{Willis1973, Green2019}.  From H$_2$ infrared emission lines, \citet{Lee2020} find a $V_{LSR}$ of $44 \pm 1$ km s$^{-1}$ and obtain a kinematic distance to be $5.4 \pm 0.1$ kpc which are consistent with values from HI.  With combined CO-HI profiles, \citet{Supan2022} provide evidence for environmental molecular clouds that are physically linked to the remnant at its centre, the lower edge, and towards the bright regions on the top-left and lower-right rims on the far side of the SNR shell.  \citet{Sun2011} observed SNR G46.8$-$0.3 at $\lambda$6 cm to be 8\% polarized with an integrated polarized flux density of $595 \pm 32$ mJy and a spectral index of $-0.54 \pm 0.02$ at $9\farcm5$ resolution.  

\begin{figure*}[htb!]
\centering
   \centerline{\includegraphics[width=1\linewidth]{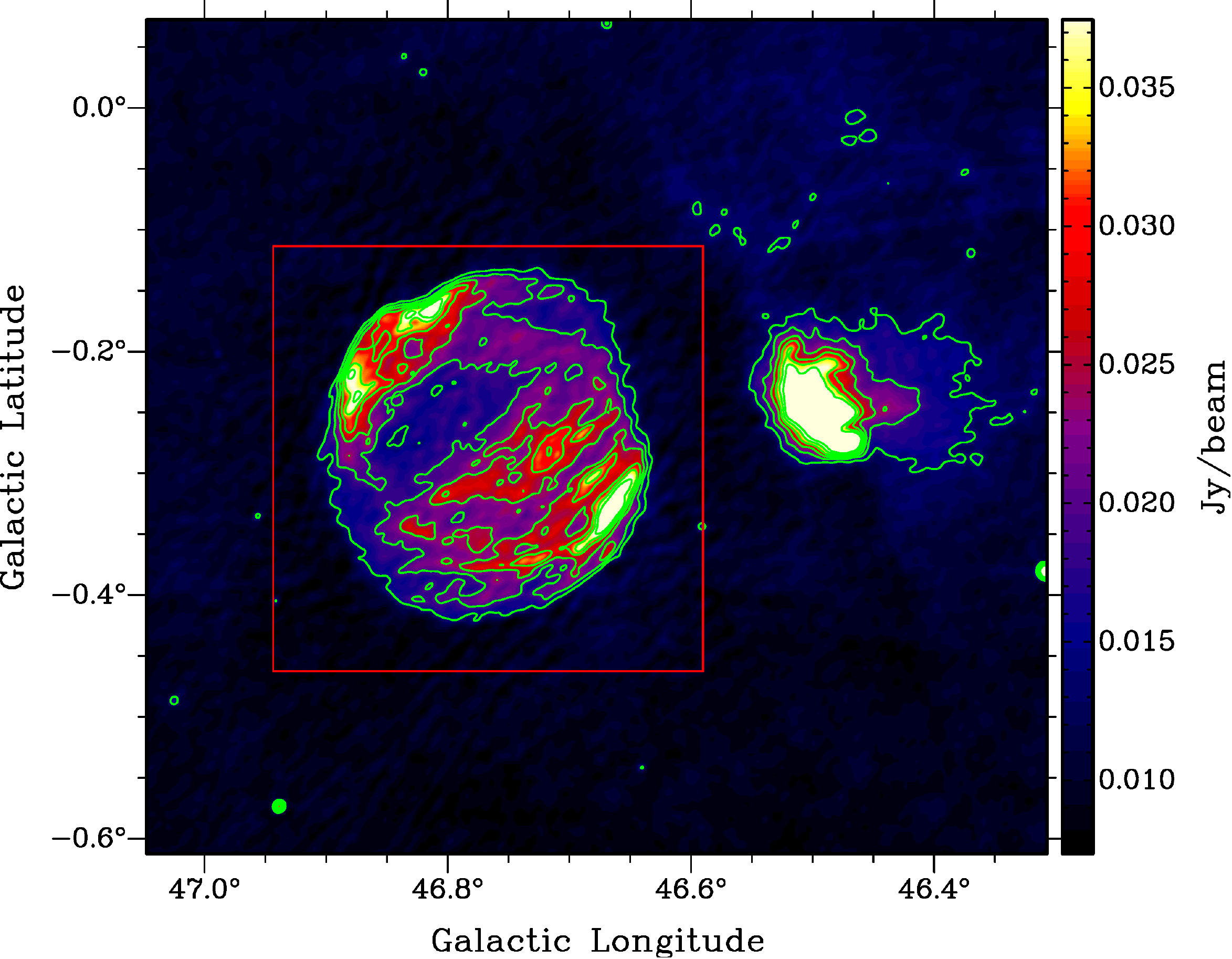}}
   \caption{Emission at 1420 MHz toward the SNR G46.8$-$0.3 reconstructed from THOR and VGPS observations.  The green contour levels correspond to 14, 19, 24, 29, 34 and 39 mJy/beam. The red box indicates the region in which we performed a grid search for polarization and Faraday rotation. The object on the right-hand side of the box is the HII region G046.495$-$00.247 \citep{Lockman1989,Anderson2018}.}
   \label{fig:G46.8-0.3_THOR+VGPS} 
\end{figure*}

\indent	Figure \ref{fig:G46.8-0.3_THOR+VGPS} presents the SNR and the HII region at $\lambda$21 cm in total intensity continuum as a combination of THOR and VGPS data.   The red box outlines the area where we search for polarization and Faraday rotation, which we separated into 7140 subregions  $16\arcsec \times 16\arcsec$, the synthesized beam at 1.2 GHz.  For each frequency, we extract the flux density by performing a sum of each pixel within a subregion and dividing by the frequency scaled beam.   Faraday rotation measure synthesis is then performed on the Stokes $IQU$ spectra from each subregion.  Each peak in the FD spectrum that passes our detection criteria (see Appendix \ref{appendix:DT}) has a Faraday depth and polarized intensity. The Faraday depth has an associated error with a median value of $\sim 5 \text{ rad m}^{-2}$.  To derive fractional polarization, we divide the polarized intensity by the background subtracted Stokes $I$ flux from THOR+VGPS.  Both refer to nearly the same frequency, so no correction for the Stokes $I$ spectral index was made. 

\begin{figure*}[htb!]
\centering
   \centerline{\includegraphics[width=0.8\linewidth, angle=270]{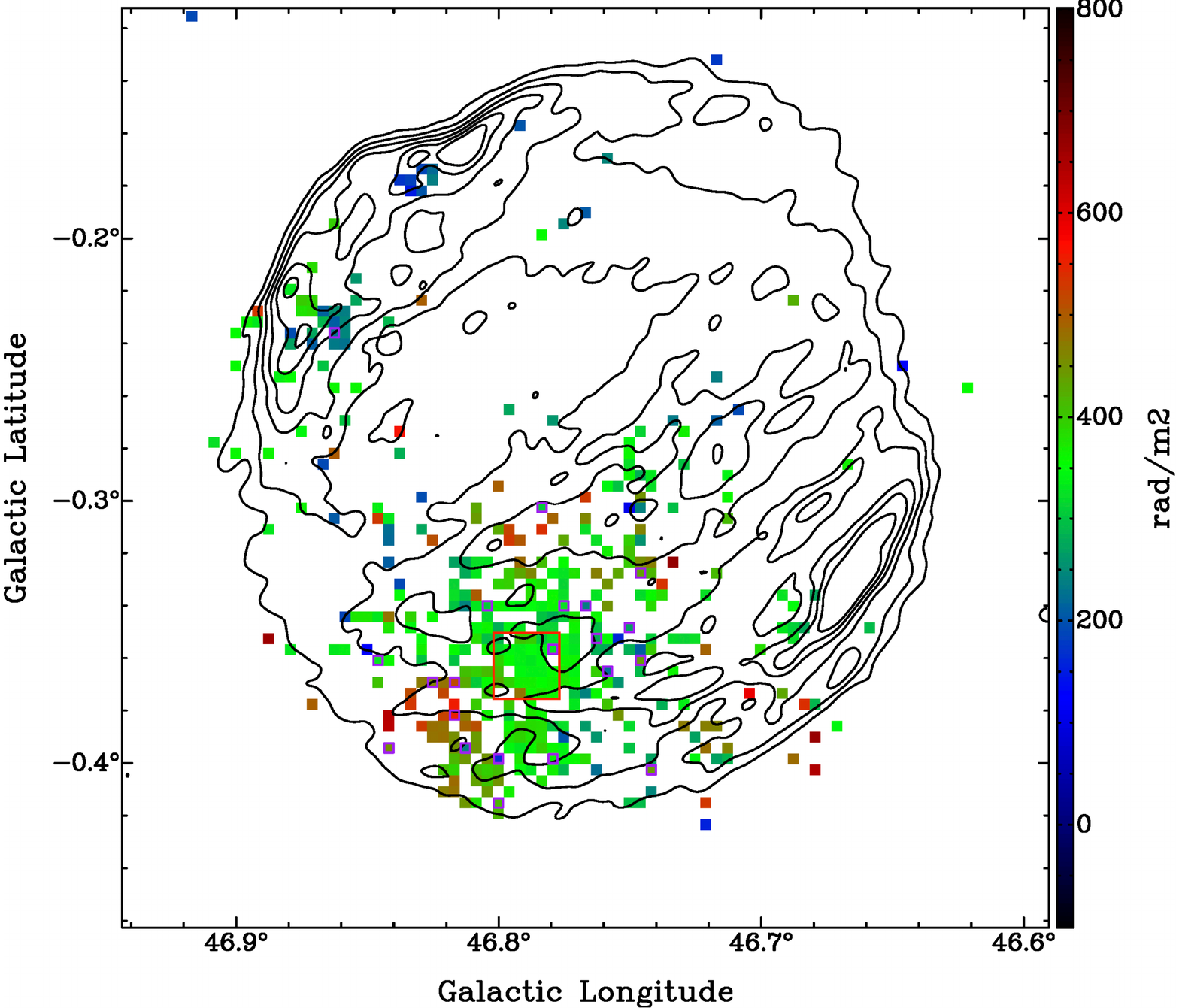}}
   \caption{Faraday depth map of SNR G46.8$-$0.3. The contour levels correspond to 14, 19, 24, 29, 34 and 39 mJy/beam. The subregions with a purple edge indicate the locations of two-component Faraday rotation.  The Faraday depth shown for subregions with two-component Faraday rotation is derived from the strongest peak.  The red box indicates the region where we present the individual Faraday depth spectra as a grid plot seen in Figure \ref{fig:G46.8-0.3_RM_Grid2}.}  
   \label{fig:G46.8-0.3_RM-Map} 
\end{figure*}

\indent	The Faraday depth map of G46.8$-$0.3 is displayed in Figure \ref{fig:G46.8-0.3_RM-Map} which shows the red box seen in Figure \ref{fig:G46.8-0.3_THOR+VGPS}.  Each coloured pixel (376 out of 7140) indicates a detection where the Faraday depth is derived from the highest peak that passes our detection criteria. 

\indent	Figure \ref{fig:G46.8-0.3_RM-Map} illustrates evidence of small scale polarization structure within the SNR.  The bottom region of the SNR shell has the highest density of detection where we observe a variation of $300 < \phi < 600 \text{ rad m}^{-2}$.  A gradient of Faraday depth is observed across the SNR from the bottom of the SNR shell to the top left.  The brightest region of the shell revealed no confident detections of polarization.  A few regions outside the SNR with no THOR+VGPS Stokes $I$ counterpart revealed detections of polarized emission and Faraday rotation.  We attribute these detections to small-scale polarization angle fluctuations from the diffuse Galactic emission that is mostly filtered out in our images cubes due to missing short-spacing from the VLA in C-array. 

\begin{figure}[htb!]
\gridline{\rotatefig{0}{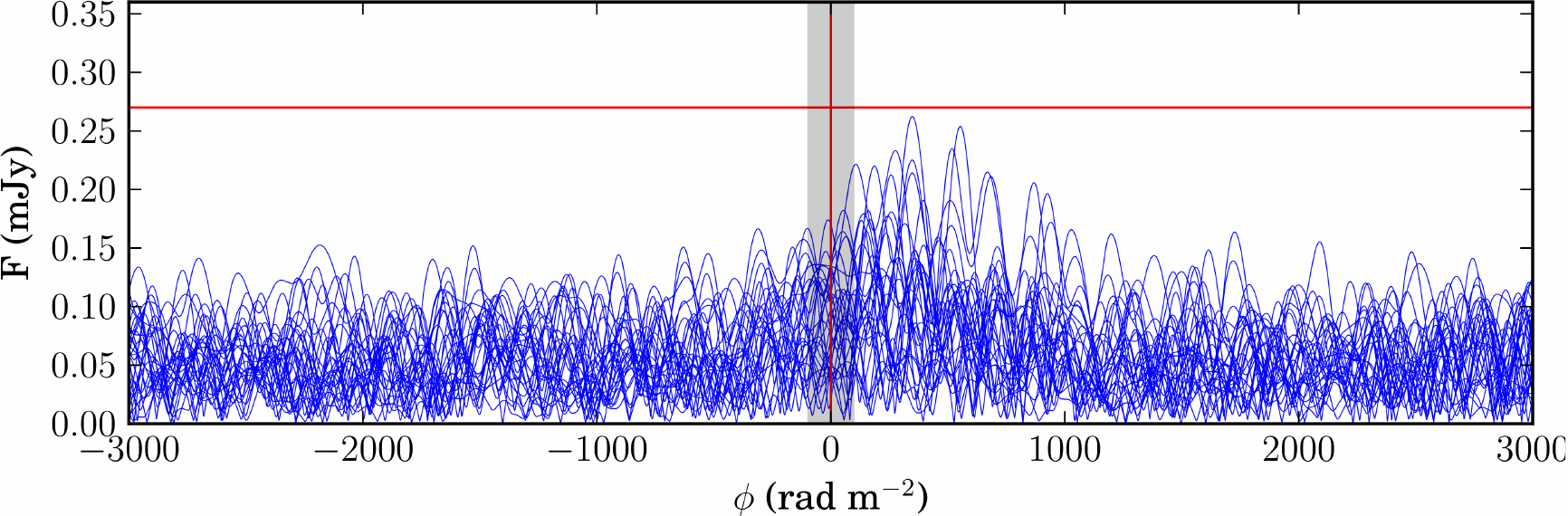}{1.0\linewidth}{(a)}}
\gridline{\rotatefig{0}{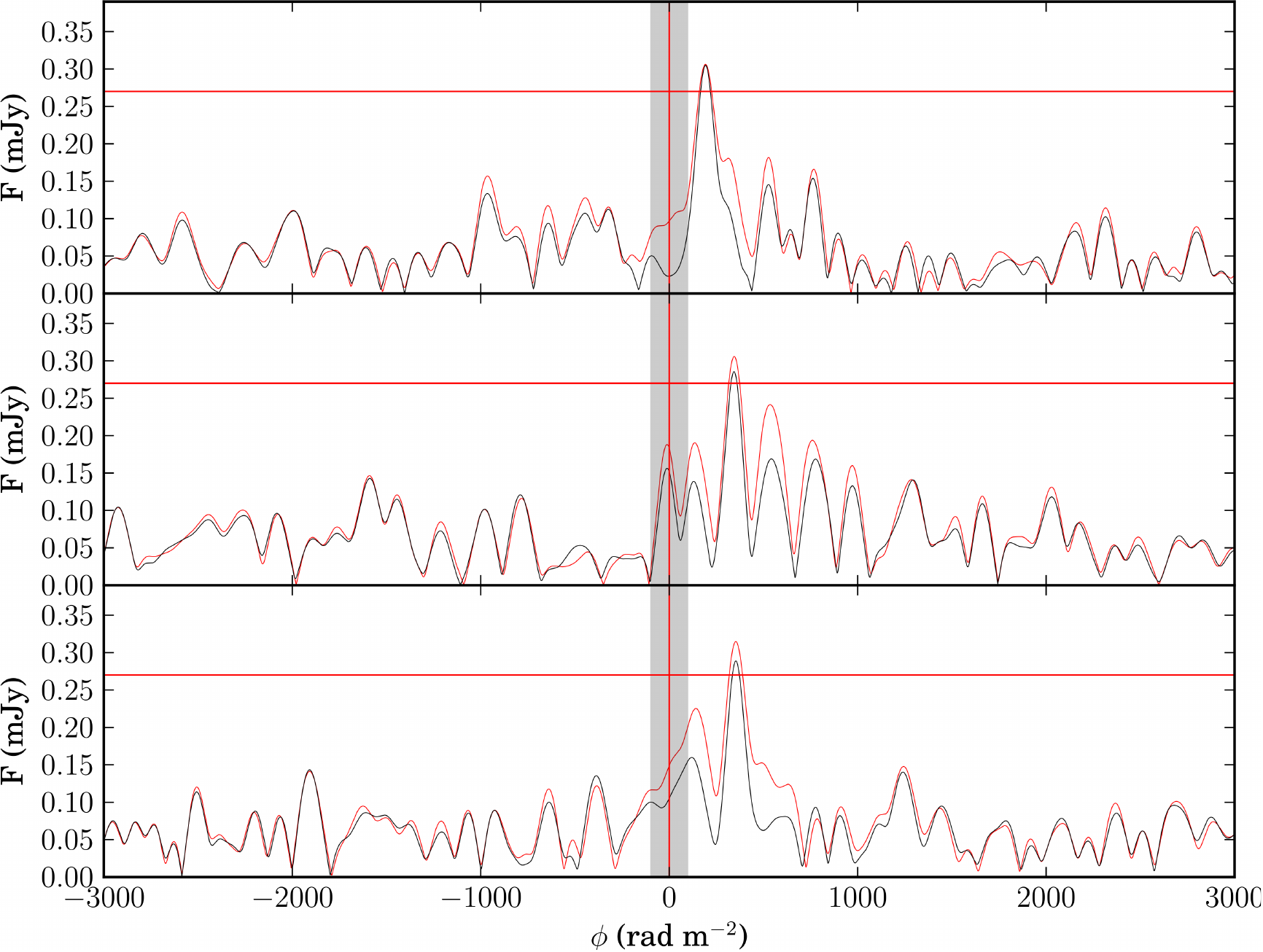}{1.0\linewidth}{(b)}}
\caption{(a) The dirty Faraday depth spectra for 30 randomly chosen profiles that are outside the SNR where no Stokes $I$ is observed at $\lambda$21 cm. (b) The red and black curves represent the dirty and clean Faraday depth spectra for three profiles that are outside the lowest contour in Figure \ref{fig:G46.8-0.3_RM-Map}.} 
\label{fig:G46_noise}
\end{figure}

\indent	In Figure \ref{fig:G46_noise} we present profiles that fall outside of the SNR.  The Faraday depth spectra in Figure \ref{fig:G46_noise} (a) are from 30 randomly chosen subregions where no peak was found to pass our detection criteria.  Here we can see that each peak falls below the detection threshold and that the average noise in Faraday space is approximately 0.05 to 0.1 mJy. These profiles are mainly noise, yet for some spectra we find elevated peaks in the Faraday depth range of $100 < \phi < 1000 \text{ rad m}^{-2}$, which is lower than the Faraday depth spread of $500 < \phi < 1500 \text{ rad m}^{-2}$ observed from extra-Galactic sources at $\ell \approx 46^\circ$ \citep{Shanahan2019}.  Figure \ref{fig:G46_noise} (b) are spectra from subregions outside the SNR which do not exhibit strong polarization, yet a peak is found to pass our detection criteria.  The polarized emission and elevation in the noise profiles could originate from diffuse polarized emission, however verification would require short-spacing observations.   

\begin{figure*}[htb!]
\gridline{\fig{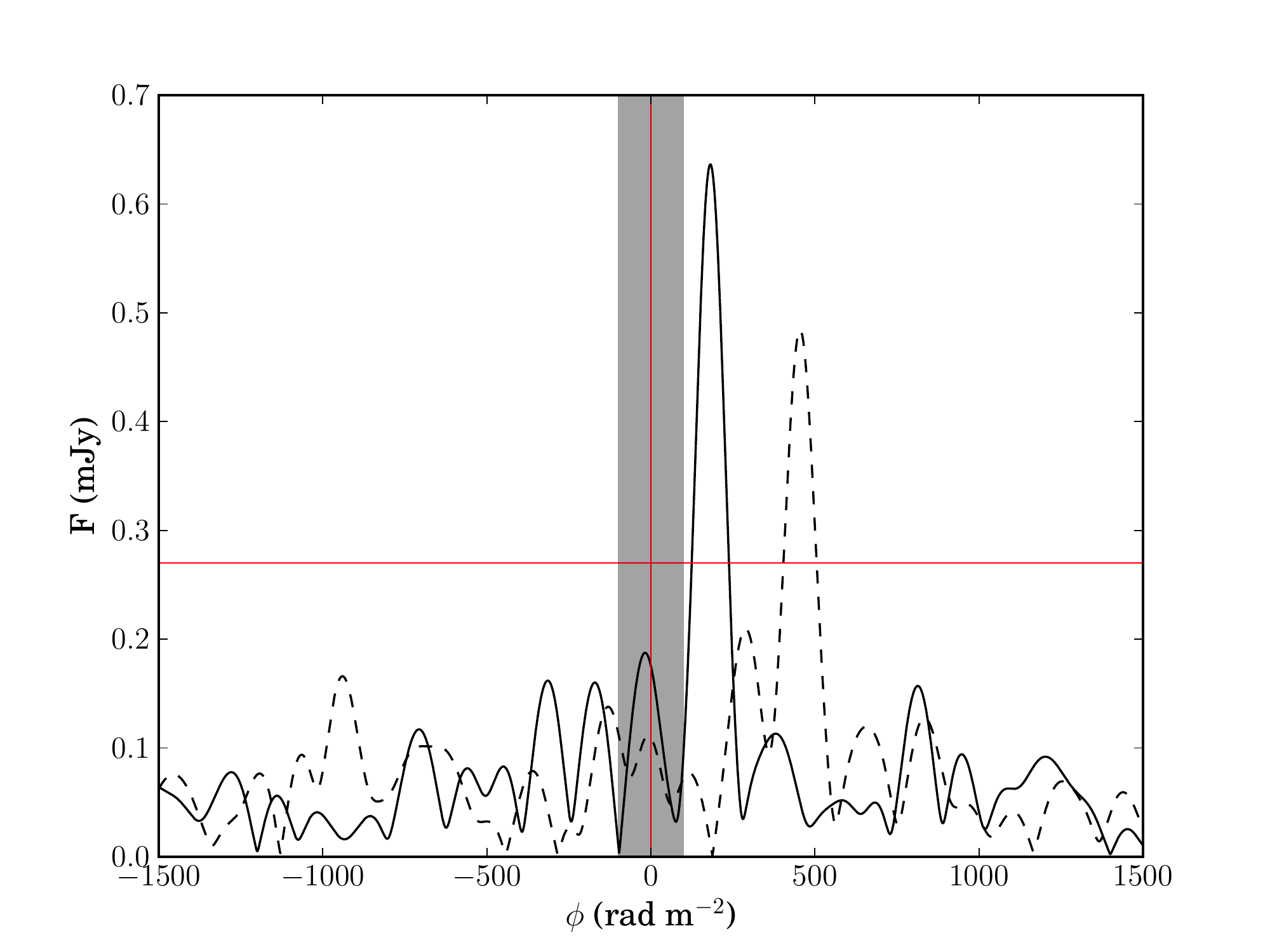}{0.5\textwidth}{(a)}
          \fig{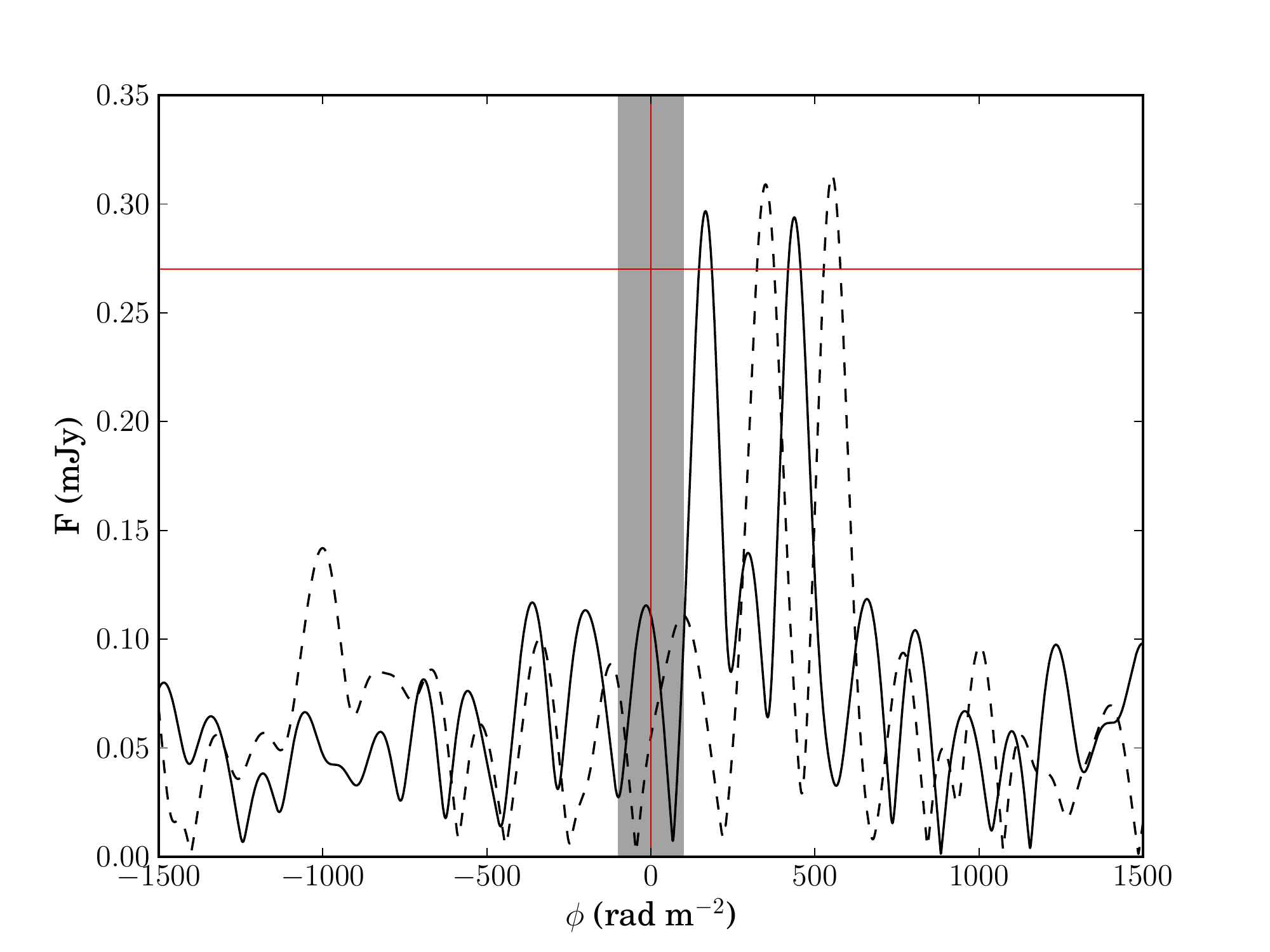}{0.5\textwidth}{(b)}}
\caption{ Sample Faraday depth spectra for G46.8-0.3.  Single component spectra are presented in (a), at $(\ell, b)=(46\fdg834, -0\fdg178)$, $\phi = 181.2 \pm 3.5 \text{ rad m}^{-2}$ (solid line), and at $(\ell,b)=(46\fdg746, -0\fdg323)$, $\phi = 454.4 \pm 5.2 \text{ rad m}^{-2}$ (dashed line).  Two component spectra are presented in (b), at $(\ell,b)=(46\fdg800, -0\fdg398)$, $\phi_{1} = 166.6 \pm 3.1 \text{ rad m}^{-2}$ and $\phi_{2} = 437.2 \pm 3.1 \text{ rad m}^{-2}$ (solid line), and at $(\ell,b)=(46\fdg817, -0\fdg382)$, $\phi_{1} = 353.6 \pm 2.3 \text{ rad m}^{-2}$ and $\phi_{2} = 552.2 \pm 2.3 \text{ rad m}^{-2}$ (dashed line).  The grey region indicates our Faraday depth rejection range of $|\phi| < 100 \text{ rad m}^{-2}$.  The red horizontal line indicates the detection threshold 0.27 mJy which is derived in Appendix \ref{appendix:DT}. } 
\label{fig:G46.8-0.3_FDspectra}
\end{figure*}

\indent	Displayed in Figure \ref{fig:G46.8-0.3_FDspectra} are examples of individual Faraday depth spectra at four locations in the SNR where we observe polarization detections.  The Faraday spectra in Figure \ref{fig:G46.8-0.3_FDspectra} (a) are from subregions where single-component Faraday rotation is observed.  If a single peak is found to pass our detection criteria, the location of the peak in Faraday space is the Faraday depth ($\phi$) and the amplitude is the polarized intensity ($P$) at the reference wavelength ($\lambda_0$) \citep{Brentjens2005}.  Figure \ref{fig:G46.8-0.3_FDspectra} (b) illustrates the Faraday depth spectra from subregions with two-component Faraday rotation, where each peak that passes our detection criteria is treated independently.  For Faraday depth spectra with two-component Faraday rotation, each peak that passes our detection criteria is treated independently in the same way as for single component Faraday rotation. 

\begin{figure*}[htb!]
\centering
   \centerline{\includegraphics[width=0.9\linewidth]{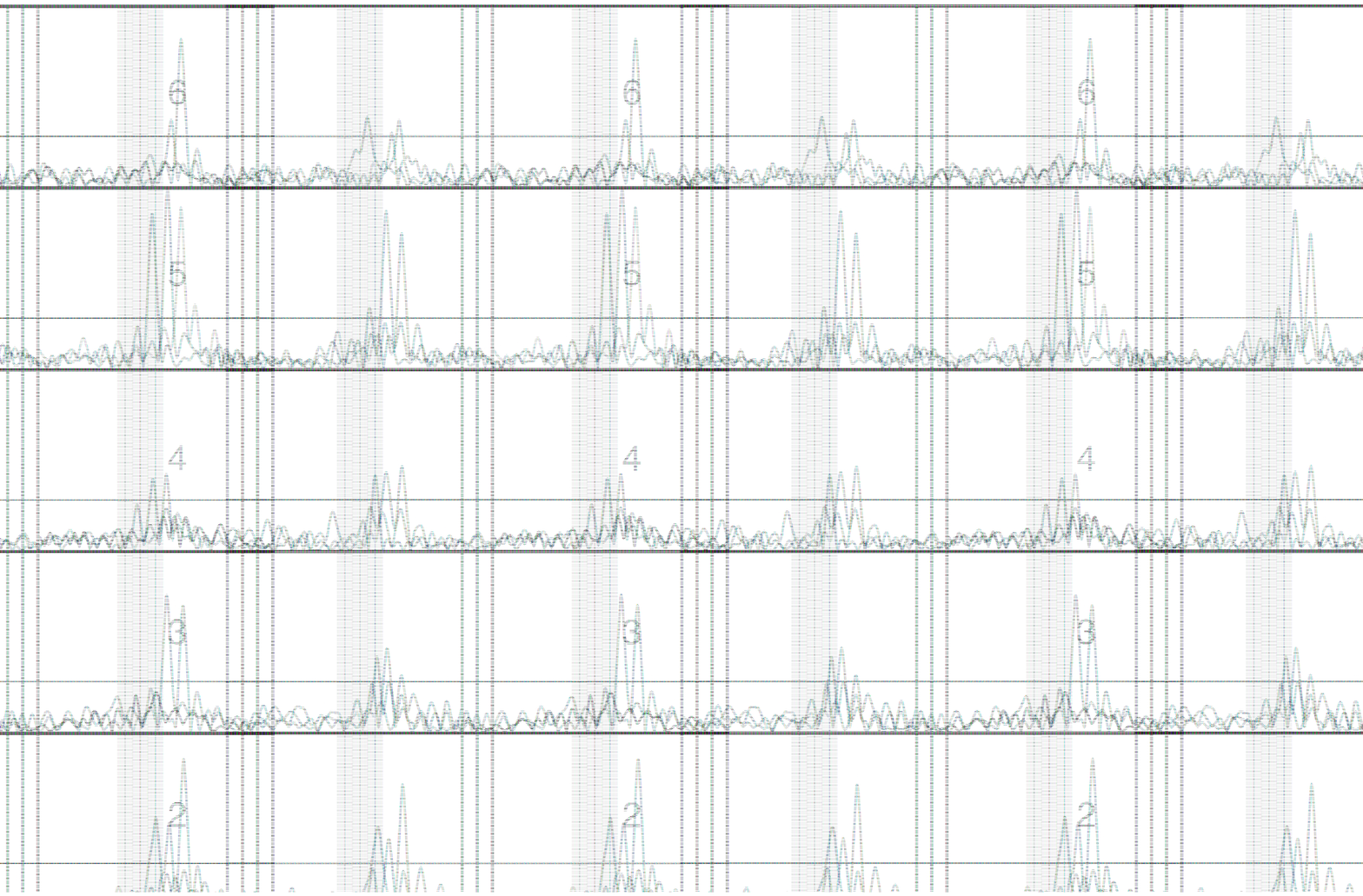}}
   \caption{Grid plot of the Faraday depth spectra for the red box seen in Figure \ref{fig:G46.8-0.3_RM-Map}.  The horizontal limits are $-1500 \text{ rad m}^{-2} < \phi < 1500 \text{ rad m}^{-2}$ and vertical axis limits are $0 < P < 0.95$ mJy.  The grey region illustrates our Faraday depth rejection range.  The red horizontal line is our detection threshold where $P = 0.27$ mJy.  Colours indicate spectra where zero (black), one (blue) or two (purple) peaks satisfy our detection criteria (see Appendix \ref{appendix:DT}).  The coordinates of the bottom left and top right corners are $(\ell, b)=(46\fdg802, 0\fdg377)$ and $(\ell, b)=(46\fdg777, -0\fdg352)$.  The physical size of this region is $2.36$ pc on each side.} 
   \label{fig:G46.8-0.3_RM_Grid2} 
\end{figure*}

\indent	To display the details of small scale polarization structure, Figure \ref{fig:G46.8-0.3_RM_Grid2} displays the individual Faraday depth spectra in a $6 \times 6$ pixel grid for the subregions within the region marked by a red square in Figure \ref{fig:G46.8-0.3_RM-Map}.  Specific spectra in the grid will be given a horizontal and vertical position label as marked in the margins of Figure \ref{fig:G46.8-0.3_RM_Grid2}.  For example the bottom right spectrum is position 6:1 and the upper left spectrum is position 1:6.  

\indent	In Figure \ref{fig:G46.8-0.3_RM_Grid2} we find a high detection density and a smooth gradient in Faraday depth across the region.  The subregions in this area display a decrease in Faraday depth from the bottom left (subregion 1:2) to the top right (subregion 6:6) with Faraday depths of $425.2 \pm 1.4 \text{ rad m}^{-2}$ and $329.2 \pm 0.9 \text{ rad m}^{-2}$, respectively, on scales of approximately 1\arcmin.  The differences in Faraday depth across larger scales are approximately $100 \text{ rad m}^{-2}$ but adjacent subregions differ by approximately $30 \text{ rad m}^{-2}$.  Although the variation in Faraday depth is smooth, large variation in polarized intensity is observed in the top rows as well as the bottom right columns.  Large variation in polarized intensity in contiguous positions is observed throughout the bottom region of the SNR.  Such variation is not observed in total intensity from the THOR+VGPS observations (see Figure \ref{fig:G46.8-0.3_THOR+VGPS}).  The observation of a Stokes $I$ counterpart to the structure in polarized intensity requires higher resolution observations.

\begin{figure*}[htb!]
\gridline{\rotatefig{0}{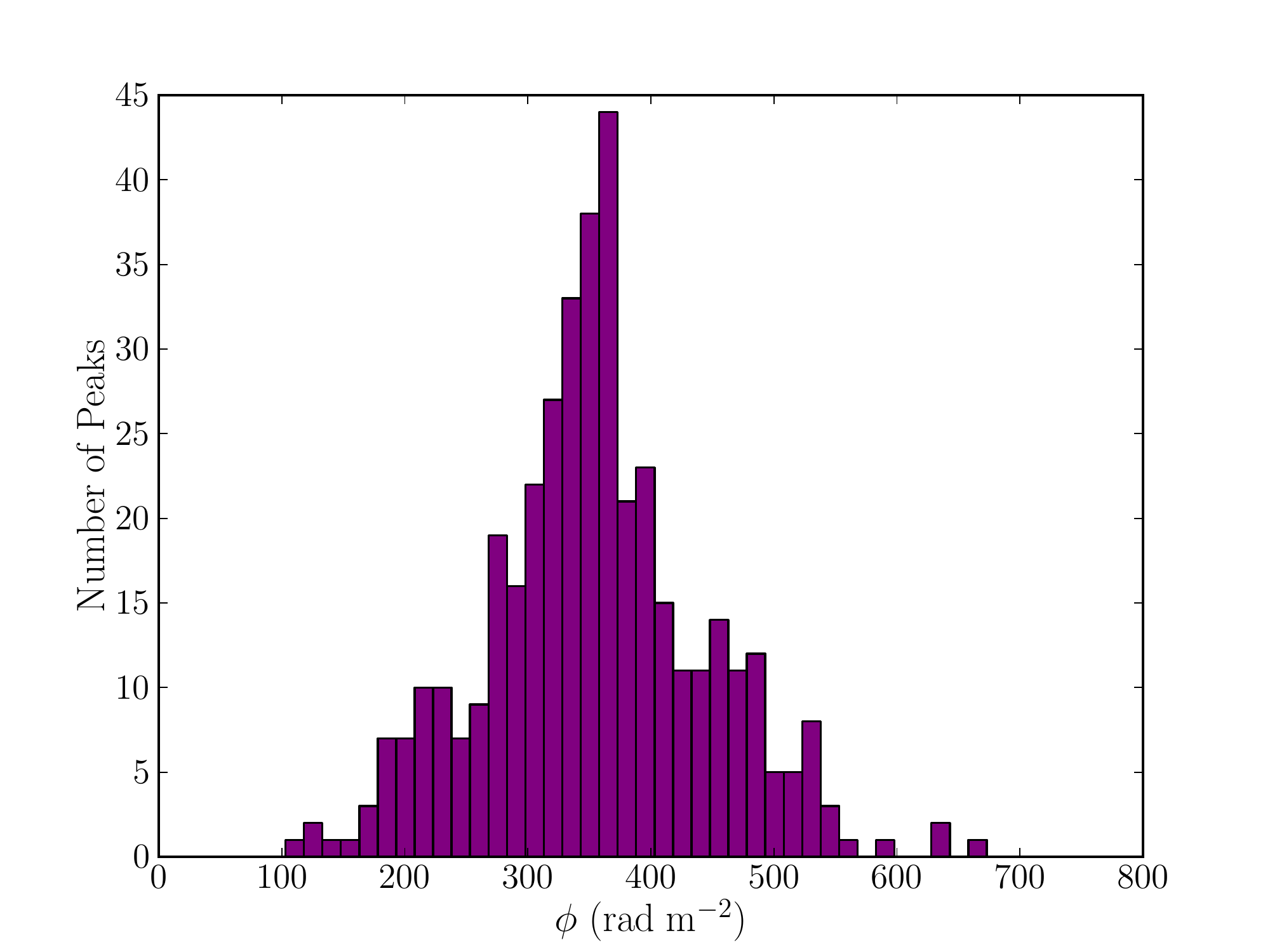}{0.5\textwidth}{(a)}
	\rotatefig{0}{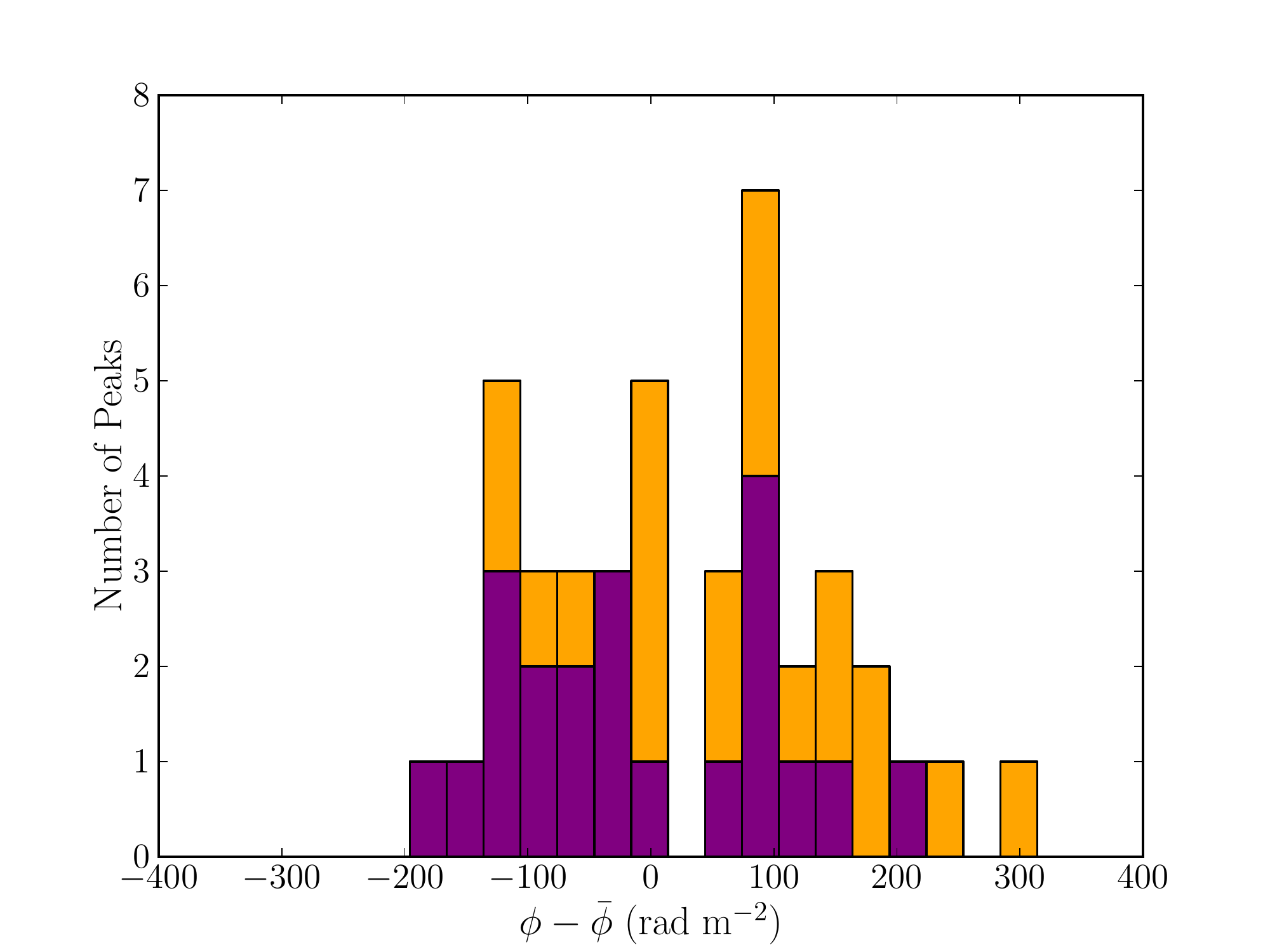}{0.5\textwidth}{(b)}}
\caption{(a) Distribution of Faraday depth for all single-component peaks that satisfy all detection criteria.  (b) Faraday depth complexity attributed to the internal structure of the SNR.  This histogram is found from subregions where two peaks are observed that satisfy all detection criteria.  The values have been shifted by the mean Faraday depth of $357 \text{ rad m}^{-2}$ found from all the subregions where single-component Faraday rotation was observed.  The purple bars represent the highest peak and the orange bars represent the secondary peak.}
\label{fig:G46.8_hist}
\end{figure*}

\indent	We present the Faraday depth distributions of single and two-component Faraday rotation in Figure \ref{fig:G46.8_hist}.  Figure \ref{fig:G46.8_hist} (a) is a histogram of peaks from single-component Faraday rotation that pass our detection criteria.  We find a mean Faraday depth of $354 \text{ rad m}^{-2}$ for all subregions with single-component Faraday rotation.  Figure \ref{fig:G46.8_hist} (b) shows the Faraday depth distribution for subregions with two-component Faraday rotation after subtracting the mean single-component Faraday rotation.  The purple bins indicate the first peaks and the orange bins indicate secondary peaks that satisfy our detection criteria.  In order to produce two-component Faraday rotation, two synchrotron emitting regions must be present with a Faraday screen separating them \citep{Brentjens2005}.  In the case of two-component Faraday rotation, a foreground Galactic Faraday screen cannot be the only medium causing Faraday rotation along the line of sight.  From Figure \ref{fig:G46.8_hist} (b) we observe a concentration of Faraday components in two ranges where the mean is separated by approximately $200 \text{ rad m}^{-2}$.  Of the 376 subregions where polarization is detected, 21 exhibit two-component Faraday rotation.

\indent	Inspection of Figure \ref{fig:G46.8-0.3_RM-Map} reveals that most of our detections are outside the brightest regions of the SNR.  We will see that here and in other SNRs in this study, the fractional polarization in bright regions is low.  We will include upper limits in areas of the SNR where these upper limits represent a fractional polarization $<1\%$.

\begin{figure*}[htb!]
\centering
   \centerline{\includegraphics[width=1\linewidth, angle=0]{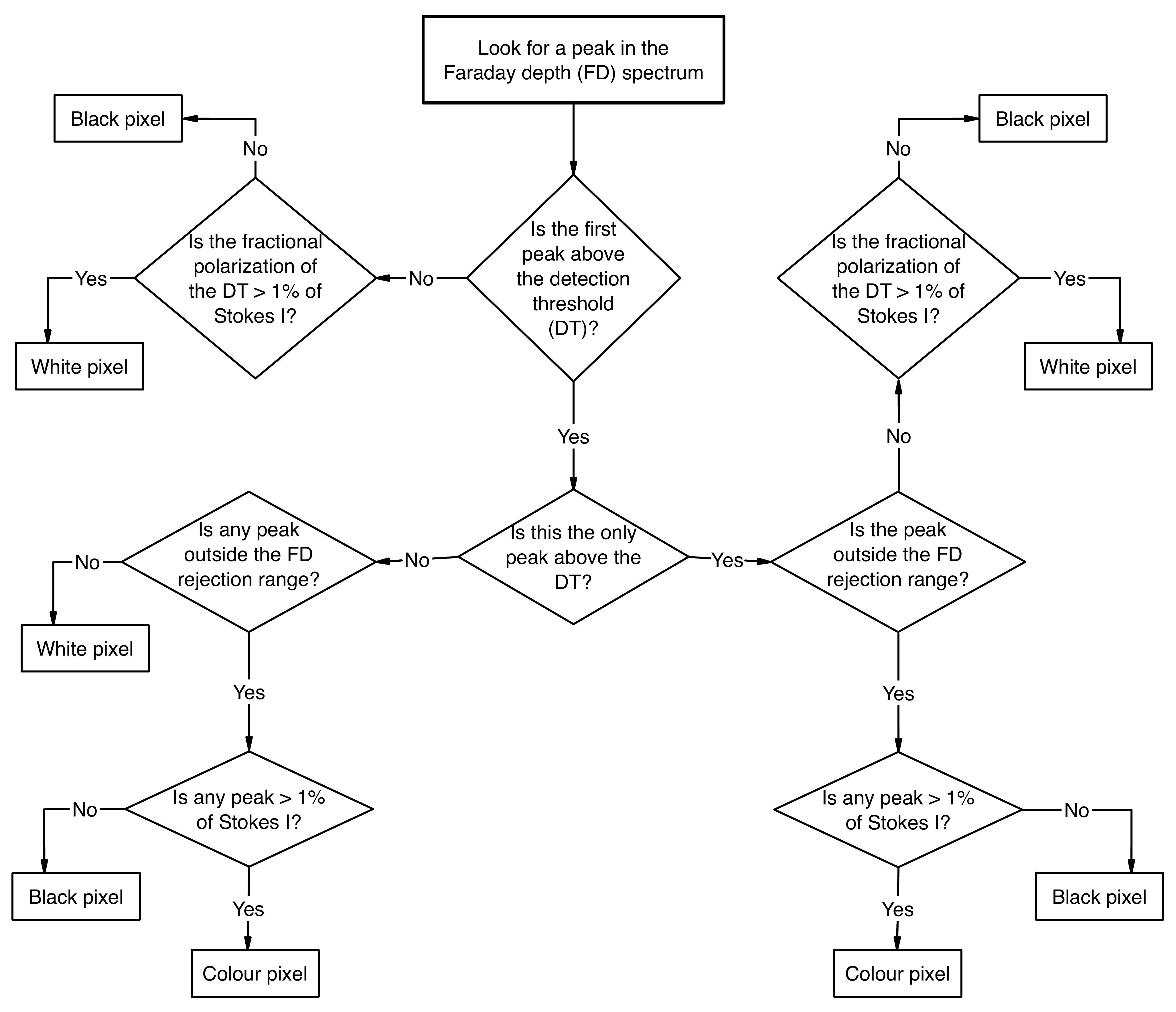}}
   \caption{The decision process used to determine if a pixel is black, white or coloured in the fractional polarization maps. A coloured pixel indicates a detection with colour indicating the fractional polarization.  A black pixel indicates an upper-limit to the fractional polarization $\le 1\%$.  A white pixel represents no information. }
   \label{fig:DecisionTree} 
\end{figure*}

\indent	The construction of a fractional polarization map that includes upper limits requires a detailed selection process, which we describe in Figure \ref{fig:DecisionTree}. A coloured pixel represents a detection where the values of fractional polarization of these subregions was calculated by $\Pi = P/(I - I_{\text{bg}})$, where $\Pi$ is the fractional polarization, $P$ is the polarized intensity from the Faraday depth spectrum, $I$ is the total intensity from the THOR+VGPS map and $I_{\text{bg}}$ is the average background Stokes $I$.  Due to missing short spacing from the image cubes, we use the Stokes $I$ from the THOR+VGPS image.  The black subregions represent either that the polarized emission or the detection threshold (which ever is higher) is $\le 1\%$ of Stokes $I$.  Therefore, these black subregions represent an upper-limit to polarization of $1\%$ of Stokes $I$.  A white pixel could either indicate a non-detection with an upper-limit above $1\%$ of Stokes $I$ or there is a peak above our detection threshold and $1\%$ of Stokes $I$ but falls within our Faraday depth rejection range (see Figure \ref{fig:DecisionTree} and Appendix \ref{appendix:DT}). 

\begin{figure*}[htb!]
\centering
   \centerline{\includegraphics[width=1\linewidth, angle=0]{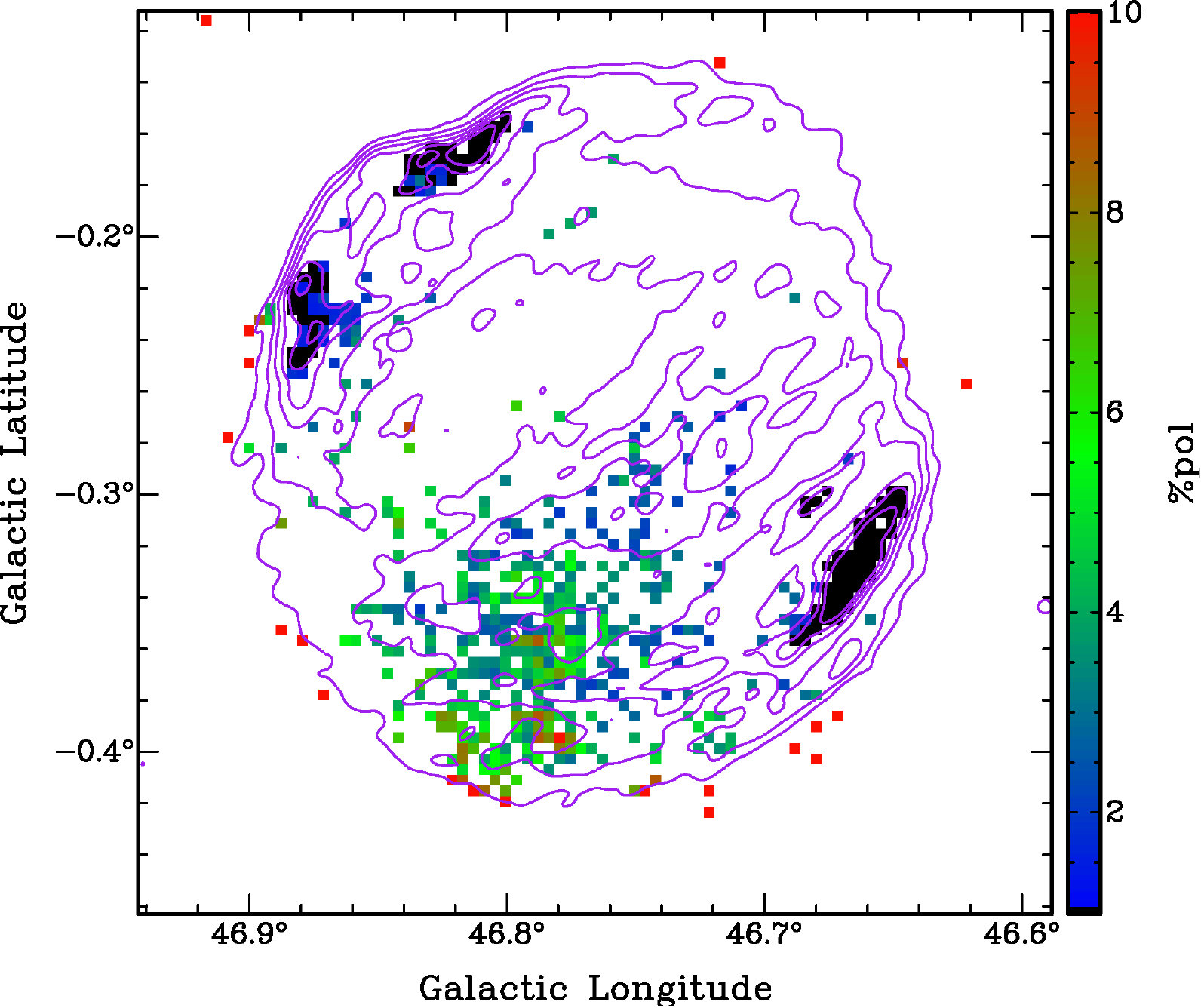}}
   \caption{Fractional polarization map of SNR G46.8$-$0.3.  Contours are from the THOR+VGPS map at 14, 19, 24, 29, 34 and 39 mJy/beam.  The coloured subregions were calculated by $\Pi = P/(I - I_{\text{bg}})$ where $\Pi$ is the fractional polarization, $P$ is the peak from the Faraday depth spectrum, $I$ is the Stokes $I$ from the THOR+VGPS map and $I_{\text{bg}}$ is the average background Stokes $I$.  The black subregions are upper limits for regions where polarized signal is heavily affected by leakage.  The criteria for these subregions is that $\Pi_{\text{upper}} = P_{\text{limit}}/(I - I_{\text{bg}}) < 1\%$, where $P_{\text{limit}}$ is the maximum from the noise profiles.}
   \label{fig:G46.8-0.3_PP-Map} 
\end{figure*}

\indent	Figure \ref{fig:G46.8-0.3_PP-Map} presents the fractional polarization map of SNR G46.8$-$0.3.  We find an average polarization of 4.2\% with a fractional polarization gradient across the SNR.  We observe $\Pi \approx 7\%$ in the bottom left part of the shell which gradually decreases towards the middle where $\Pi \approx 4 \%$.  In the lower part of the shell, where the detection density is highest, we observe no correlation between structure in Stokes $I$ and $\Pi$.  In Figure \ref{fig:G46.8-0.3_PP-Map} we observe that $\Pi \approx 2 \%$ surrounding regions of bright Stokes $I$ emission, where the brightest regions are dominated by upper limits.  We note in passing that the bright region in the lower right corner has the brightest polarized intensity in a $\lambda$6 cm image made with the Effelsberg telescope (W. Reich private communication), suggesting strong wavelength-dependent depolarization in this part of the SNR. 

\indent	We find polarization detections with high $\Pi$ around the edges of the SNR.  These subregions have weak polarized intensity, similar to examples shown in Figure \ref{fig:G46_noise} (b), but due to $I \approx I_{bg}$ we derive a high $\Pi$ using the method described in the caption of Figure \ref{fig:G46.8-0.3_PP-Map}.  Detections on the edge or just outside the SNR could be caused by variation in polarization angle over small-scales.  This implies that the subregions with large $\Pi$ near the edge of the SNR (see Figure \ref{fig:G46.8-0.3_PP-Map}) may not be directly related to polarized emission of the SNR, but rather an effect of the SNR acting as a Faraday screen on the diffuse polarized emission in the background.  We will return to this when discussing SNRs G43.3$-$0.2 and G39.2$-$0.3 in Sections \ref{sec:G43} and \ref{sec:G39}, respectively. 

\indent	As part of the Sino-German survey, \citet{Sun2011} analyzed the polarization of the SNR G46.8$-$0.3 at $\lambda6$ cm with an angular resolution of $9\farcm5$, where they find a polarized flux density of $595 \pm 32$ mJy with a fractional polarization of 8\%.  The polarization angles from \citet{Sun2011} took north to have an angle of $0^\circ$ with positive rotation in the clockwise direction.  We keep the same definitions used in \citet{Sun2011}.  We use $\phi$ from each subregion to derive the polarization angle at $\lambda6$ cm using $\theta_{\lambda6 \text{ cm}} = \theta_{\lambda21 \text{ cm}} + \phi (0.06^2 - 0.21^2)$.  The B-vector in the plane of the sky is derived by rotating the E-vector at $\lambda6$ cm by $90^\circ$.  Since RM-tools derives the polarization angle in RA and DEC, we rotate each vector by $62\fdg12$ to express the polarization angle in Galactic coordinates.  The same is done for the B-vectors shown in \citet{Sun2011}.

\begin{figure*}[htb!]
\centering
   \centerline{\includegraphics[width=0.85\linewidth, angle=270, bb=0 0 580 830]{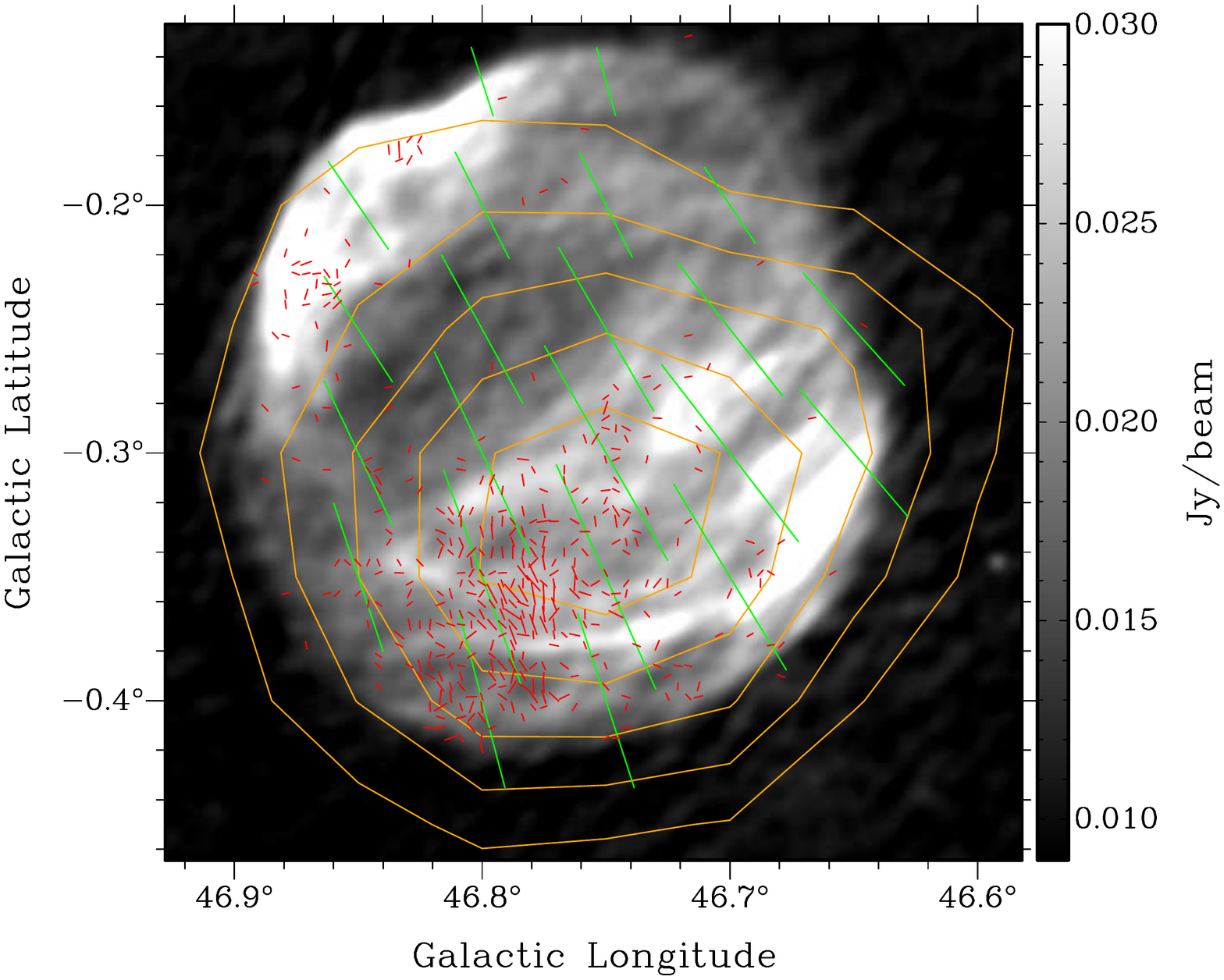}}
   \caption{THOR and VGPS total intensity map of the SNR G46.8$-$0.3.  Red vectors indicate B-vectors in the plane of the sky at $\lambda$21 cm wavelength rotated to $\lambda$6 cm using the observed Faraday depth. Green vectors represent the B-vectors at $\lambda$6 cm from the Sino-German survey.  The length of the vectors is proportional to the polarized intensity observed in the associated subregion.  Orange contours are the polarized intensity from the Sino-German survey at 23, 35, 45, 55 and 65 mK.}
   \label{fig:G46.8-0.3_B-vec} 
\end{figure*}

\indent	The peak of polarized emission at $\lambda$6 cm is offset from the geometrical centre of the SNR toward the lower shell where more polarized emission is observed at $\lambda$21 cm.  We have confirmed the astrometry of the Sino-German $\lambda$6 cm polarization map to be offset with respect to the geometrical centre in Stokes $I$ as shown in \citet{Sun2011} (X. Sun \& W. Reich 2022, private communication).  The mean angle of the B-vectors from the Sino-German and THOR survey are $152\fdg18 \pm 1\fdg69$ and $144\fdg78 \pm 2\fdg97$, respectively.  The polarization angles at $\lambda$6 cm derived from THOR observations show substantial spatial variation in polarization angle, indicating beam depolarization in the $\lambda$6 cm data.  The B-vectors shown in Figure \ref{fig:G46.8-0.3_B-vec} are rotated by $\phi\lambda^2 \approx 1$ radian for $\phi=300 \text{ rad m}^{-2}$ from their true angle at 0 wavelength. Thus, they are not a true representation of the magnetic field direction in the SNR.

\subsection{SNR G43.3$-$0.2}
\label{sec:G43}

\indent	G43.3-0.2 (W49B) is located in the W49 complex, including W49A which is a collection of numerous smaller HII regions \citep{Brogan2001}.  \citet{Moffett1994} found the SNR G43.3$-$0.2 to have a fractional polarization of $0.44\% \pm 0.06\%$ at $\lambda6$ cm with a resolution of $4\arcsec$ and suspect low polarization to be caused by some combination of internal Faraday depolarization, due to thermal gas densities, and beam depolarization, if the magnetic field is largely disordered.  \citet{Sun2011} do not report any polarization for SNR G43.3$-$0.2.  \citet{Lee2020} revise the velocity and distance to $63 \pm 2$ km s$^{-1}$ and $7.5 \pm 0.2$ pc, respectively.   \citet{Lacey2001} observe spatially resolved thermal absorption at 74 MHz and attribute significant attenuation towards the upper-right of the SNR to foreground absorption by the intervening HII regions.  \citet{Castelletti2021} adopt the electron density for the remnant to be $\sim500 \text{ cm}^{-3}$. 

\begin{figure*}[htb!]
\centering
   \centerline{\includegraphics[width=0.75\linewidth, angle=270, bb=0 0 570 850]{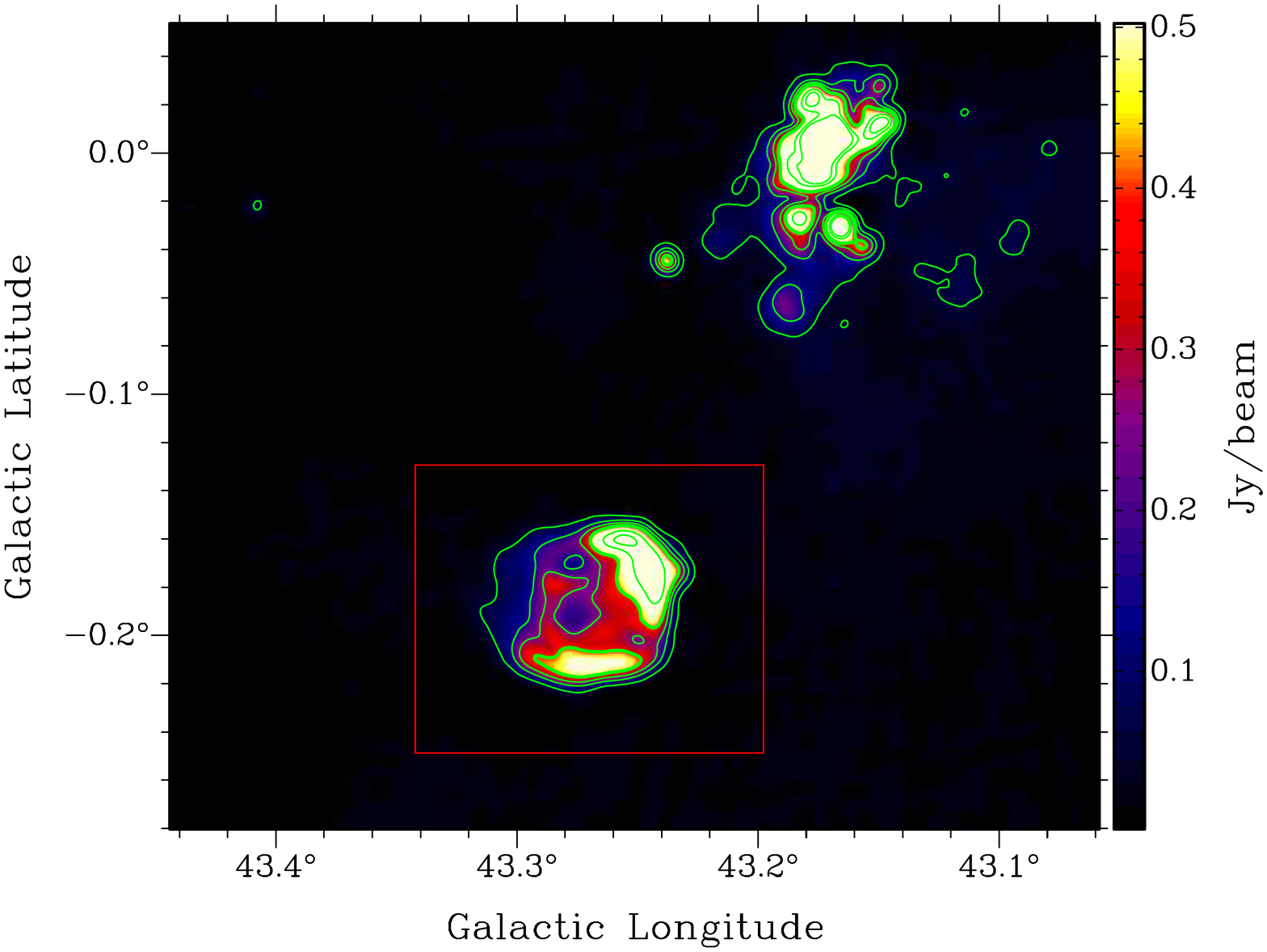}}
   \caption{THOR+VGPS map of SNR G43.3$-$0.2 at 1420 MHz.  Contour levels are Stokes $I$ (black) at 60, 160, 260, 400, 700 and 900 mJy/beam.  The red box indicates the region that a grid search was done for polarization and Faraday rotation.  Included in this image is the HII region W49N at G043.1658+00.0118 \citep{Dreher1984, Anderson2018}.  W49A is included in this figure for reference because it is used in the analysis of polarization leakage outlined in Appendix \ref{appendix:DT}.}
   \label{fig:G43.3-0.2_THOR+VGPS} 
\end{figure*}

\indent	In Figure \ref{fig:G43.3-0.2_THOR+VGPS} we show SNR G43.3$-$0.2 and the nearby HII region at $\lambda21$ cm in Stokes $I$ from the THOR+VGPS data.  The red box illustrates the region over which we performed RM synthesis in 1015 subregions. 

\begin{figure*}[htb!]
\gridline{\rotatefig{0}{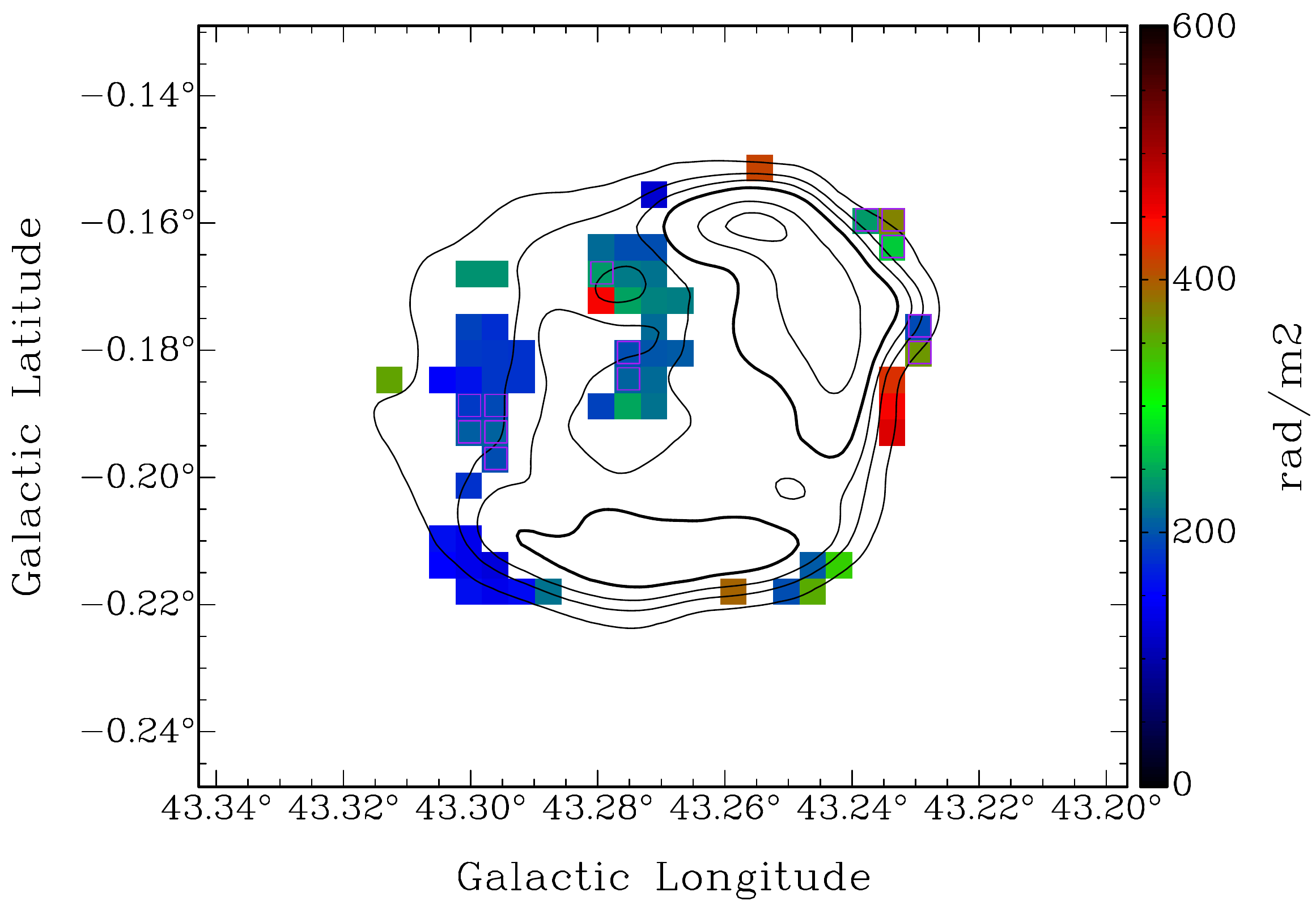}{0.8\textwidth}{(a)}}
\gridline{\rotatefig{0}{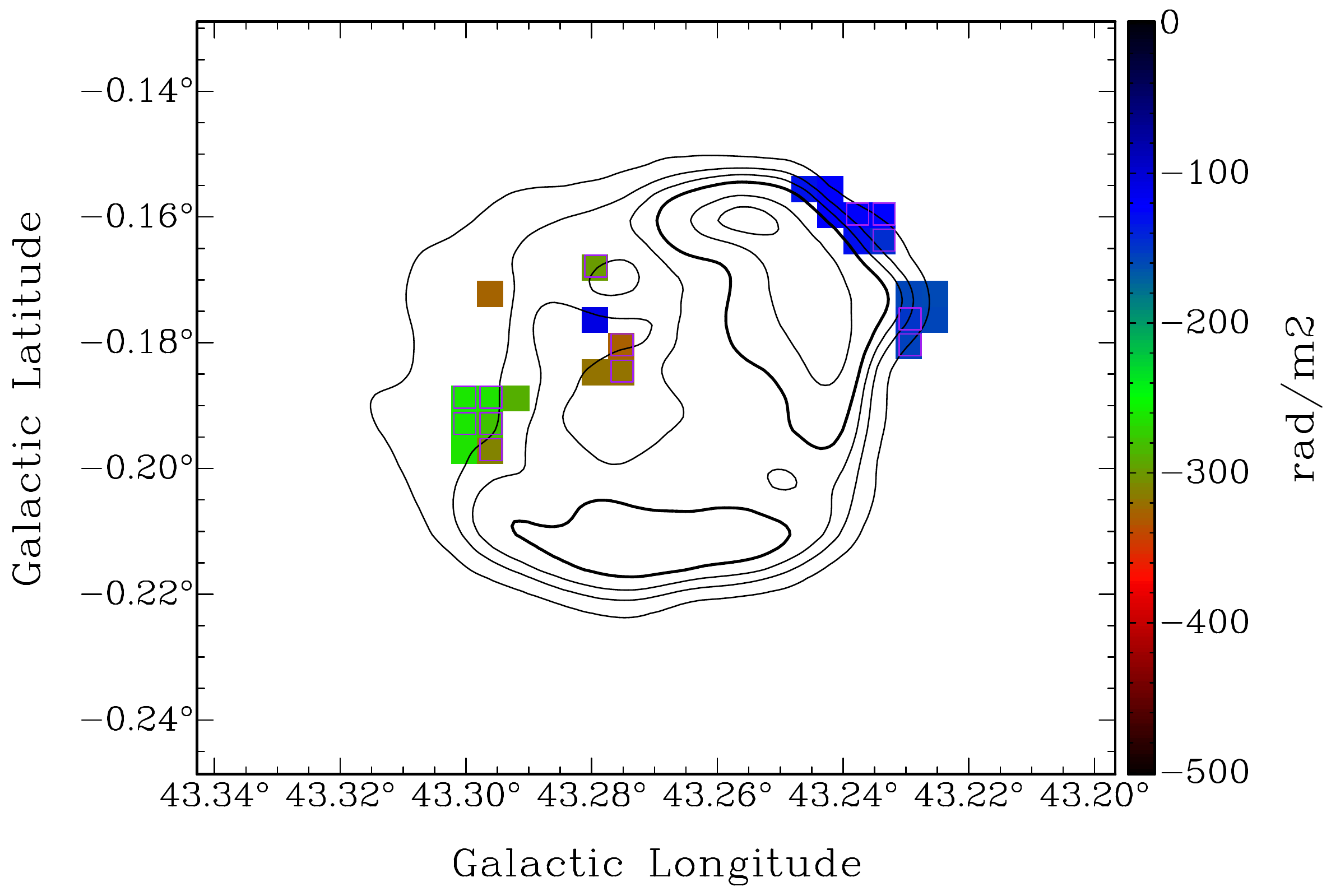}{0.8\textwidth}{(b)}}
\caption{Faraday depth map of the SNR G43.3$-$0.2 for the positive components (a) and negative components (b). Contour levels are 60, 160, 260, 400, 700 and 900 mJy/beam. The subregions with a purple edge indicate the locations with two-component Faraday rotation.  The Faraday rotation shown for subregions with two-component Faraday rotation are derived from the strongest peak.}
\label{fig:G43.3-0.2_RM-Map}
\end{figure*}

\indent	In Figure \ref{fig:G43.3-0.2_RM-Map} we present a Faraday depth map of this area.  Figures \ref{fig:G43.3-0.2_RM-Map} (a) and (b) present Faraday depth maps for positive and negative components respectively.  G43.3$-$0.2 is the only SNR in this study where peaks at negative Faraday depth were found to satisfy our detection criteria.  Of the 1015 subregions, 61 have a positive Faraday depth and 25 have a negative Faraday depth.  We do not observe negative Faraday depths in the polarization analysis of W49A, suggesting they are not a result of instrumental polarization (see Appendix \ref{appendix:DT}).  The negative Faraday depth components have an average fractional polarization of $1.5$\%.  The positive and negative components have average Faraday depths $ 231 \text{ rad m}^{-2}$ and $-212\text{ rad m}^{-2}$, respectively.  Most of our detections are found in lower surface brightness regions where the intensity is less than 260 mJy/beam (see Figure \ref{fig:G43.3-0.2_RM-Map}).

\begin{figure*}[htb!]
\centering
   \centerline{\includegraphics[width=0.8\linewidth, angle=0]{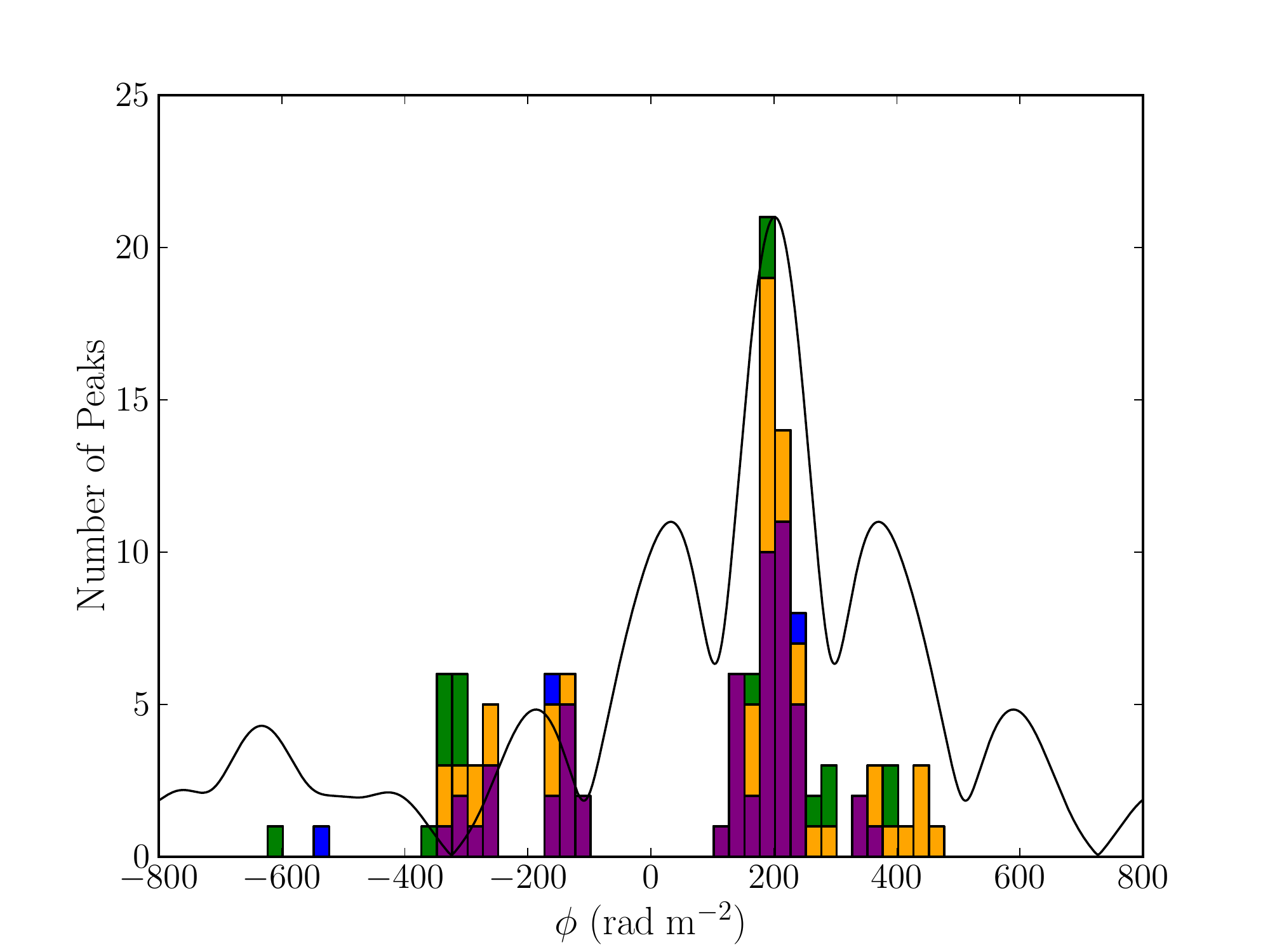}}
   \caption{Distribution of Faraday depth for all peaks above our detection threshold.  The purple, orange, green and blue bars represent the first, second, third and fourth highest peaks.  The black curve is the RMSF shifted to the average Faraday depth of the cluster with the highest peak and scaled to the peak value.}
   \label{fig:G43.3-0.2_histogram} 
\end{figure*}

\indent	The Faraday depth distribution shown in Figure \ref{fig:G43.3-0.2_histogram} indicates multi-component Faraday rotation within the SNR.  The average Faraday depths for these clusterings from right to left are 397 $\text{ rad m}^{-2}$, 201 $\text{ rad m}^{-2}$, -139 $\text{ rad m}^{-2}$ and -304 $\text{ rad m}^{-2}$.  Our detection criteria eliminate most instrumental polarization (see Appendix \ref{appendix:DT}), but weak polarized signal in the SNR could be affected.  

\indent	When investigating Faraday depth spectra with faint secondary components, the location of these components with respect to the sidelobes of RMSF should be considered.  In Figure \ref{fig:G43.3-0.2_histogram}, the brightest components (purple) cluster around $\phi = 200 \text{ rad m}^{-2}$.  The components around $\phi = 400 \text{ rad m}^{-2}$ coincide with the first sidelobes of the RMSF, the negative components do not coincide with any RMSF sidelobes.  In general, we do not find a correlation between secondary components and sidelobes of the RMSF.  

\begin{figure*}[htb!]
\centering
   \centerline{\includegraphics[width=0.8\linewidth, angle=0]{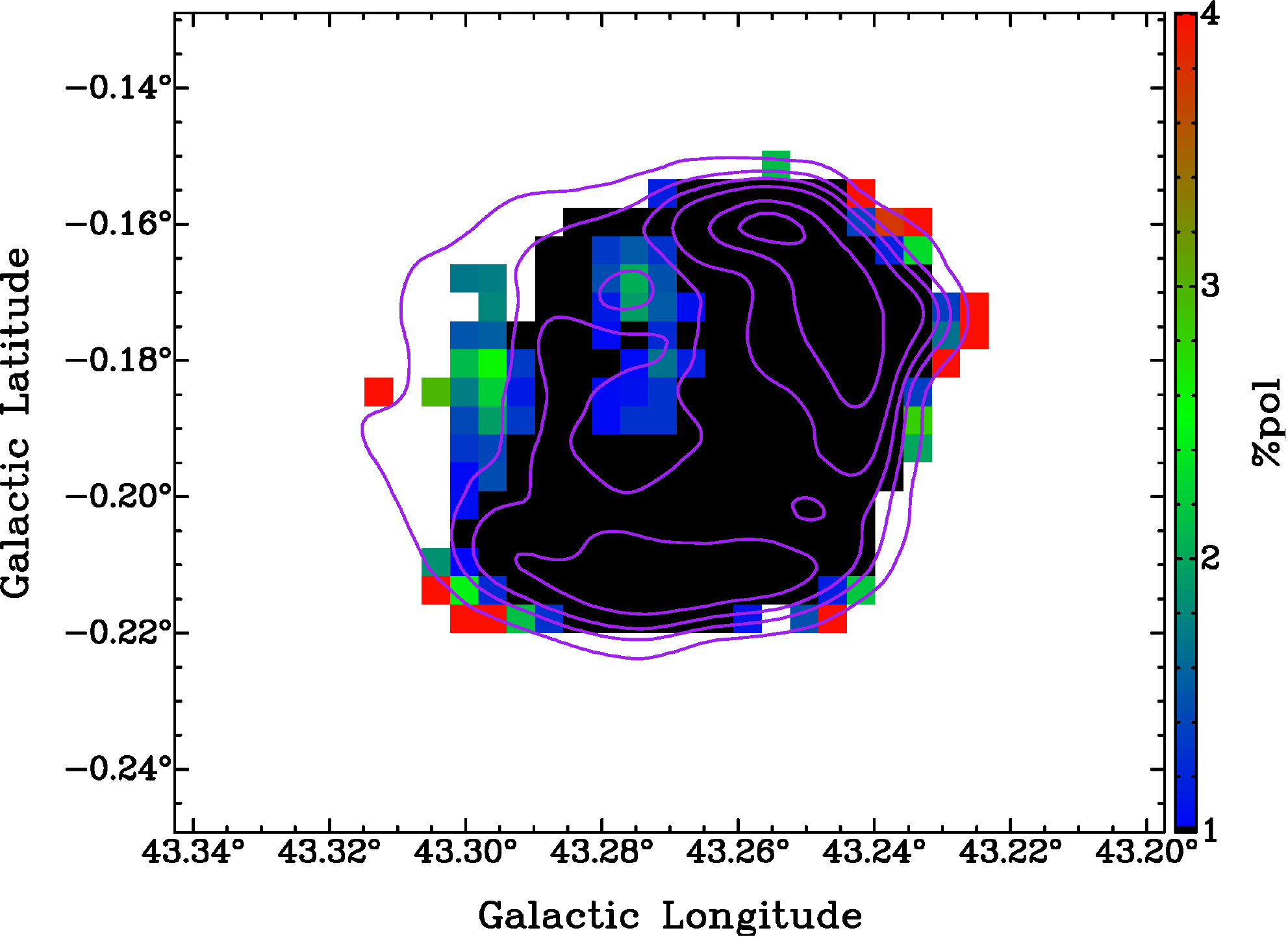}}
   \caption{Fractional polarization map of the SNR G43.3$-$0.2 with contour levels (green) at 60, 160, 260, 400, 700 and 900 mJy/beam from THOR+VGPS.  The process by which coloured and black subregions are defined is shown in Figure \ref{fig:DecisionTree}.  The pixel colour corresponds to the fractional polarization found from the highest peak in the Faraday depth spectrum for each subregion.}
   \label{fig:G43.3-0.2_PP-Map} 
\end{figure*}

\indent	In Figure \ref{fig:G43.3-0.2_PP-Map} we present the fractional polarization map of G43.3$-$0.2 derived from the highest peak that passes our detection criteria in the Faraday depth spectrum of each subregion.  The average fractional polarization for the SNR is found to be 2.9\%.  However, these averages include the subregions along the edges which have higher values, as discussed previously for G46.8$-$0.3. 

\subsection{SNR G41.1$-$0.3}
\label{sec:G41}

\begin{figure*}[htb!]
\centering
   \centerline{\includegraphics[width=1\linewidth, angle=0]{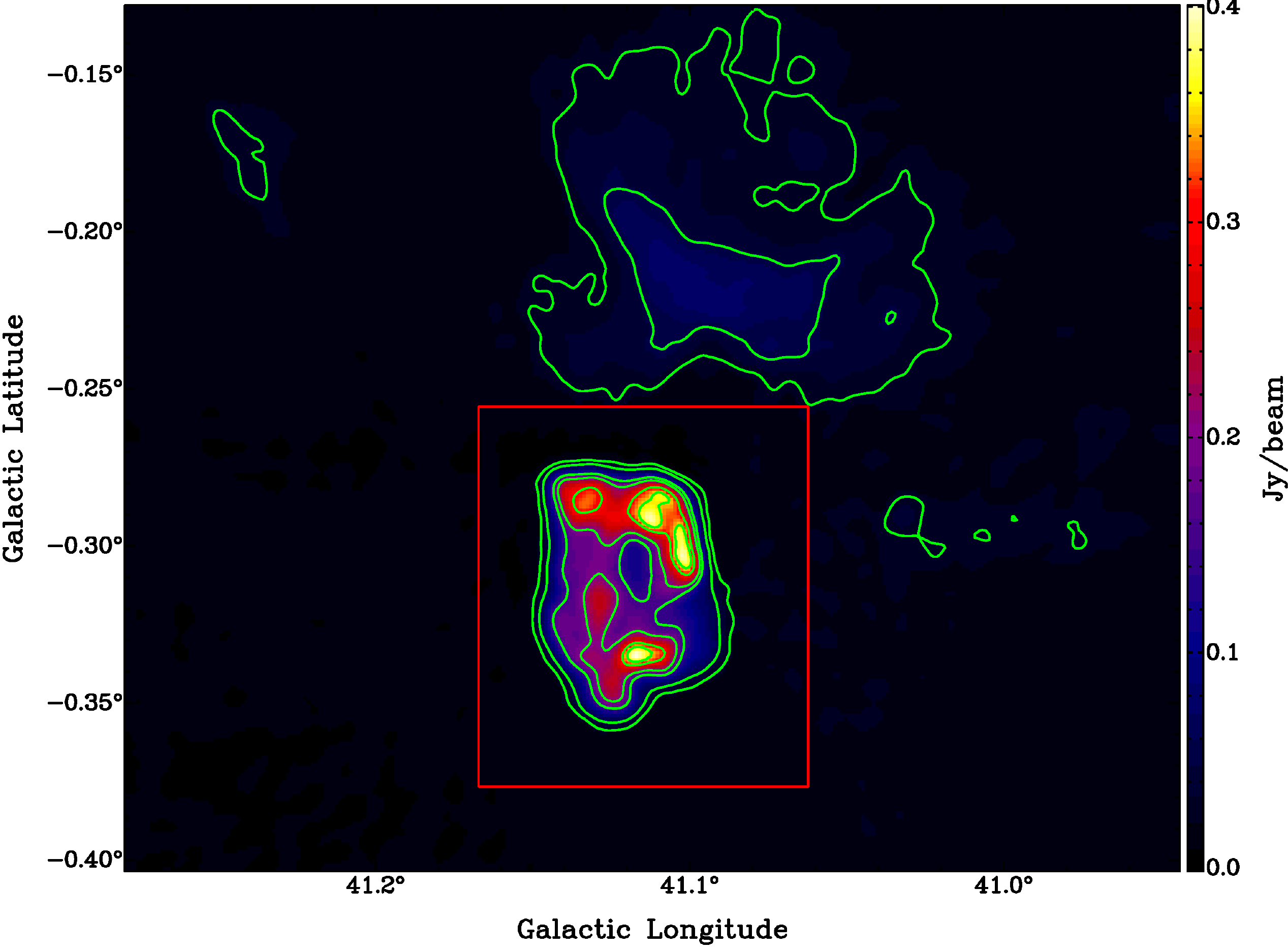}}
   \caption{THOR+VGPS map of SNR G41.1$-$0.3 at 1420 MHz.  Contour levels are for Stokes $I$ (green) at 25, 50, 150, 200, 300 and 350 mJy/beam.  The red box indicates the region that a grid search was done for polarization and Faraday rotation. The source outside of the red box is identified from THOR as the HII region G041.101-00.221 \citep{Anderson2018, Wang2018}. } 
   \label{fig:G41.1-0.3_THOR+VGPS} 
\end{figure*}

\indent	The SNR G41.1$-$0.3 (3C397) was first observed separately from the nearby HII region G$41.1-0.2$ by \citet{Caswell1975}, both of which can be seen in Figure \ref{fig:G41.1-0.3_THOR+VGPS}.  \citet{Castelletti2021} find thermal absorption attributed to ionized gas along the line of sight, possibly from extended HII region envelopes, along with a spectral index $\alpha = -0.356 \pm 0.013$. 

\begin{figure}[htb!]
\centering
   \centerline{\includegraphics[width=1\linewidth, angle=0]{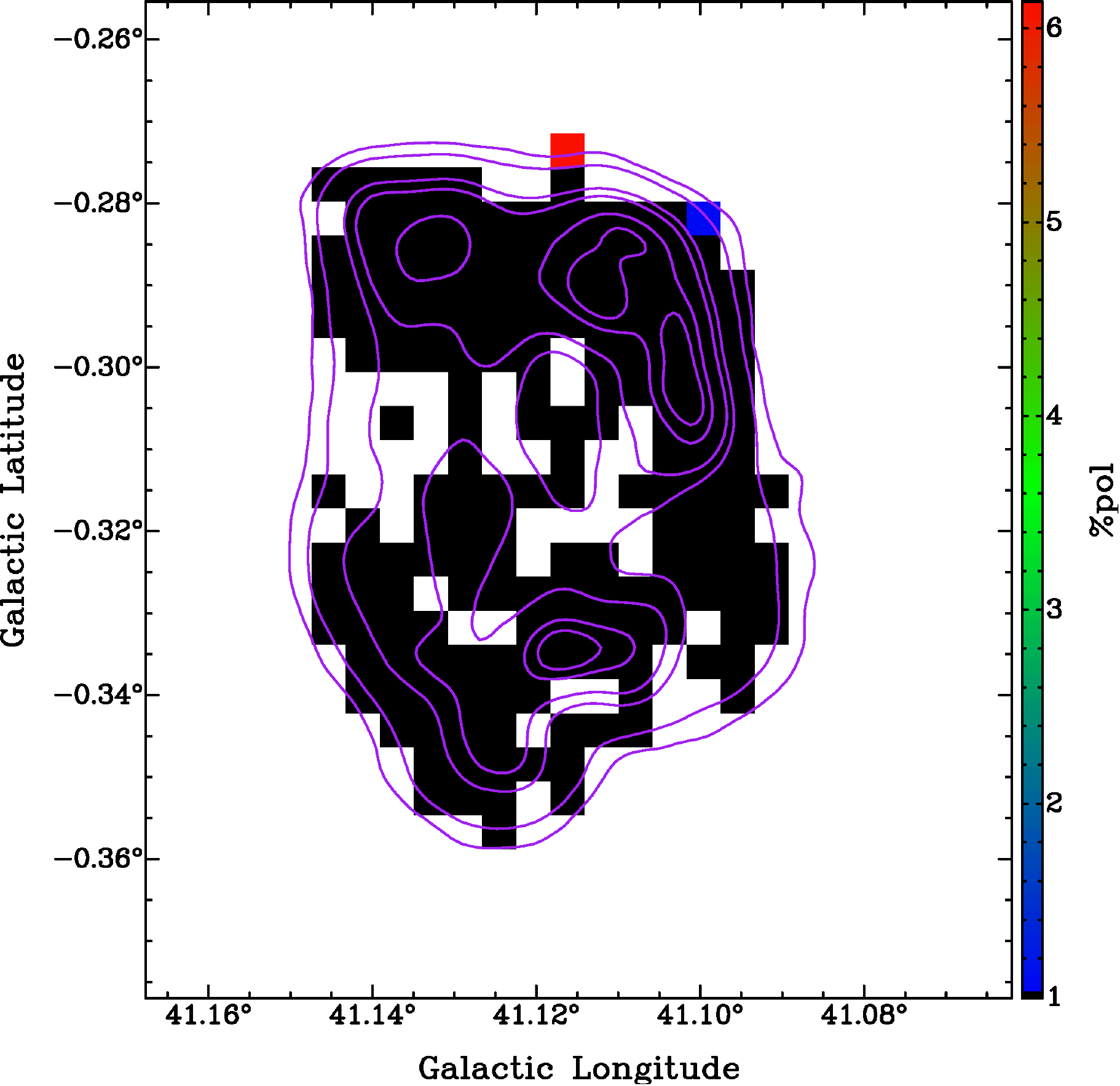}}
   \caption{Fractional polarization map of SNR G41.1$-$0.3.  The purple contours are from the THOR+VGPS map at 25, 50, 150, 200, 300 and 350 mJy/beam.  The coloured and black subregions are defined in the caption of Figure \ref{fig:G46.8-0.3_PP-Map}.} 
   \label{fig:G41.1_PP+SINO} 
\end{figure}

\indent	Our search for polarization within the red rectangle shown in Figure \ref{fig:G41.1-0.3_THOR+VGPS}, revealed only two subregions with polarized signal and mostly upper limits at 1\% of Stokes $I$ (Figure \ref{fig:G41.1_PP+SINO}).  The Sino-German survey reveals no polarized emission at $\lambda$6 cm \citep{Sun2011}.

\pagebreak

\subsection{SNR G39.2$-$0.3}
\label{sec:G39}

\begin{figure*}[htb!]
\centering
   \centerline{\includegraphics[width=0.7\linewidth, angle=270, bb=0 0 570 850]{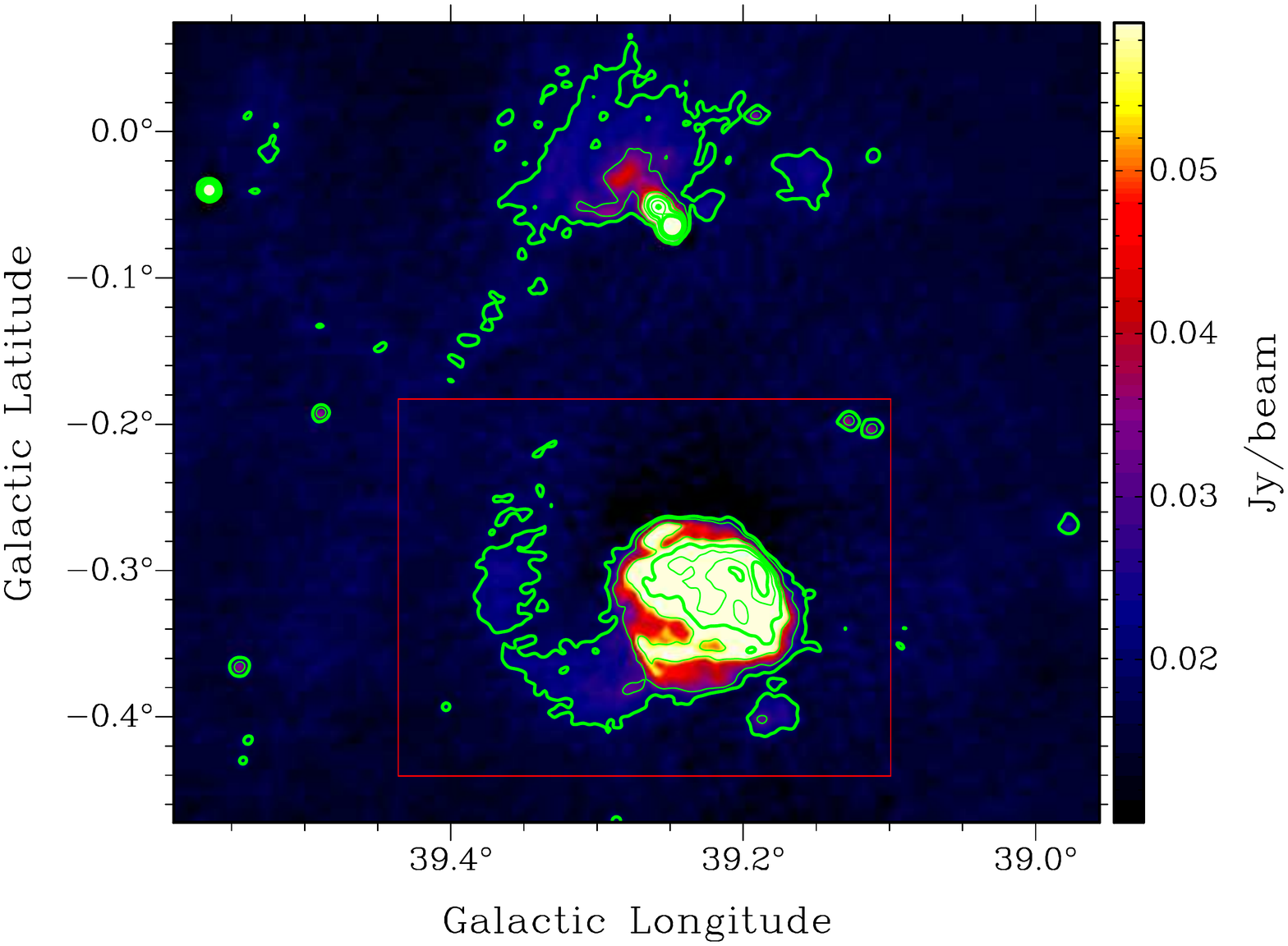}}
   \caption{THOR+VGPS map of SNR G39.2$-$0.3 at 1420 MHz.  Contour levels (green) at 20, 30, 55, 80, 110 and 140 mJy/beam.  The red box indicates the region in which a grid search was done for polarization and Faraday rotation. The extended source outside of the red box is identified from THOR as the HII region G039.280$-$00.003 \citep{Anderson2018,Wang2018}. }
   \label{fig:G39.2-0.3_THOR+VGPS} 
\end{figure*}

\indent	Shown in Figure \ref{fig:G39.2-0.3_THOR+VGPS} is G39.2$-$0.3 (3C396) and the surrounding area in Stokes $I$ from the combined THOR and VGPS data.  Using H$_2$ emission lines, \citet{Lee2020} find the distance to be $9.5 \pm 0.1$ pc and a $V_{LSR}$ of $56 \pm 2$ km s$^{-1}$.  At the top of Figure \ref{fig:G39.2-0.3_THOR+VGPS} is the bright HII complex NRAO 591 with a distance of 11.6$\pm$0.6 kpc and $V_{LSR}$ of 23 km s$^{-1}$ \citep{Watson2003} which is not related to the SNR.  \citet{Anderson2011} classifies the low surface brightness emission on lower right of the SNR to be the HII region G39.176-00.399 with a distance of 9.4$\pm$0.5 kpc and $V_{LSR}$ of 55.5 km s$^{-1}$.  The arc of extended emission from the lower left of the SNR is part of the HII region G39.294-00.311 which has an unknown distance, but \citet{Anderson2018} find a radio recombination line (RRL) velocity $V_{LSR} = $ 53.7 km s$^{-1}$. These two HII regions are possibly related to the SNR.  \citet{Olbert2003} observes non-thermal X-ray emission near the geometric centre of the SNR they attribute to a central pulsar wind nebula powered by an unobserved pulsar.  \citet{Castelletti2021} finds thermal absorption between the SNR shock and the surrounding medium which indicates local plasma in the foreground of the SNR. 

\begin{figure*}[htb!]
\centering
   \centerline{\includegraphics[width=0.9\linewidth, angle=0, bb=0 200 610 600]{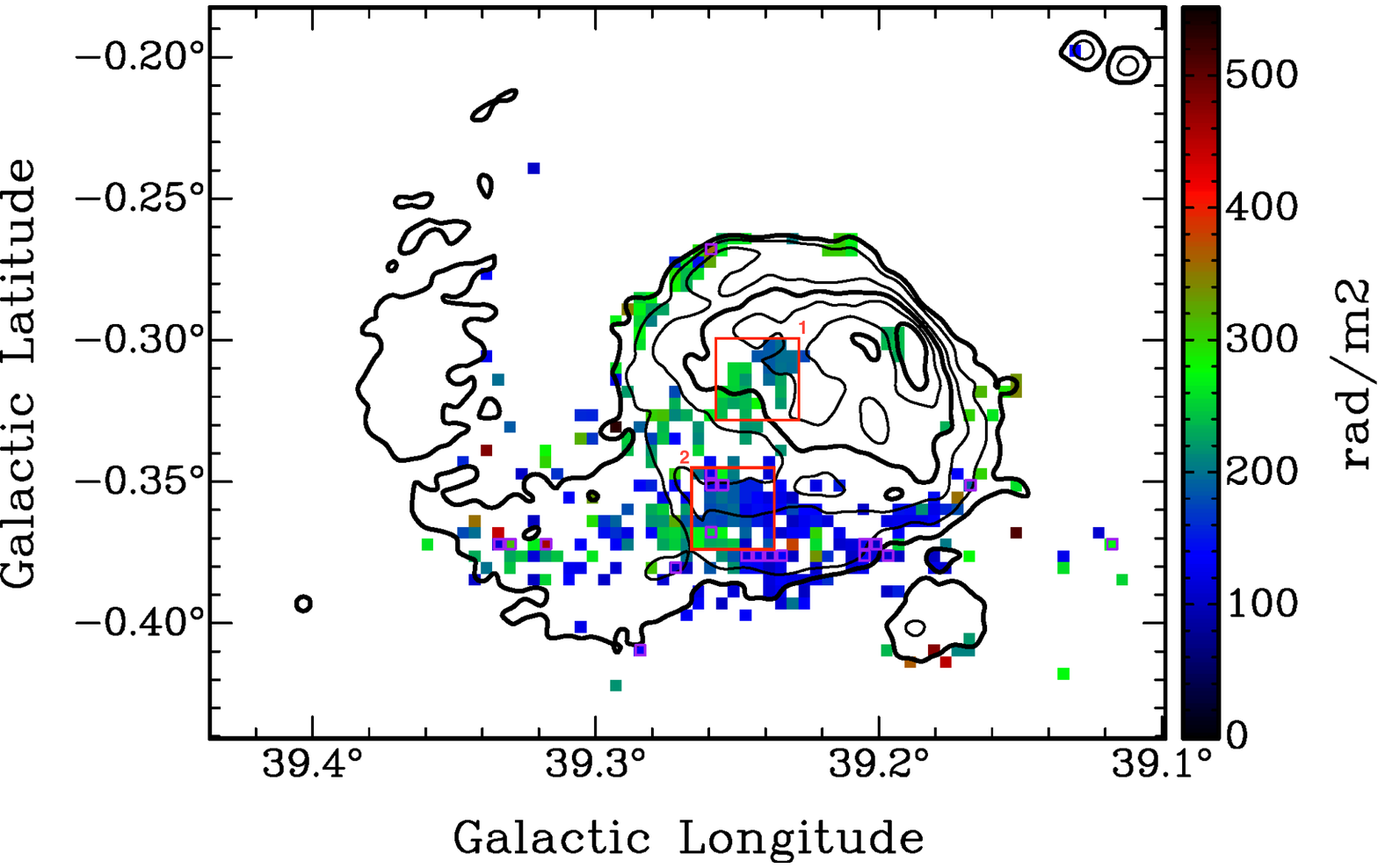}}
   \caption{Faraday depth map of SNR G39.2$-$0.3.  Contours are from the THOR+VGPS map at 20, 30, 55, 80, 110 and 140 mJy/beam.  The red boxes are identified as Region 1 and Region 2 where the Faraday depth spectra are shown as grid plots in Figures \ref{fig:G39.2-0.3_RM_Grid1} and \ref{fig:G39.2-0.3_RM_Grid2}. The subregions with a purple edge indicate the locations of two-component of Faraday rotation.  The Faraday rotation shown for subregions with two-component Faraday rotation is derived from the strongest peak.} 
   \label{fig:G39.2-0.3_RM-Map} 
\end{figure*}

\indent	In Figure \ref{fig:G39.2-0.3_THOR+VGPS}, the region outlined in red is where we performed our search for polarization.  The Faraday depth map for the SNR G39.2$-$0.3 is shown in Figure \ref{fig:G39.2-0.3_RM-Map} where the area covered in this map is the region outlined by the red rectangle in Figure \ref{fig:G39.2-0.3_THOR+VGPS}.  Of the 5022 subregions analyzed, 330 subregions exhibited signal that passed our detection criteria.  Most of the detections lie within the shell of the SNR, yet we observe subregions outside the SNR that pass our detection criteria.  Detections outside the SNR are mainly concentrated around the bottom side of the SNR and within the HII region G39.294-00.311.  We observe a gradient in Faraday depth from the bottom part of the SNR shell, with average Faraday depth $\bar{\phi} \approx 180 \text{ rad m}^{-2}$, to the upper left edge of the SNR where $\bar{\phi} \approx 250 \text{ rad m}^{-2}$.

\indent	Region 1 in Figure \ref{fig:G39.2-0.3_RM-Map} encompasses two patches of detections where the left has $\bar{\phi} \approx 250 \text{ rad m}^{-2}$ the right has $\bar{\phi} \approx 200 \text{ rad m}^{-2}$. In each patch we observe a variation of $\sim 30 \text{ rad m}^{-2}$ from the top subregions to the bottom.  In the patch on the right we observe a decrease in $\phi$ from top to bottom and in the patch on the right $\phi$ increases from top to bottom.

\begin{figure*}[htb!]
\centering
   \centerline{\includegraphics[width=0.9\linewidth]{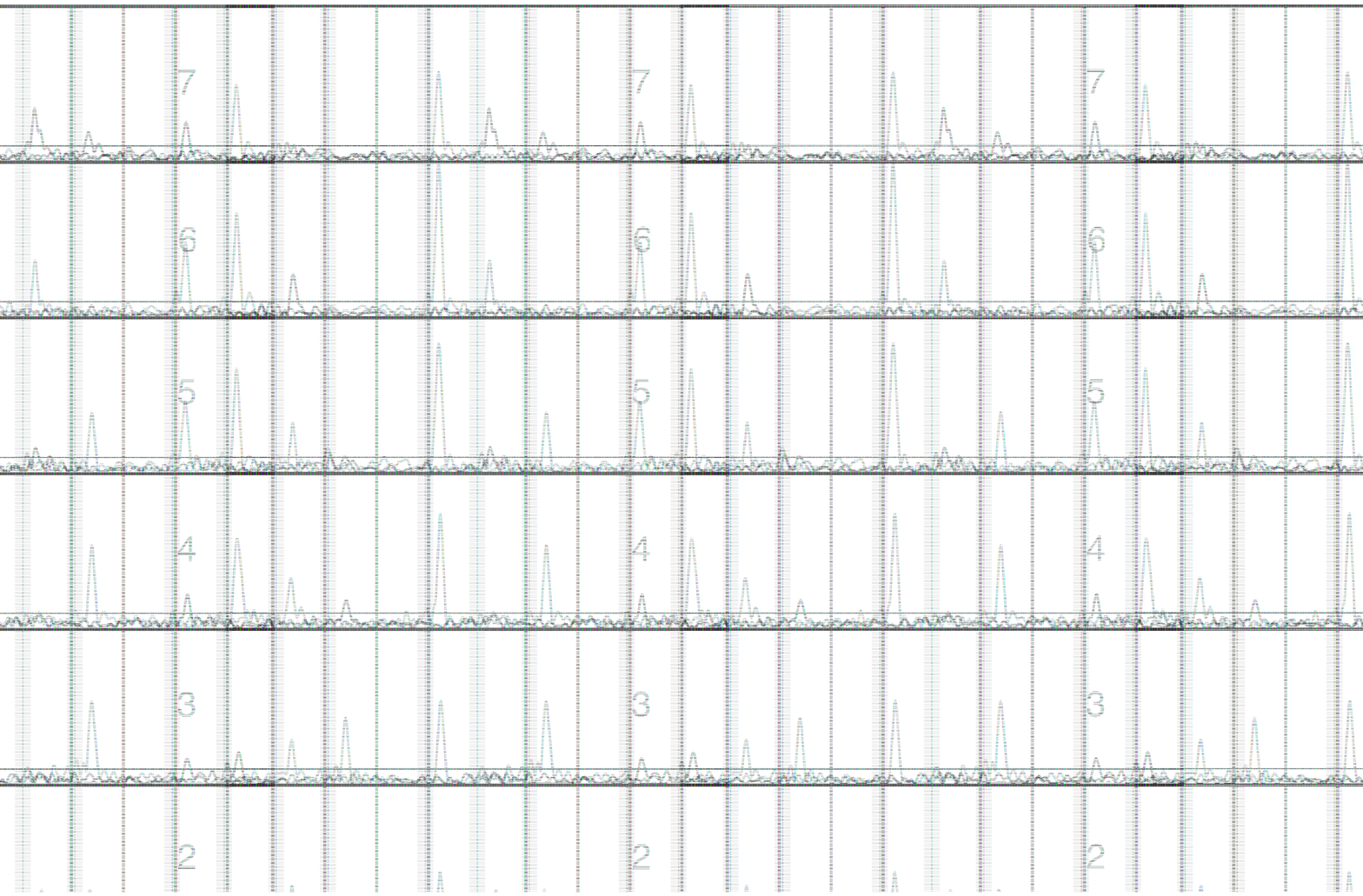}}
   \caption{Grid plot of the Faraday depth spectra for Region 1 seen in Figure \ref{fig:G39.2-0.3_RM-Map}.  The horizontal limits and grey region are the same as described in Figure \ref{fig:G46.8-0.3_RM_Grid2}. The red horizontal line is our detection threshold $P = 0.30$ mJy.  See Figure \ref{fig:G46.8-0.3_RM_Grid2} for description of the blue and black profiles.  The coordinates of the bottom left and top right corners are $(\ell, b)=(39\fdg257, -0\fdg329)$ and $(\ell, b)=(39\fdg227, -0\fdg299)$.  Using the distance of $9.5 \pm 0.1$ kpc derived in \citet{Lee2020}, the physical size of the region we show as a grid plot is $4.97 \pm 0.05$ pc on each side.}
   \label{fig:G39.2-0.3_RM_Grid1} 
\end{figure*}

\indent	In Figure \ref{fig:G39.2-0.3_RM_Grid1} we present a grid plot of the individual Faraday dispersion functions for the subregions in Region 1.  The subregions at location 3:5 and 4:6 have a difference of 65 rad m$^{-2}$ in Faraday depth over scales of $\sim 23 \arcsec$.  Figure \ref{fig:G39.2-0.3_RM_Grid1} also visualizes the contrast in polarized intensity between the two patches and the boundary region that separates them.  Polarized intensity is strongest in the centre of each patch and decreases to either side.  We observe no Stokes $I$ counterpart to these polarized structures. 

\begin{figure*}[htb!]
\centering
   \centerline{\includegraphics[width=0.9\linewidth]{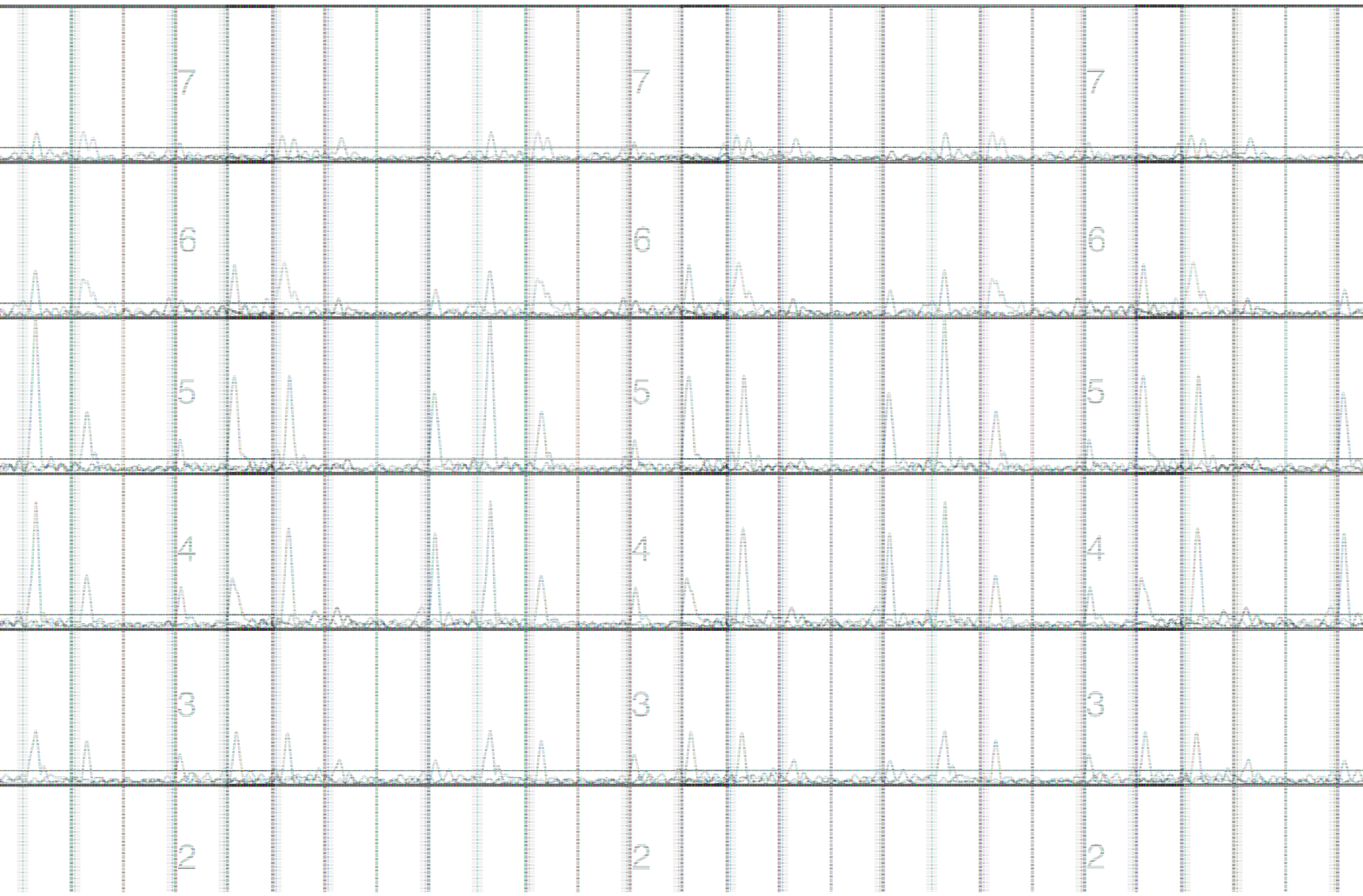}}
   \caption{Grid plot of the Faraday depth spectra for Region 2 seen in Figure \ref{fig:G39.2-0.3_RM-Map}.   The horizontal limits and grey region are the same as described in Figure \ref{fig:G46.8-0.3_RM_Grid2}. The red horizontal line is our detection threshold $P = 0.30$ mJy.  See Figure \ref{fig:G46.8-0.3_RM_Grid2} for description of the blue, purple and black profiles. The coordinates of the bottom left and top right corners are $(\ell, b)=(39\fdg265, -0\fdg374)$ and $(\ell, b)=(39\fdg235, -0\fdg344)$.  Using the distance of $9.5 \pm 0.1$ kpc derived in \citet{Lee2020}, the physical size of the region we show as a grid plot is $4.97 \pm 0.05$ pc on each side.} 
   \label{fig:G39.2-0.3_RM_Grid2} 
\end{figure*}

\indent	The individual Faraday depths for the subregions in Region 2 of Figure \ref{fig:G39.2-0.3_RM-Map} are shown in Figure \ref{fig:G39.2-0.3_RM_Grid2}.  This region covers a section of the southern arm of the SNR shell observed in Stokes $I$.  The Faraday depth spectra for this region show higher polarized intensity that correlates with the approximate middle of the southern arm where the local Stokes $I$ is brightest.  This is noteworthy because the fractional polarization structure is not correlated with Stokes $I$ structure in other parts of the SNR.  In this lower part of the SNR, a smooth variation of Faraday depth is observed over arc minute scales within the southern arm, yet abrupt variation of $\sim100\text{ rad m}^{-2}$ is observed on the edges of the arm on $16\arcsec$ scales.  Within the arm, the Faraday depth is the highest where the polarized intensity is brightest with an average value of 192 rad m$^{-2}$.  To the left of this region of local maxima (panel 4:5 in Figure \ref{fig:G39.2-0.3_RM_Grid2}), the Faraday depth and polarized intensity decrease until the polarized signal vanishes.  Within the arm on the right side, as the polarized intensity decreases, so does the Faraday depth by $\sim$70 rad m$^{-2}$.  This decrease within the southern arm happens gradually over scales of $\sim1\arcmin$.  Outside of the arm on the lower side, a change in Faraday depth by $\sim$100 rad m$^{-2}$ is observed in adjacent subregions on scales of $16\arcsec$ (see panels 2:3, 3:3, 2:2 and 3:2 in Figure \ref{fig:G39.2-0.3_RM_Grid2}).  

\indent	Two-component Faraday rotation is observed in four subregions at positions 2:2, 2:6, 2:7 and 3:6 in Figure \ref{fig:G39.2-0.3_RM_Grid2}.  These subregions are all located at positions on the edge of the southern arm.  The first and second peaks from the Faraday depth spectra at 2:2 are at 250 rad m$^{-2}$ and 145 rad m$^{-2}$.  The separation between peaks of three subregions on the upper side of the arm at 2:6, 2:7 and 3:6 are 64 rad m$^{-2}$, 145 rad m$^{-2}$ and 153 rad m$^{-2}$ respectively.  The peaks at lower Faraday depths are similar to the values within the arm and the peaks with higher Faraday depths are similar to values outside the arm.  

\begin{figure*}[htb!]
\gridline{\rotatefig{0}{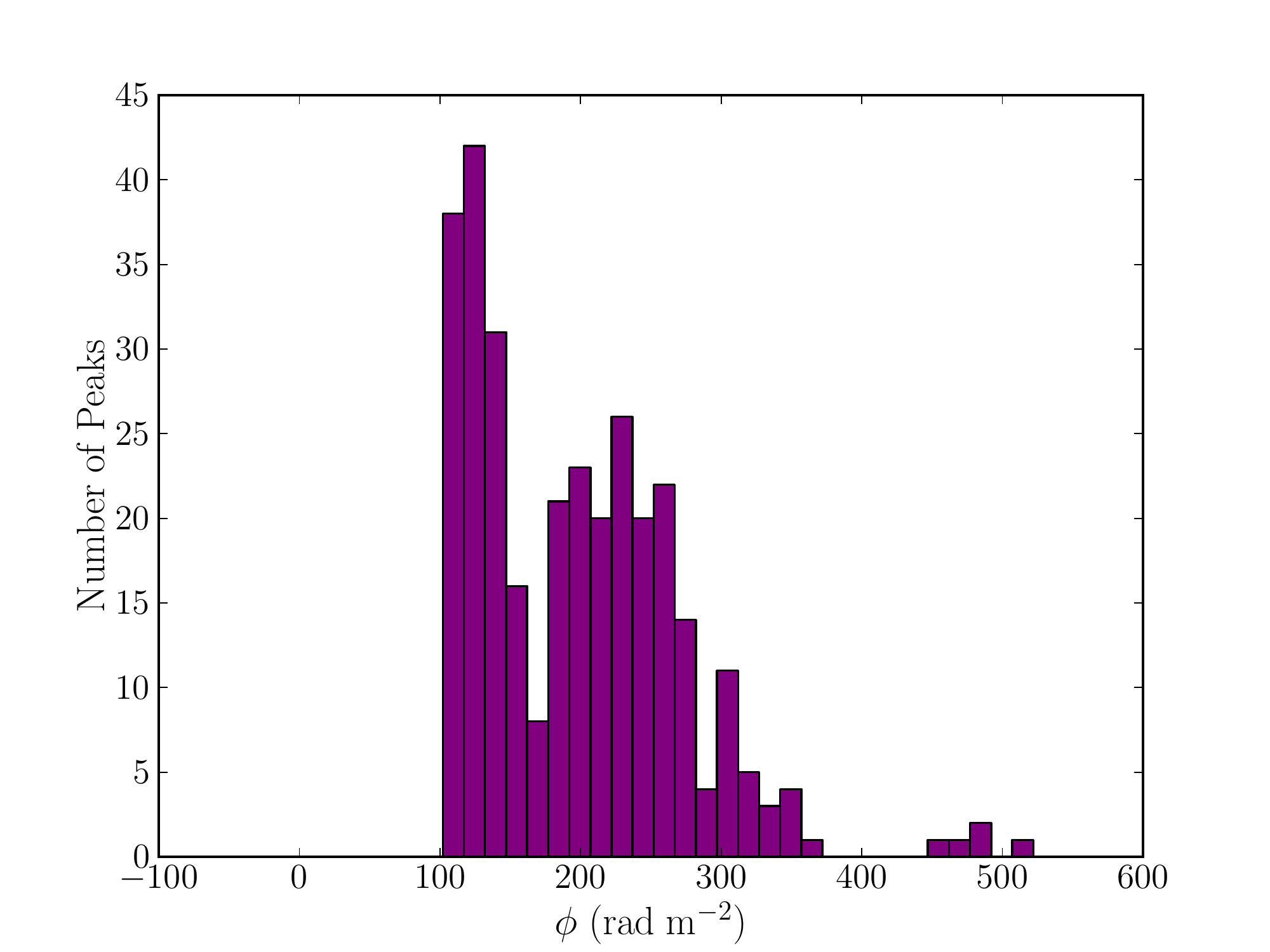}{0.5\textwidth}{(a)}
	\rotatefig{0}{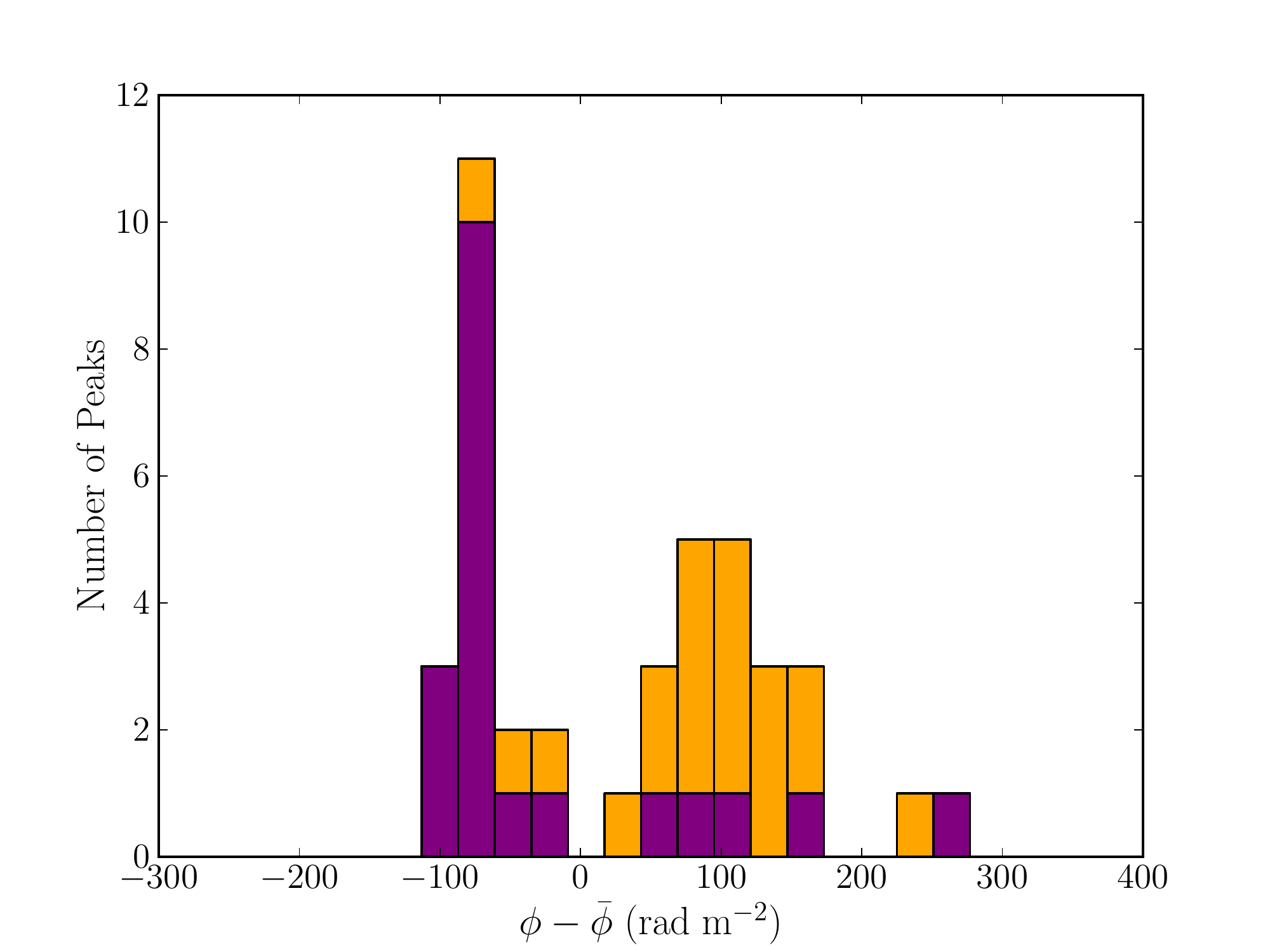}{0.5\textwidth}{(b)}}
\caption{(a) Faraday depth distribution for all subregions with single-component Faraday rotation.  This distribution has a mean Faraday depth $\bar{\phi} =  197 \text{ rad m}^{-2}$. (b) Faraday depth distribution for subregions where two-component Faraday rotation is observed.  The values have been shifted by $\bar{\phi}$  The purple bars represent the highest peak and the orange bars represent the secondary peak.}
\label{fig:G39.2_hist}
\end{figure*}

\indent	Figure \ref{fig:G39.2_hist} (a) presents the Faraday depth distribution for single-component peaks that satisfy our detection criteria.  From this distribution there are two ranges in Faraday depth where the density of detections are highest which are centred around the Faraday depths $\sim$130 rad m$^{-2}$ and $\sim$230 rad m$^{-2}$.  The majority of the subregions that contribute to the large number of peaks found around $\sim$130 rad m$^{-2}$ are located in the southern part of the SNR shell.  The subregions around the edges of the SNR shell and the middle (Region 1 in Figure \ref{fig:G39.2-0.3_RM-Map}) are the main contributors to the concentration of peaks around $\sim$230 rad m$^{-2}$.  The subregions with higher Faraday depths are scattered in areas outside the SNR.

\indent	Of the 330 subregions where polarization is detected, 20 exhibited two-component Faraday rotation.  The mean Faraday depth of all single-component peaks is $\bar{\phi} =197\text{ rad m}^{-2}$ and the mean Faraday depth of all two-component peaks is $228\text{ rad m}^{-2}$.  Figure \ref{fig:G39.2_hist} (b) presents the Faraday depth distribution of subregions with two-component Faraday rotation.  These detections have been shifted by the mean Faraday depth of subregions with single-component Faraday rotation $\bar{\phi}$.  

\indent	We observe a bimodal Faraday depth distribution of a narrow peak at lower Faraday depths and broad peak at higher Faraday depths for both single-component and two-component subregions. The similar Faraday depth distribution found for subregions with single-component and two-component Faraday depth spectra signifies that the physical environment causing the concentration of peaks at a lower and higher Faraday depth are the same.  

\begin{figure*}[htb!]
\gridline{\rotatefig{0}{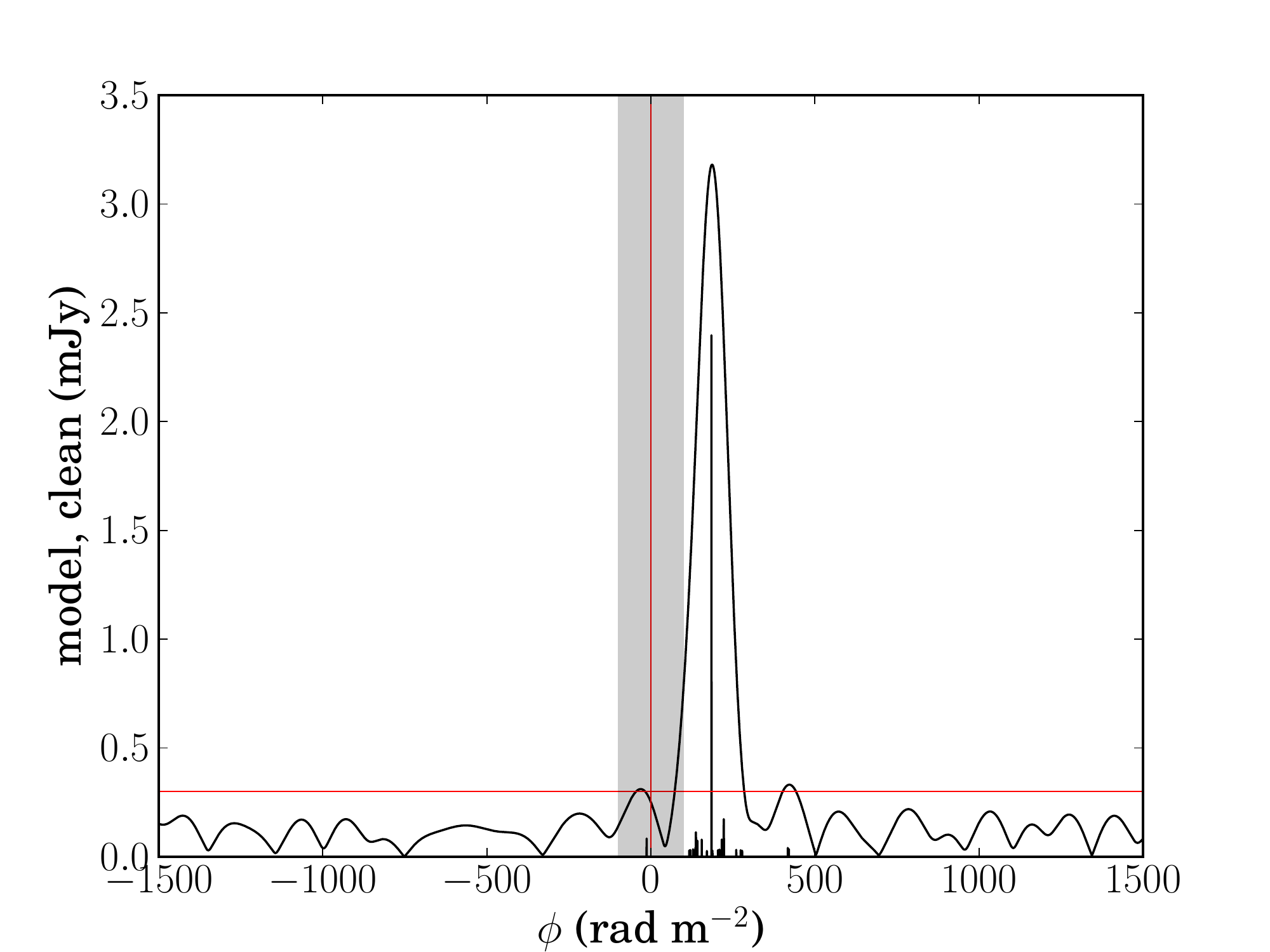}{0.5\textwidth}{(a)}
	\rotatefig{0}{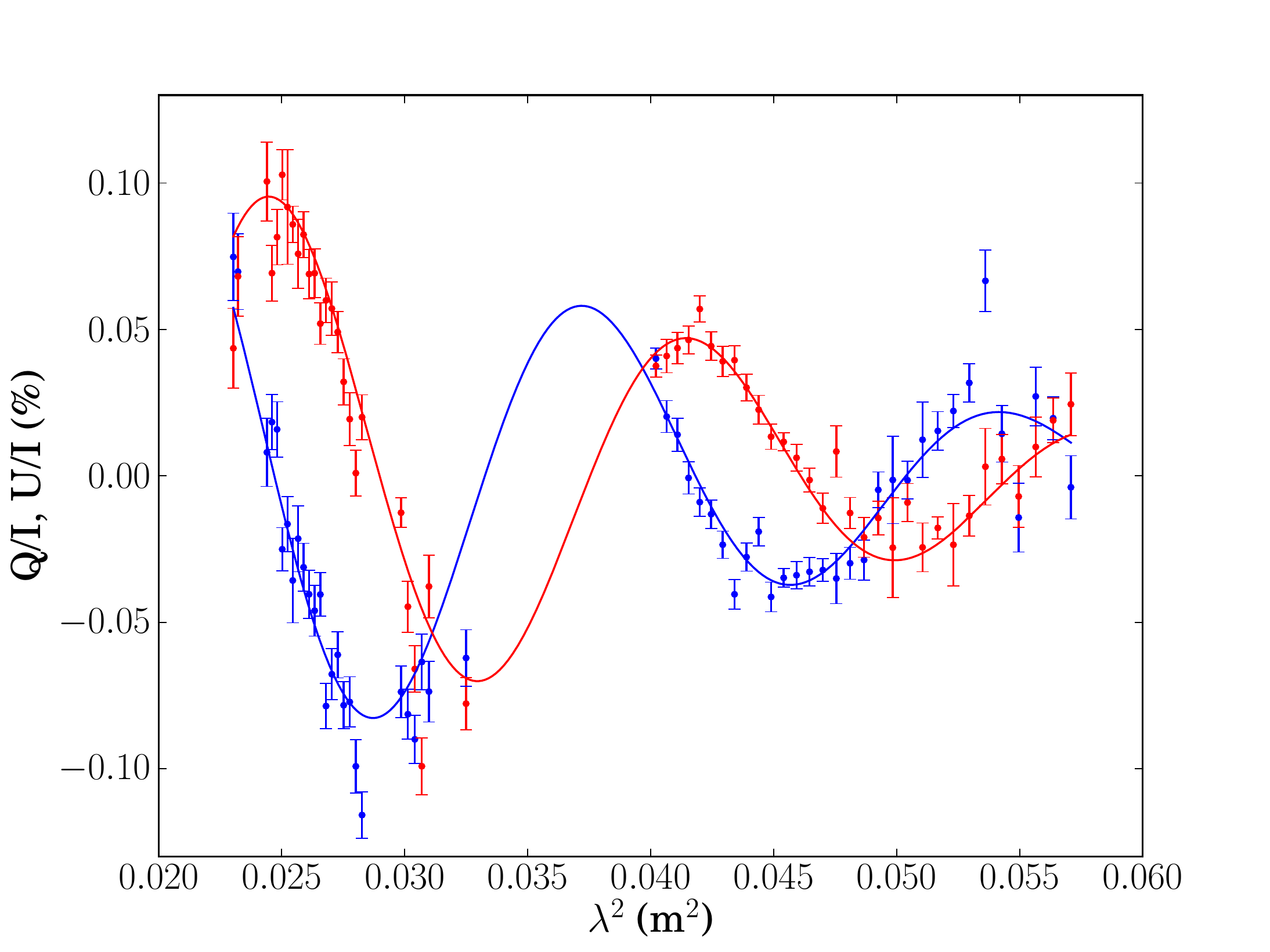}{0.5\textwidth}{(b)}}
\caption{(a) Enlarged Faraday depth spectrum of the profile with a red border in Figure \ref{fig:G39.2-0.3_RM_Grid2}. (b) Results of Stokes $QU$ fitting.  The blue and red data are Stokes $Q/I$ and $U/I$ respectively.  The blue and red curves are the fitted model functions for single-component Burn depolarization. }
\label{fig:G39.2-depol}
\end{figure*}

\indent We also observe subregions to exhibit a signature of single-component depolarization as described in \citet{Burn1966} by the factor $\text{e}^{-2\sigma_{\phi}^2\lambda^4}$, where $\sigma_{\phi}$ is the depolarization metric.  Figure \ref{fig:G39.2-depol} shows an example of a subregion where Burn depolarization across the band is observed.  We find that for this subregion $\sigma_{\phi} = 17.69 \pm 0.29 \text { rad m}^{-2}$.  For Stokes $QU$ fitting, we took the Stokes $I$ value from the THOR+VGPS data at $\lambda$21 cm and applied the mean spectral index found by \citet{Sun2011} (see Table \ref{tab:summary} for spectral indices).  Figure \ref{fig:G39.2-depol} (b) illustrates the results of Stokes $QU$ fitting for this subregion.  A further investigation of this type of depolarization for the SNRs covered in this study will be presented in a follow-up paper (Shanahan et al. in prep). 

\begin{figure*}[htb!]
\centering
   \centerline{\includegraphics[width=0.9\linewidth]{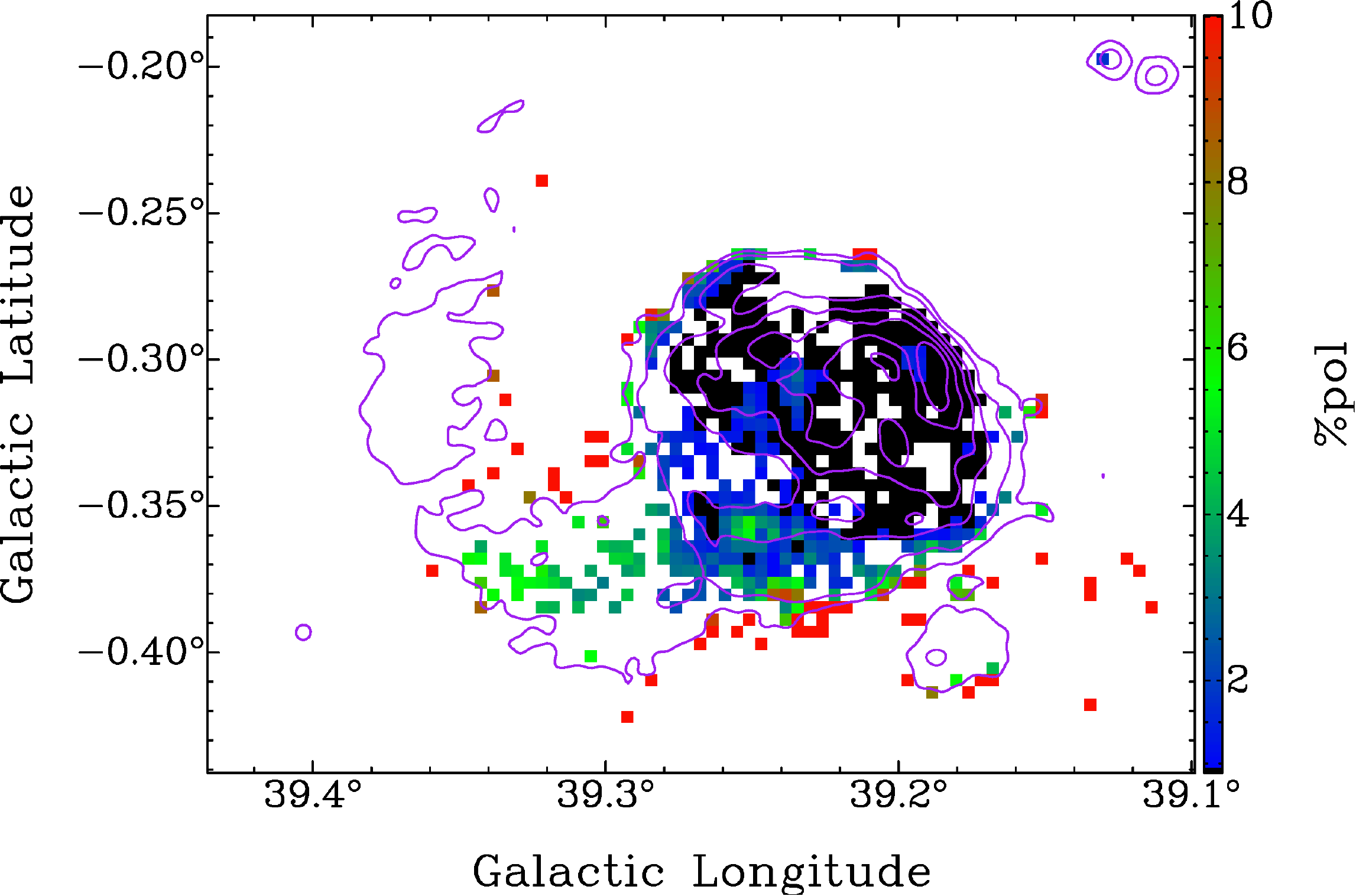}}
   \caption{Polarization map of SNR G39.2$-$0.3.  Contours are from the THOR+VGPS map at 20, 30, 55, 80, 110 and 140 mJy/beam.  The coloured and black subregions are defined in the caption of Figure \ref{fig:G46.8-0.3_PP-Map}. }
   \label{fig:G39.2_PP-Map} 
\end{figure*}

\indent 	In Figure \ref{fig:G39.2_PP-Map} we present a map of fractional polarization.  Within the central regions of the SNR there is minimal variation in the fractional polarization for subregions where polarized emission is observed.  In the bottom part of the shell we identify an increase in fractional polarization as well as filamentary structure as a Stokes $I$ counterpart.  Along the edge of this filament we observed two-component Faraday rotation as well as single-component Burn depolarization within the filament.  Within the lower left section of the SNR, the fractional polarization is seen to increase from $\sim 2\%$ to $\sim 4\%$ and continues to increase to $\sim 6\%$ within the filament.  Outside the SNR, fractional polarization increases from $\sim 3\%$ to $\sim 6\%$ in a region that traces the centre line of total intensity associated with the HII region G39.294-00.311. Since we detect polarization within this region, it is possible that this source is not a simple HII region.

\begin{figure*}[htb!]
\gridline{\rotatefig{0}{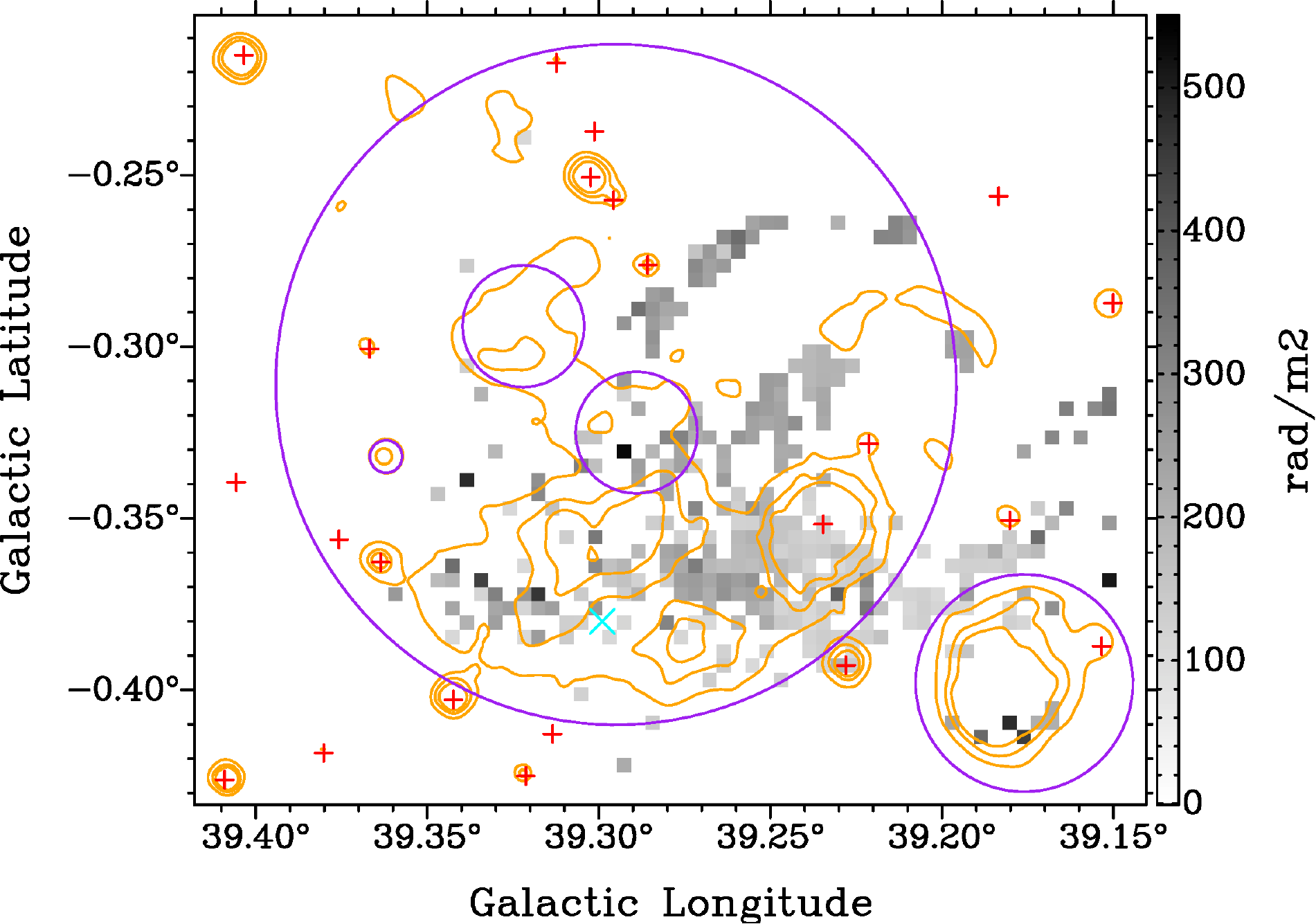}{0.502\textwidth}{(a)}
	\rotatefig{0}{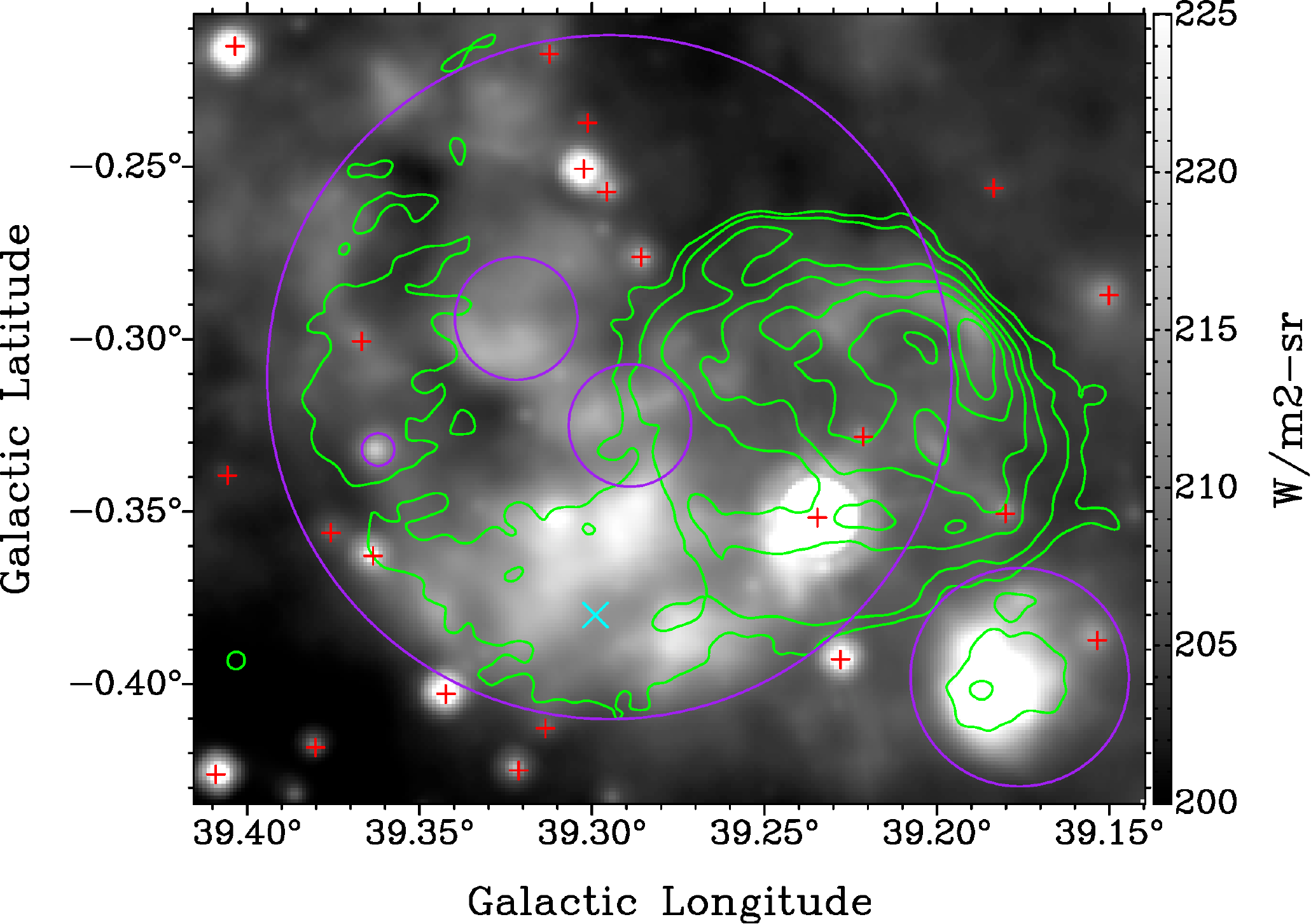}{0.50\textwidth}{(b)}}
\caption{(a) The Faraday depth map (in grey scale) with WISE $\lambda$22 $\mu\text{m}$ intensity (orange contours at 210, 215 and 220 W m$^{-2}$ sr$^{-1}$). (b) WISE $\lambda$22 $\mu\text{m}$ image (in grey scale) with THOR+VGPS 21 cm continuum intensity (green contours at 20, 30, 55, 80, 110 and 140 mJy/beam).  The purple circles indicate the locations of HII regions from the WISE catalog with the size of the circle corresponding to the published size.  The red $+$ symbols correspond to stars that were identified from the WISE 3.4 $\mu\text{m}$ map.  The cyan $\times$ symbol identifies the location where \citet{Anderson2018} detected RRLs from the HII region G39.294-00.311 at $V_{LSR} = 53.7 \text{ km s}^{-1}$ with a beam of 150\arcsec.  The SNR was found to have $V_{LSR} = 56 \pm 2 \text{ km s}^{-1}$ \citep{Lee2020}.}
\label{fig:G39.2_wise}
\end{figure*}

\indent	The locations of cataloged HII regions in relation to the SNR are shown in Figure \ref{fig:G39.2_wise}.  We present the Faraday depth map of the polarized emission observed from THOR with WISE $\lambda$22 $\mu\text{m}$ contours in Figure \ref{fig:G39.2_wise} (a).   The WISE $\lambda$22 $\mu\text{m}$ image is shown along with contours of THOR+VGPS Stokes $I$ in Figure \ref{fig:G39.2_wise} (b).  The \citet{Anderson2018} HII region catalog places five HII regions within the vicinity of the SNR.  The largest circle marks the HII region G39.294-00.311 that is associated with the extended emission from THOR+VGPS.  We see two regions of enhanced $\lambda$22 $\mu\text{m}$ emission that are not classified as separate HII regions on either side of our polarization detections.  Our polarization detections do not align perfectly with the local minimum in $\lambda$22 $\mu\text{m}$ intensity.

\indent	\citet{Anderson2018} measures the HII region, G39.294-00.31, to have an RRL velocity of $V_{LSR} = 53.7 \text{ km s}^{-1}$ at the location of the cyan $\times$ symbol in Figure \ref{fig:G39.2_wise} with $150\arcsec$ resolution.  From H$_2$ emission, \citep{Lee2020} derives a systemic velocity for the SNR to be $V_{LSR} = 56 \pm 2$ km s$^{-2}$.  The similarity in velocity between the HII region and the SNR suggests a likely association.  We see no evidence in the $\lambda$22 $\mu\text{m}$ intensity that the HII region is in the foreground and overlaps the SNR.  A curved ridge of $\lambda$22 $\mu\text{m}$ emission at $(\ell, b) = (39\fdg22, -0\fdg29)$ resembles radio emission from the SNR and coincides with shock heated H$_2$ \citep{Lee2019b}.

\indent	On the edges of HII regions we have found detections of polarized emission that may arise from rapid variation in polarization angle on small scales.  We observe a higher concentration of polarized emission detections within the extended $\lambda$22 $\mu\text{m}$ structure than what is seen surrounding HII regions.  This suggests that rapid variation in polarization angle is not likely to be cause of polarization detections. 
 
\begin{figure*}[htb!]
\centering
   \centerline{\includegraphics[width=0.8\linewidth, angle=0]{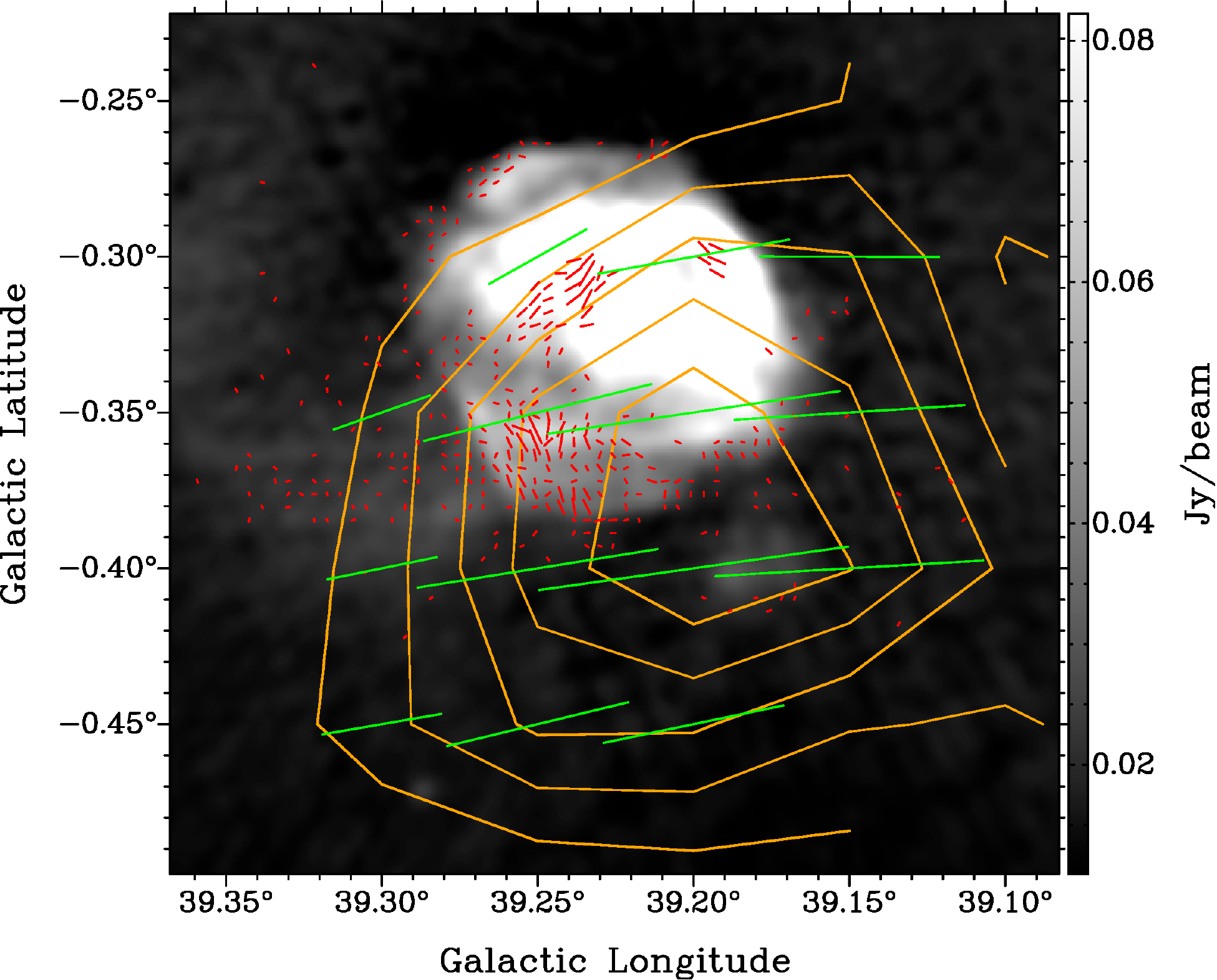}}
   \caption{THOR and VGPS total intensity map of the SNR G39.2$-$0.3.  The red vectors are from THOR polarization at $\lambda21 \text{ cm}$ rotated to $\lambda6 \text{ cm}$ using the observed Faraday depth and the green vectors from Sino-German survey.  The length of the vectors is proportional to the fractional polarization observed in the associated subregion.  The orange contours from the polarization map of the Sino-German survey at levels 10, 15, 20, 25 and 30 mK. }
   \label{fig:G39.2_B-vec} 
\end{figure*}

\indent	Figure \ref{fig:G39.2_B-vec} shows polarization from the $\lambda$6 cm Sino-German survey \citep{Sun2007, Sun2011b} as orange contours and long green polarization vectors on a THOR+VGPS Stokes $I$ image with short red polarization vectors rotated to $\lambda$6 cm.  We observe that the centroid of the $\lambda$6 cm polarized intensity is offset to the lower right edge of the SNR Stokes $I$ emission.  This offset was also observed in \citet{Sun2011}, however it is more obvious when comparing the Sino-German polarization to THOR+VGPS Stokes $I$ (Reich, W. \& Sun, X. 2022 private communication).  In the region where $\lambda$6 cm polarization is brightest we detect polarization outside of the SNR.  Further investigation of the polarized emission and Faraday rotation observed from this region require higher resolution observations with spectral indices to confirm the type of emission and the origin of polarized emission.

\indent 	The locations of polarization detections within the SNR shown in Figure \ref{fig:G39.2_B-vec} match the strongest $\lambda$6 cm polarization emission in \citet{Patnaik1990}.  In comparison to Figure 9 of \citet{Patnaik1990}, we observe similar patches of polarized emission within the SNR as well as B-vector orientation at $\lambda$6 cm.  However, the polarization detections we find outside the SNR are not observed in \citet{Patnaik1990}.  The centroid of peak polarized emission at $\lambda$6 cm from the Sino-German survey falls on the edge of the SNR where THOR and \citet{Patnaik1990} observe minimal polarized emission.  This suggests that the single-dish observations might be dominated by a diffuse component that is suppressed by spatial filtering in the interferometer observations.  Polarization detections outside the lowest contour could suggest that the SNR acts as a Faraday screen to background diffuse polarized emission.   


\pagebreak

\section{Discussion}
\label{sec:Disc}

\begin{figure*}[htb!]
\centering
   \centerline{\includegraphics[width=1\linewidth, angle=0]{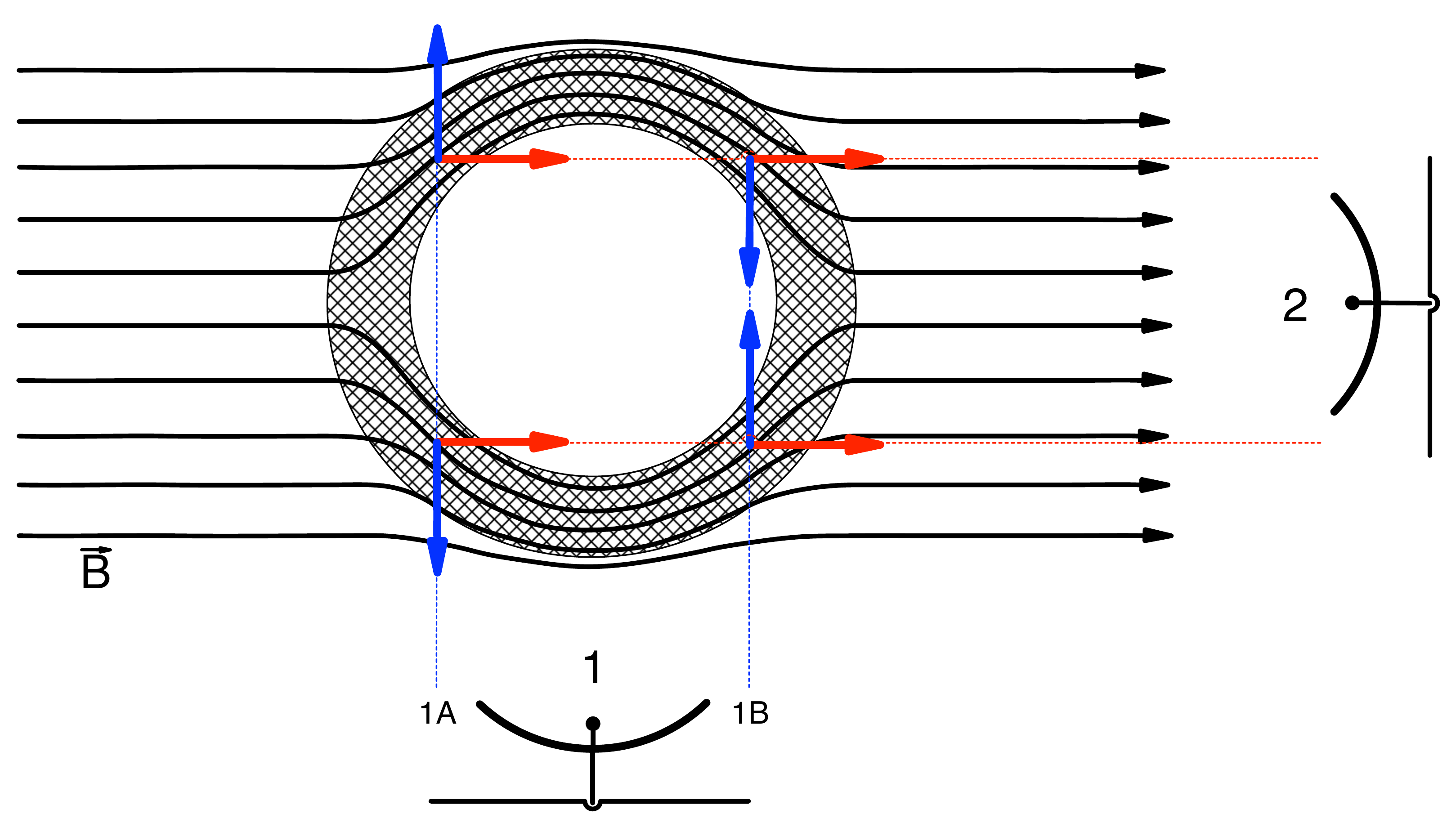}}
   \caption{A simple model of the magnetic field within an expanding SNR shell and the $\vec{B}_\parallel$ magnetic field components along two lines-of-sight.  The black lines represent the Galactic magnetic field.  The cross-hatched region indicate the location of synchrotron emission within the SNR shell.  The blue vectors represent $\vec{B}_\parallel$ along line of sight 1. The red vectors represent $\vec{B}_\parallel$ along line of sight 2.  The length of the vectors do not represent the strength of $\vec{B}_\parallel$, only the vector orientation being toward or away from the observer is represented.}
   \label{fig:SNR_LOS} 
\end{figure*}

\indent	In the SNRs G46.8$-$0.3, G43.3$-$0.2, and G39.2$-$0.3 we found peaks from two-component Faraday rotation that are separated by $\Delta\phi>100 \text{ rad m}^{-2}$, where $\Delta\phi$ is the Faraday depth separation of the two peaks.  These peaks originate from separate synchrotron emitting regions subject to different Faraday rotation.

\indent	The morphology of a SNR is determined in part by the direction of the local magnetic field and the line of sight \citep{West2015, West2017}.  We model each line of sight through the SNR as two Faraday rotating slabs depicted in Figure \ref{fig:SNR_LOS}.  For any line of sight intersecting the front and the back of the shell, we can consider them as two emitting Faraday screens.  For Observer 1, as defined in Figure \ref{fig:SNR_LOS}, the symmetry of the model SNR implies that the total Faraday depth for any line of sight perpendicular to the Galactic magnetic field (line of sight 1A and 1B in Figure \ref{fig:SNR_LOS}) is zero.  We introduce $\phi_0$ as the absolute value of the Faraday depth of one side of the SNR shell.  If we model the back side of the SNR shell as a uniform slab, its net internal Faraday rotation is equal to $\pm \frac{1}{2}\phi_0 \lambda^2$ \citep{Sokoloff1998}, with the sign determined by $B_\parallel$ in the \textit{back} side of the SNR.  The back side of the SNR is also Faraday rotated by the amount $\mp \phi_0 \lambda^2$ by the front side, resulting in the total Faraday rotation of the back side of $\mp \frac{1}{2} \phi_0 \lambda^2$, with the sign defined by the direction of $B_\parallel$ in the \textit{front} side of the SNR shell.  The emission of the front side of the shell has net Faraday rotation $\mp \frac{1}{2} \phi_0 \lambda^2$, which is the same amount as the total Faraday rotation of the back side.  Therefore, the resulting Faraday depth spectrum would consist of a single Faraday component centred at $\frac{1}{2} \phi_0$ with width $\phi_0$.   According to this model, Observer 1 will see a gradient of Faraday rotation across the SNR in the direction of the Galactic magnetic field in the plane of the sky \citep{Kothes2009}, but cannot separate the front and back in Faraday depth space. 

\indent	For Observer 2, as defined in Figure \ref{fig:SNR_LOS},  the back side of the SNR has net internal Faraday rotation equal to $\frac{1}{2}\phi_0 \lambda^2$ and is Faraday rotated by the amount $\phi_0 \lambda^2$ by the front side, resulting in the total Faraday rotation of the back side of $\frac{3}{2} \phi_0 \lambda^2$.  The emission of the front side of the shell has net Faraday rotation $\frac{1}{2} \phi_0 \lambda^2$.  With this model the direction of $\vec{B}_\parallel$ is the same for all lines-of-sight parallel to the Galactic magnetic field, therefore the sign of the Faraday rotation for the front side and the back side of the shell will be the same.  Similar to what is shown in \citet{Kothes2009}, we would expect to see relatively uniform Faraday rotation throughout the SNR shell.  In a Faraday depth spectrum, the front side Faraday component would be centred at $\frac{1}{2} \phi_0$ and the back side Faraday component centred at $\frac{3}{2} \phi_0$.  Since each component has a width of $\phi_0$ the front side and back side would appear to be a single Faraday component centred at $\phi_0$ with width $2\phi_0$.  Much like in the case of Observer 1, for lines-of-sight parallel to the Galactic magnetic field the front side and the back side of the SNR cannot be observed as two separated Faraday components. 

\indent	Our observations of two-component Faraday rotation within each polarized SNR suggest that a model of Faraday rotation of SNR shell must include Faraday rotation between the two side of the shell.  Due to a low density in the inner part of the SNR shell, a new model would require a thick shell that does Faraday rotation inside a thinner shell of synchrotron emission.  However, under these conditions for lines-of-sight perpendicular to the Galactic magnetic field, if the shell is symmetric the back side would become de-rotated by the front side and two-component Faraday rotation would again not be observed.  In order to observe two-component Faraday rotation, our model requires small-scale structure within the SNR shell so that the back side and front side have different values of Faraday rotation or viewing angle with respect to the Galactic magnetic field.  A break in global symmetry can create the proper conditions for two-component Faraday rotation.  \citet{Xiao2008} present fine filamentary structure in optical observations of the SNR S147 that are unresolved in their $\lambda$11 cm radio observations.  For G46.8$-$0.3, G43.3$-$0.2 and G39.2$-$0.3 we suspect a similar case where these SNRs have structure unresolved at 16\arcsec.  The presence of such structure would break symmetry in the SNR shell and thus could be the cause of two-component Faraday rotation.  

\indent	In each polarized SNR, large variation in single-component Faraday rotation in adjacent subregions is not accompanied by a counterpart observed in Stokes $I$.  This suggests that small scale variations in polarized intensity are related to Faraday rotation, which implies small scale structure in $n_e$ or $B_\|$. Since the Galaxy does not exhibit these large variations in Faraday depth on this scale \citep[Figure~2]{Leahy1987}, the effect must be internal to the SNR where we are observing two physically separated emitting Faraday screens along the line of sight.  The result of Faraday depth variation on small-scales is not unique to a single SNR in our study.  A physical interpretation of this could be fine filamentary structure that is unresolved within our beam.  

\indent	The distribution of Faraday depth for SNR G39.2$-$0.3 reveals a possible interpretation as to the nature of the two-component Faraday rotation.  The dispersion in Faraday depth for the narrow peak at a lower Faraday depth is $\sim$40$\text{ rad m}^{-2}$ for both the total and two-component distribution.  Variation in the Galactic Faraday screen will account for $\sim$20$\text{ rad m}^{-2}$ to $\sim$30$\text{ rad m}^{-2}$ on $16\arcsec$ angular scales \citep{Leahy1987}.  Therefore, the large Faraday depth variation observed in these SNRs may be caused by internal effects from the SNR itself.  Considering a SNR as a spherical shell, as the shell expands, variation in ISM densities will cause non-uniform expansion and variations in density in the SNR shell through different lines-of-sight \citep{Villagran2020}.  When considering the internal environment of a SNR, a complex structure of free electrons and magnetic field is expected inside the SNR shell \citep{Orlando2019}.  These two scenarios are consistent with our interpretation that the narrow peak at a lower Faraday depth and the wide peak at a higher Faraday depth originate from the front and back sides of the SNR shell respectively.  We therefore propose that the front and back side of a shell SNR can be separated in Faraday depth under the right conditions.

\indent	We suspect that the Faraday depth distribution of the SNR G46.8-0.3 is also bimodal, but each component has similar mean Faraday depths.  From Figure \ref{fig:G46.8_hist} (b) the narrow peak is more obvious with a broad peak offset slightly towards a lower Faraday depth.  For SNR G43.3$-$0.2, from Figure \ref{fig:G43.3-0.2_histogram}, a narrow peak is obvious, but a broad distribution is less obvious, due to the $\pm 100 \text{ rad m}^{-2}$ rejection range.  That being said, the grouping at negative Faraday depths could be interpreted as part of a broad distribution and possibly a signature of the far side of the SNR shell.  The compression and bulging of the Galactic magnetic field by the expanding SNR shell could cause the back side of the shell to have either a positive, negative or the same Faraday depth as the front side of the shell depending on the line of sight angle with respect to the large scale Galactic magnetic field direction.  

\begin{figure*}[htb!]
\gridline{\fig{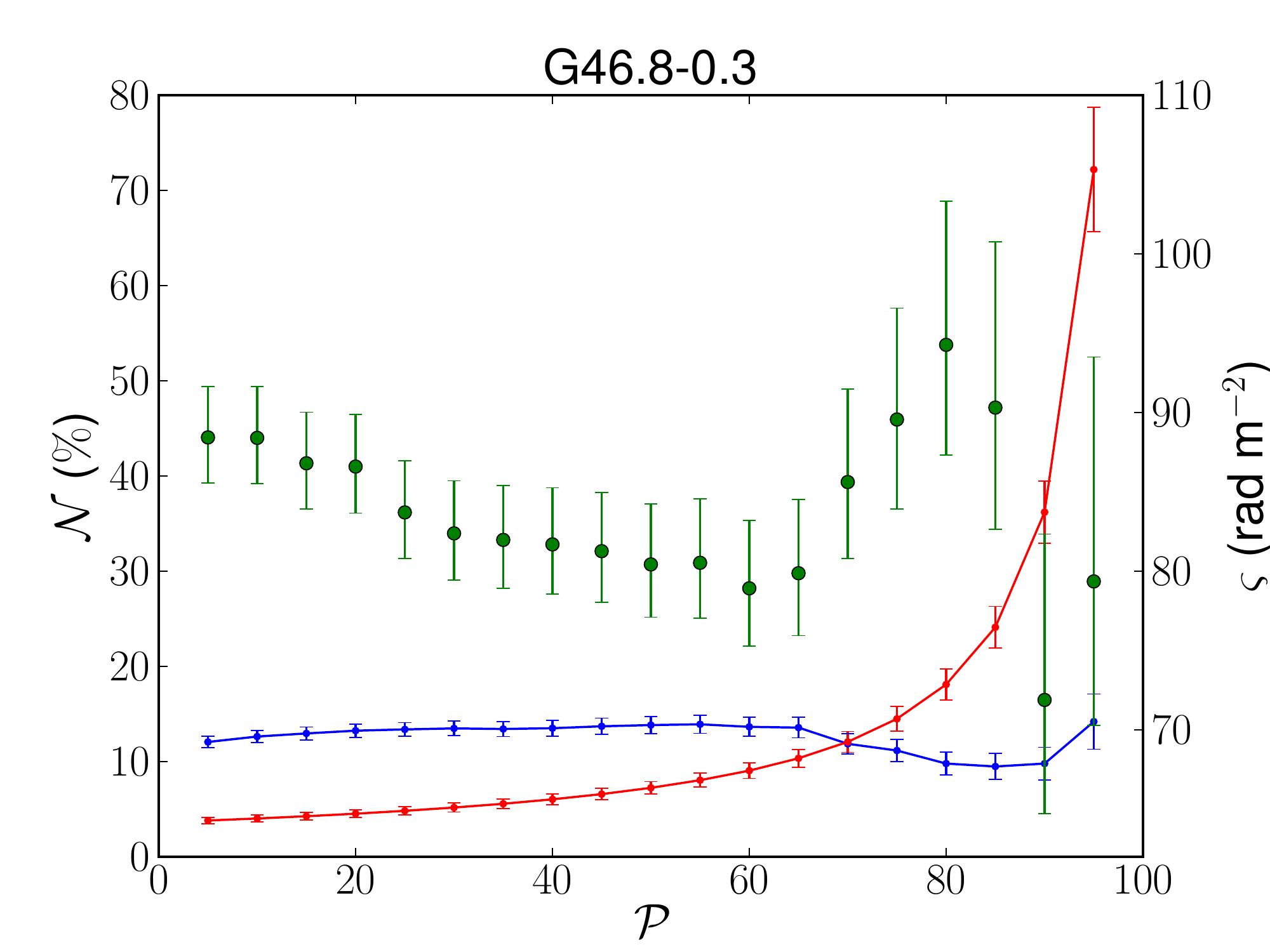}{0.5\textwidth}{(a)}
	\fig{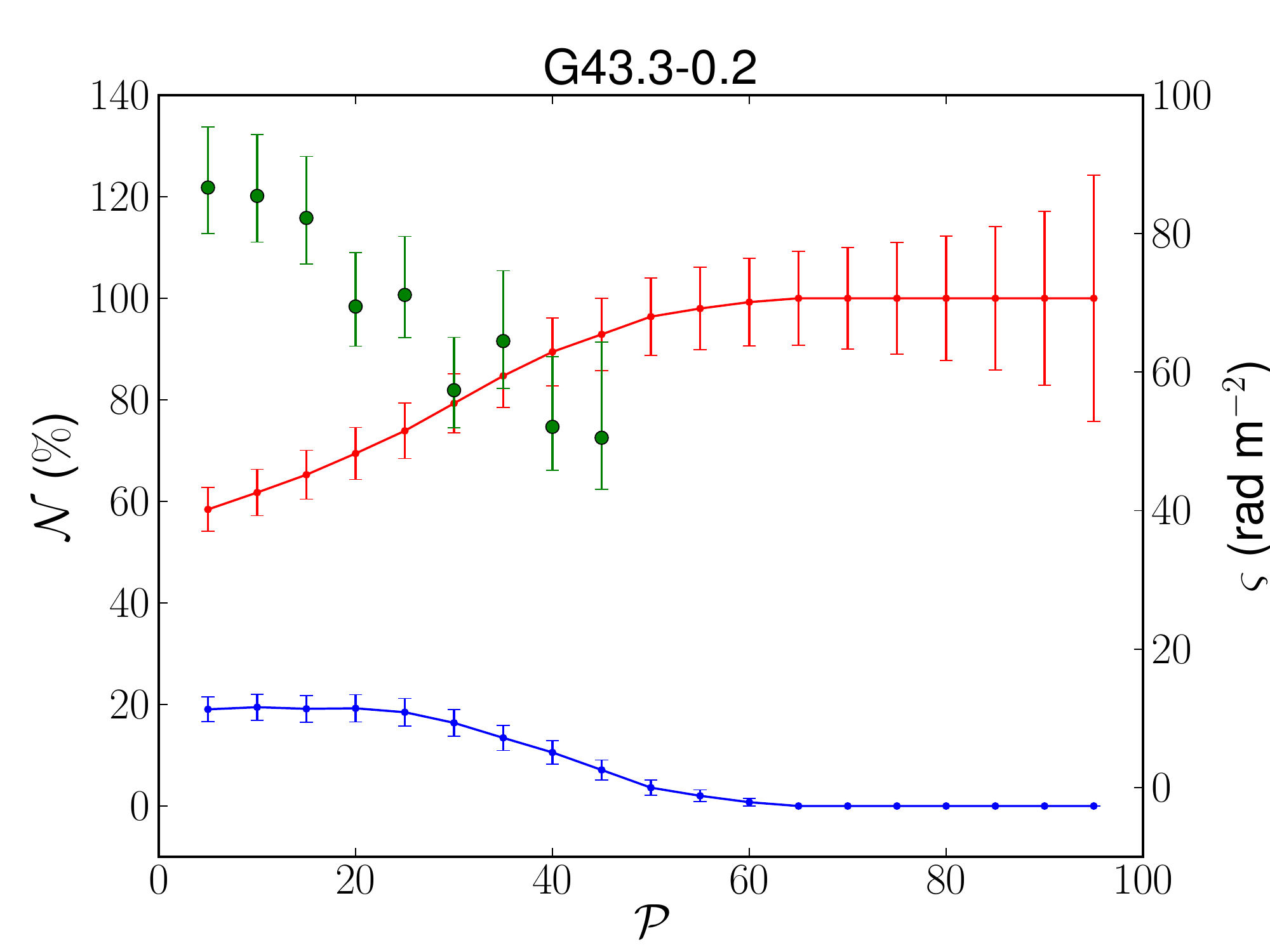}{0.5\textwidth}{(b)}}
\gridline{\fig{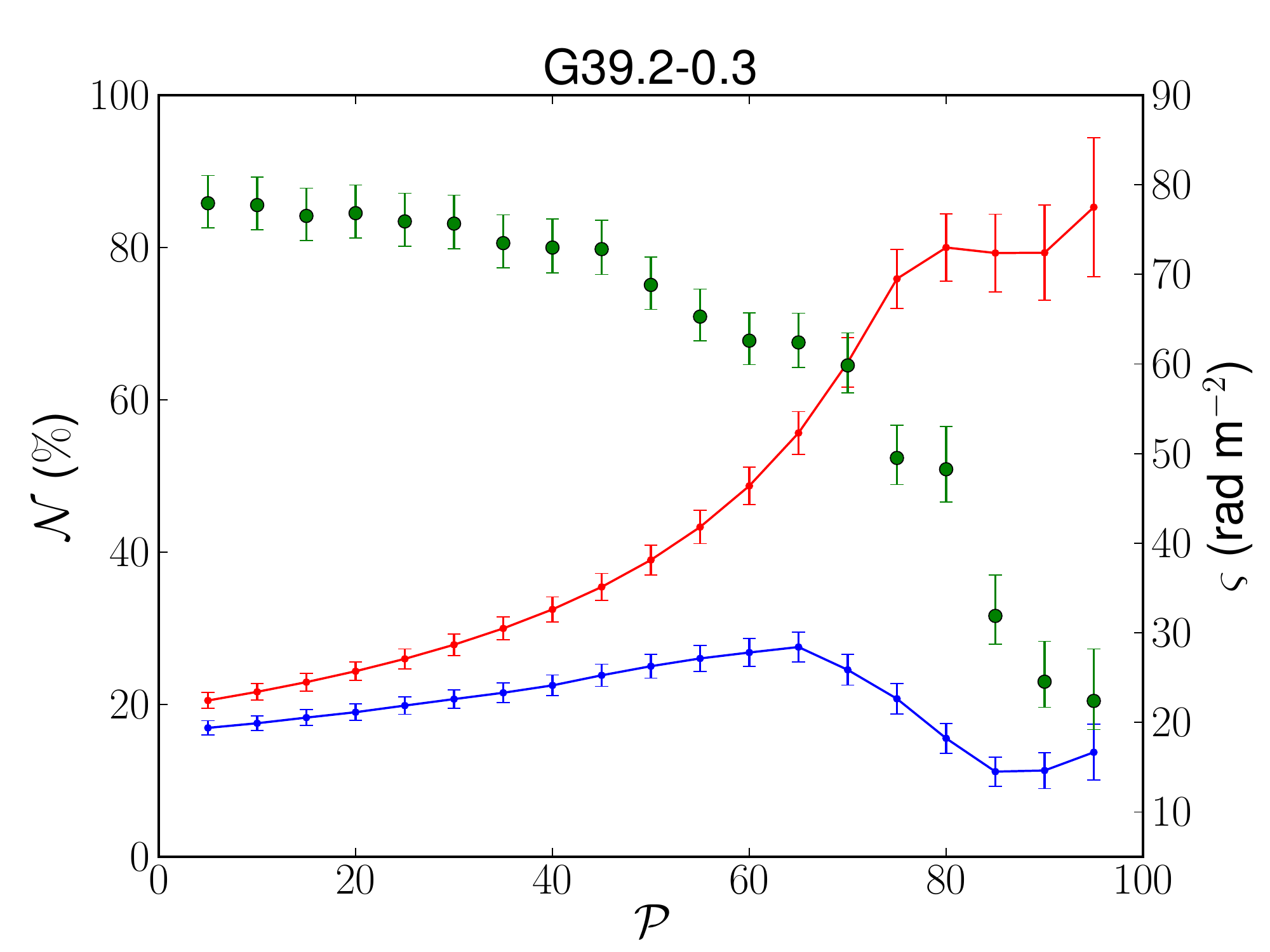}{0.5\textwidth}{(c)}}
\caption{The distribution of detections of polarized emission and upper limits with respect to Stokes $I$.  $\mathcal{P}$ is the percentile of the maximum Stokes $I$ surface brightness within the SNR.  The red and blue curves correspond to $\mathcal{N}=N_{UL}/N_{T}$ and $\mathcal{N}=N_{Det}/N_{T}$, respectively,  where $N_{UL}$ is the number of subregions with an upper-limit of fractional polarization, $N_{Det}$ is the number of subregions where polarization is detected and $N_{T}$ is the total number of subregions.  Each of these values represent how many of each are above the corresponding $\mathcal{P}$.  The green data points correspond to $\varsigma$ which represents the standard deviation of the Faraday depths for each subregion in $N_{Det}$.}
\label{fig:percentile}
\end{figure*}

\indent	Figure \ref{fig:percentile} depicts how the detections of polarized emission and upper limits are distributed as a function of the surface brightness, expressed as a percentile of the peak.  If one were to consider a SNR with constant detectable fractional polarization, there would be a Stokes $I$ contour (a percentile of the maximum intensity as in Figure \ref{fig:percentile}) at which the percentage of detections would be 100\%.  At lower intensities the fraction of detections would be rising monotonically.  On the contrary, the fraction of detections remains constant or declines as the Stokes $I$ threshold increases (see the blue curves in Figure \ref{fig:percentile}). A decline in the blue curves, accompanied by a rise in the fraction of strong upper limits (red curves), indicates that the bright parts of the SNRs are less polarized.  

\indent	Each polarized SNR has a similarly shaped blue curve (Figure \ref{fig:percentile}) where at lower percentiles the percent of detections is roughly constant with a monotonically decreasing relation after a certain percentile.  At $\mathcal{P} \approx 65$, a decline is observed for G46.8$-$0.3 and G39.2$-$0.3, which illustrates that there is a lack of polarization in these SNRs where Stokes $I$ is bright.  For G43.3$-$0.2 a decline of percent detections is observed at $\mathcal{P} \approx 20$.

\indent	In order to investigate the lower polarization of the bright parts of each SNR we define the parameter $\varsigma$ as the standard deviation of Faraday depth for all subregions with a detection above a certain Stokes $I$ threshold.  The increase in $\varsigma$ for G46.8$-$0.3 at $\mathcal{P} \approx 65$ occurs at the Stokes $I$ brightness where a large difference in the mean Faraday depth of the brightest lobes on opposite sides of the SNR dominates $\varsigma$.  The decrease in $\varsigma$ at brighter Stokes $I$ that we observe in each polarized SNR illustrates that variation in Faraday depth is lower where the SNRs are bright.  At higher percentiles, low polarization and low $\varsigma$ could indicate a more disordered magnetic field from turbulence within the SNR shell.  It should be noted that the curves for $\varsigma$ are not structure functions, because these relations are related to Stokes $I$ brightness and not spatial separation.

\begin{deluxetable*}{ccccccccccc}
\tablenum{1}
\tablecaption{Metrics for polarized SNRs \label{tab:snr}}
\tablewidth{0pt}
\tablehead{
\colhead{SNR Name}   & \colhead{$\bar{\text{I}} - \text{I}_{\text{bg}}$}  & \colhead{Size} & \colhead{$\alpha^{\text{a}}$}  & \colhead{Distance} &  \multicolumn{3}{c}{$\mathcal{P}_{50}$} & \multicolumn{3}{c}{$\mathcal{P}_{90}$} \\
\cline{6-8} \cline{9-11}
\colhead{} & \colhead{(mJy)} & \colhead{(pc)} & \colhead{} & \colhead{(kpc)} & \colhead{$N_T$} & \colhead{$N_{Det}$} & \colhead{$N_{UL}$} & \colhead{$N_T$} & \colhead{$N_{Det}$} & \colhead{$N_{UL}$}
}
\decimalcolnumbers
\startdata
G46.8$-$0.3 & 21.64 $\pm$ 0.43    & 23.65 $\pm$ 0.44 & -0.54 $\pm$ 0.02  & 5.4 $\pm$ 0.1$^{\text{b}}$ & 1684 & 233 & 122 &  377 & 33 & 122 \\
G43.3$-$0.2 & 832.49 $\pm$ 1.47  & 13.15 $\pm$ 0.47 & -0.46 $\pm$ 0.01  & 11.3 $\pm$ 0.4$^{\text{c}}$ & 166 & 6 & 160 &  34 & 0 & 34 \\
G39.2$-$0.3 & 117.72 $\pm$  0.25 & 19.78 $\pm$ 1.16& -0.34 $\pm$ 0.01  & 8.5 $\pm$ 0.5$^{\text{c}}$ & 1011 & 253 & 394 &  203 & 23 & 161 \\
\enddata
\tablecomments{Here $N_T$ is the total number of subregions in the SNR, $N_{Det}$ is the number of detections of polarized emission, and $N_{UL}$ is the number of upper limits.  $\mathcal{P}_{50}$ and $\mathcal{P}_{90}$ indicate the number of subregions at the 50th and 90th percentile respectively.\\
$^{\text{a}}$ \citet{Sun2011}\\
$^{\text{b}}$ \cite{Lee2020}\\
$^{\text{c}}$ \citet{Ranasinghe2018}}
\label{tab:summary}
\end{deluxetable*}

\indent	We performed Stokes $QU$ fitting for each subregion in a single SNR where we detect polarized emission and Faraday rotation using the single-component Burn depolarization model, where the depolarization factor $DF = \exp(-2\sigma_{\phi}^2 \lambda^4)$ \citep{Burn1966}.  Wavelength-dependent depolarization has been observed in subregions of the polarized SNRs investigated in this study (see Figure \ref{fig:G39.2-depol}).  We find that $\sigma_{RM} \approx 5 - 25 \text{ rad m}^{-2}$ for each polarized SNR we investigated.  We can ascertain from this comparison that wavelength-dependent depolarization is not as significant at $\lambda$6 cm as it is at $\lambda$21 cm.  Since we do not detect polarized emission from the brightest parts of the SNRs at $\lambda$21 cm, polarization observations at $\lambda$6 cm would confirm that depolarization in the bright regions is wavelength-dependent.  \citet{Lee2019b} observe H$_2$ line emission associated with the edges of regions of bright synchrotron emission for G43.3$-$0.2 and G39.2$-$0.3, which indicate a collision between the SNR shock and a dense molecular cloud. Similarly for G46.8$-$0.3, \citet{Supan2022} observe an interaction between the forward shock and dense molecular clouds in the centre of the SNR and the brightest regions along the edges.  For each polarized SNR in this study, we associate depolarization in regions with bright Stokes $I$ with an interaction between the SNR shock and nearby molecular clouds.  A more detailed analysis of $\sigma_{\phi}$ and small-scale structure in Faraday depth is deferred to a second paper.  


\section{Conclusions}
\label{sec:conclusions}

\indent	We present maps of Faraday rotation and fractional polarization of four SNRs in the THOR survey.  We detect polarization from G46.8$-$0.3, G43.3$-$0.2 and G39.2$-$0.3 but derive upper limits at the 1\% level for G41.1$-$0.3.  

\indent	We find significant variation in polarized intensity on scales down to the resolution limit of $16\arcsec$ with no associated variation in total intensity.  The lack of correlation between Stokes $I$ emission and polarized emission suggests a complicated internal structure of electron density and magnetic field in each SNR, causing small-scale variations in Faraday depth and polarized intensity. 

\indent	The supernova remnants G46.8$-$0.3 and G39.2$-$0.3 display large-scale rotation measure gradients on the order of $\sim 200\text{ rad m}^{-2}$ on the scale of the SNR and small-scale variations of $\sim 100\text{ rad m}^{-2}$ down to the resolution limit of 16\arcsec.  These rotation measure values are above the expected effect of the foreground ISM, which is estimated to be around 80 and 30 rad m$^{-2}$ on angular scales of $90\arcsec$ and $16\arcsec$, respectively \citep{Leahy1987}. These values are much lower than the observed large-scale gradient and small-scale variations in each SNR, therefore indicating that Faraday rotation in the foreground ISM does not account for the observed structure in Faraday rotation.  

\indent	Several locations in G46.8$-$0.3 and G39.2$-$0.3 showed two-component Faraday rotation while multi-component Faraday rotation is seen in G43.3$-$0.2.  In addition we find Faraday depth dispersion in the range 5 rad m$^{-2}$ to 25 rad m$^{-2}$. 

\indent	The L-band polarization shows higher fractional polarization in regions of intermediate surface brightness.  Regions of high surface brightness in SNRs G46.8$-$0.3, G43.3$-$0.2 and G39.2$-$0.3 have low fractional polarization with upper limits at 1\% of Stokes $I$.  These regions coincide with shock-heated molecular gas seen by \citet{Lee2019b}.  We do not find higher Faraday depth dispersion in regions of higher surface brightness, thus indicating that wavelength dependent Faraday depolarization is not the main cause of lower polarization in bright regions.  

\indent	The distribution of Faraday depth in G39.2$-$0.3 is bimodal for all detections of Faraday rotation but also for the subset of two-component Faraday rotation.  In both cases we find a narrow range around approximately 110 rad m$^{-2}$ and a broader range around 220 rad m$^{-2}$.  The width of the narrow peak is comparable to published ISM Faraday depth dispersion of $\sim$30$\text{ rad m}^{-2}$ on arcminute scales.  We identify the narrow peak in the Faraday depth distribution with emission from the front side of the SNR and the broad peak with emission from the back side of the SNR affected by internal Faraday rotation.  A net shift in Faraday depth of the back side with respect to the front side is not expected in a symmetric model that represents the line of sight through the SNR with two emitting Faraday screens.  Association of the bimodal distribution of Faraday depth in G39.2$-$0.3 implies significant Faraday rotation between the front side and back side synchrotron emitting regions.  The small fraction of two-component Faraday depth spectra with respect to the total number of detections suggests that small scale structure in the shell determines whether we detect a two-component Faraday depth spectrum.  

\indent	The high-resolution polarization observations in the THOR survey reveal the complex structure of the magnetic field in Galactic supernova remnants.  We find that the SNRs G46.8$-$0.3, G43.3$-$0.1 and G39.2$-$0.3 have similarities including small-scale variations in Faraday rotation and fractional polarization as well as evidence for internal Faraday rotation.  Our results indicate tangled magnetic fields in the brighter regions of each SNR where the expanding shock is running into denser regions of the ISM.  


\begin{acknowledgments}

The authors thank the anonymous referee for comments that helped improve the manuscript along with W. Reich and X. Sun for detailed discussion about the interpretation of the Sino-German survey data.  The National Radio Astronomy Observatory is a facility of the National Science Foundation operated under cooperative agreement by Associated Universities, Inc. The authors acknowledge the use of the RMtools package written by Cormac Purcell. JMS acknowledges the support of the Natural Sciences and Engineering Research Council of Canada (NSERC), 2019-04848. HB acknowledges support from the European Research Council under the Horizon 2020 Framework Program via the ERC Consolidator Grant CSF-648505.  HB also acknowledges support from the Deutsche Forschungsgemeinschaft in the Collaborative Research Center SFB 881 - Project-ID 138713538 - “The Milky Way System” (subproject B1). This research was carried out in part at the Jet Propulsion Laboratory, which is operated by the California Institute of Technology under a contract with the National Aeronautics and Space Administration (80NM0018D0004).

\end{acknowledgments}


\appendix
\section{Detection Threshold}
\label{appendix:DT}

\indent	Here we discuss the detection criteria for polarization we apply in this paper.  These detection criteria were applied to all SNRs in order to separate real signal from noise and instrumental polarization.  Firstly we will discuss how the detection threshold is derived by investigating the effect of noise in Stokes $Q$ and Stokes $U$ on the Faraday depth spectra and secondly how effects of instrumental polarization are accounted for. 

\subsection{The Effect of Noise}
\begin{figure*}[htb!]
\gridline{\rotatefig{0}{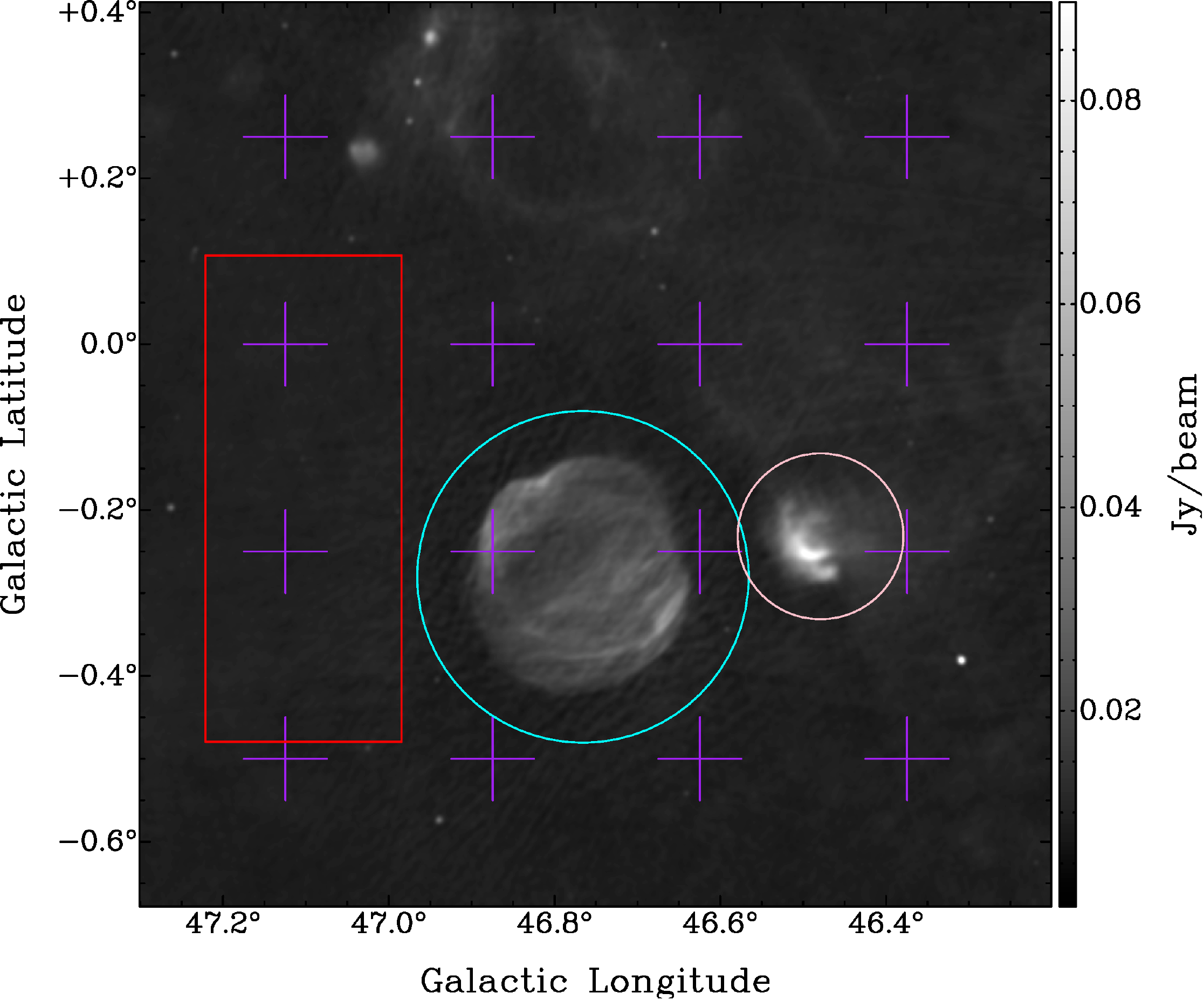}{0.5\textwidth}{(a) SNR G46.8$-$0.3}
          \rotatefig{0}{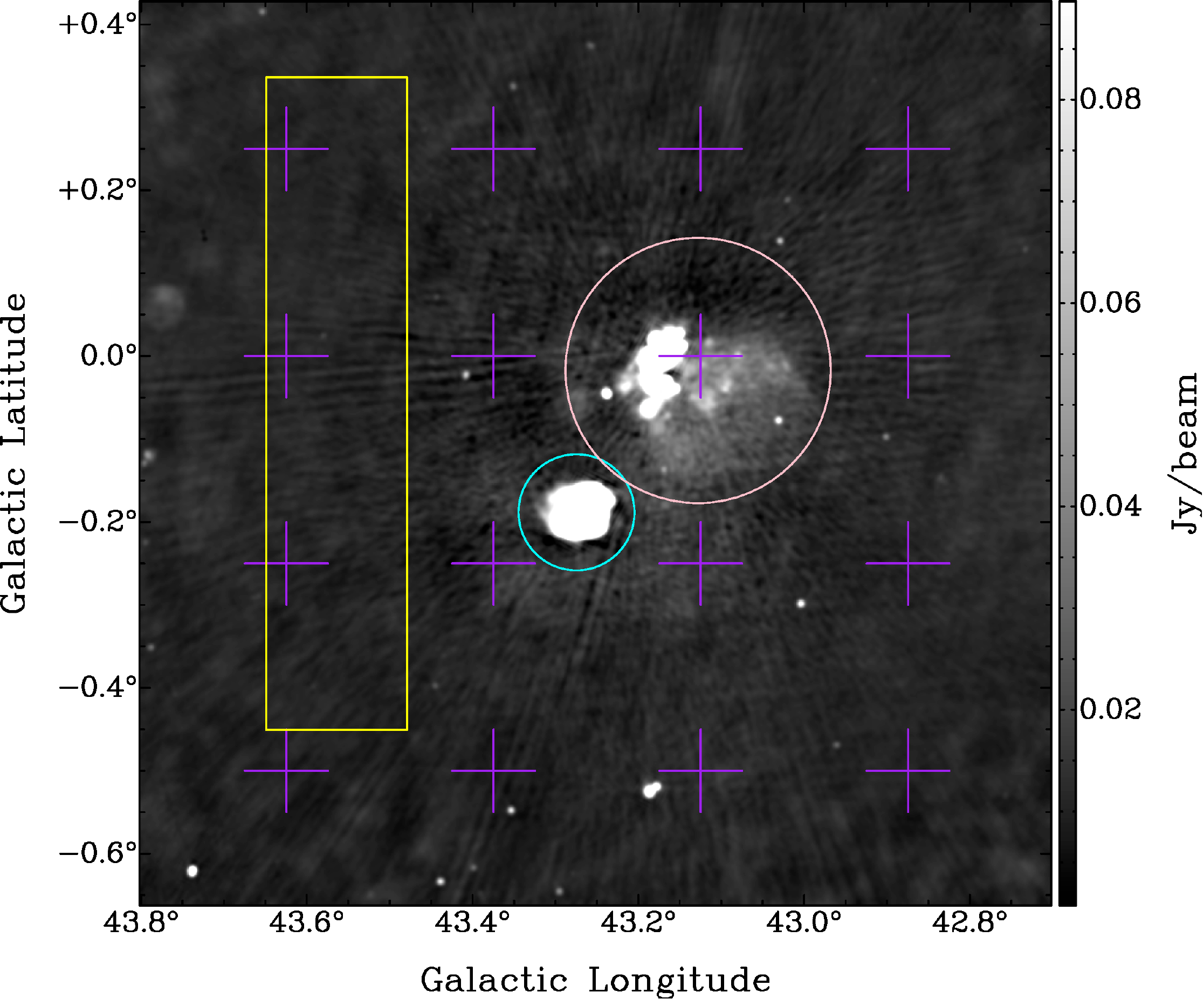}{0.5\textwidth}{(b) SNR G43.3$-$0.2}}
\gridline{\rotatefig{0}{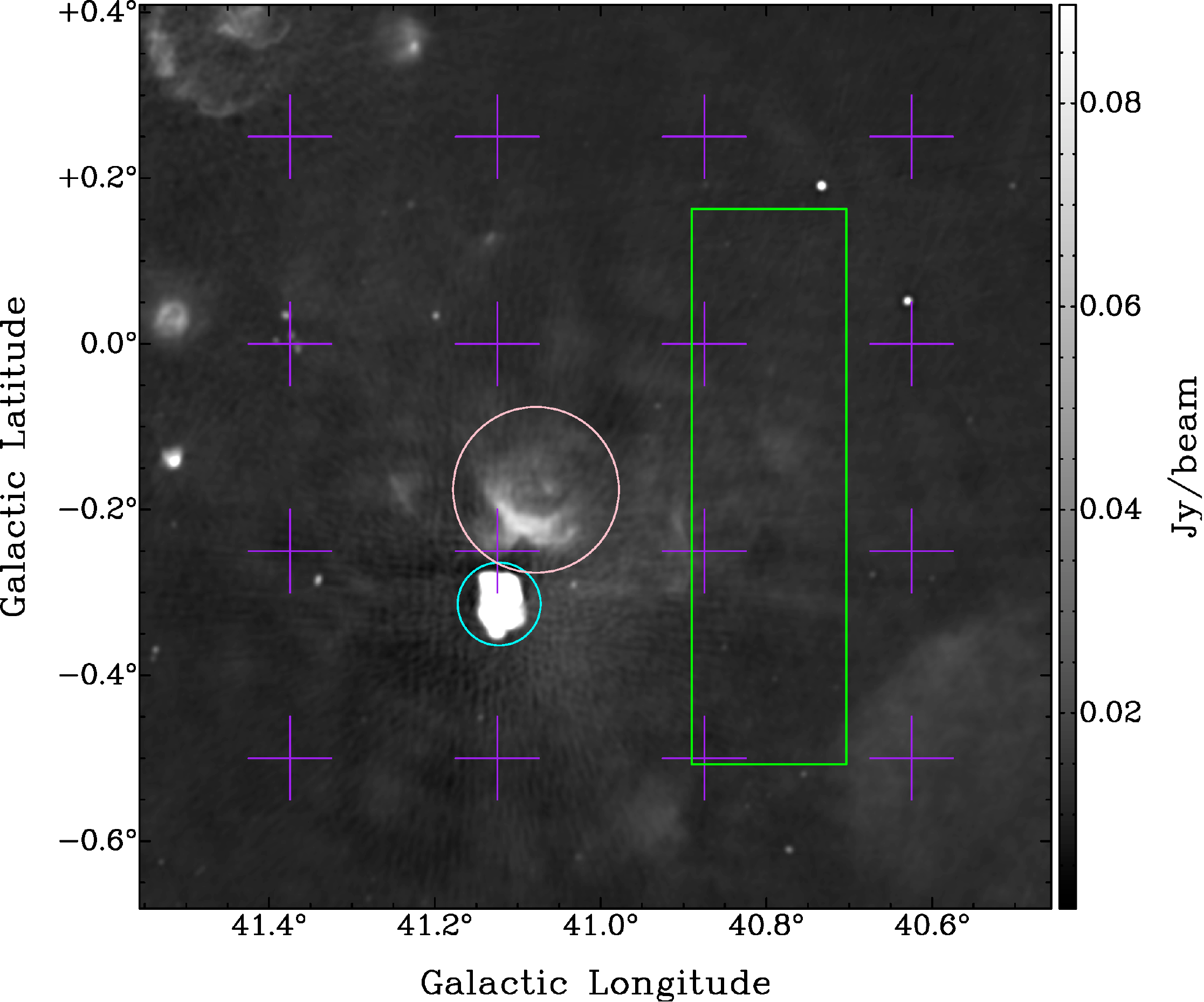}{0.5\textwidth}{(c) SNR G41.1$-$0.3}
          \rotatefig{0}{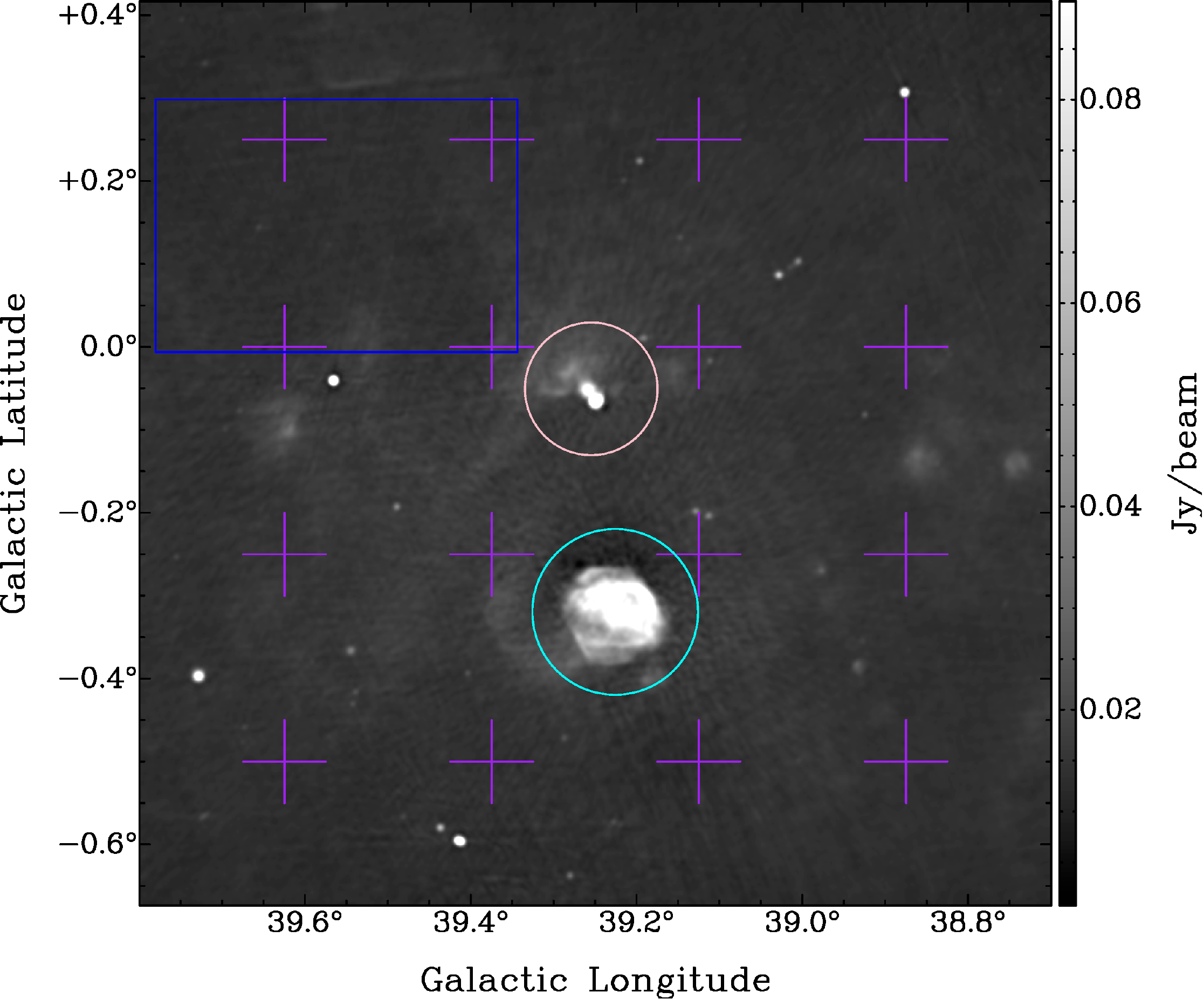}{0.5\textwidth}{(d) SNR G39.2$-$0.3}}
\caption{This figure identifies the location where the off profiles were extracted to analyze the effect the noise has on the Faraday depth profiles.  Each coloured rectangle outlines the region where the 500 off-profiles were found.  The purple crosses indicate the location of the individual field centres within the mosaic.  The cyan and pink coloured circles indicate the SNR and HII region respectively. }
\label{fig:offregions}
\end{figure*}

\indent	To investigate how image noise affects Faraday depth spectra we analyzed a sample of 500 boxes the size of our beam that are located near each SNR in regions that are dominated by noise and each box is separated from adjacent boxes by four beams to ensure noise within each box is independent.  Shown in Figure \ref{fig:offregions} are purple crosses that mark the field centres if the individual primary beams for these mosaics.  The regions selected for off-position extraction have a few field centres within the region. 

\begin{figure}[htb!]
\centering
   \centerline{\includegraphics[width=0.6\linewidth, angle=0]{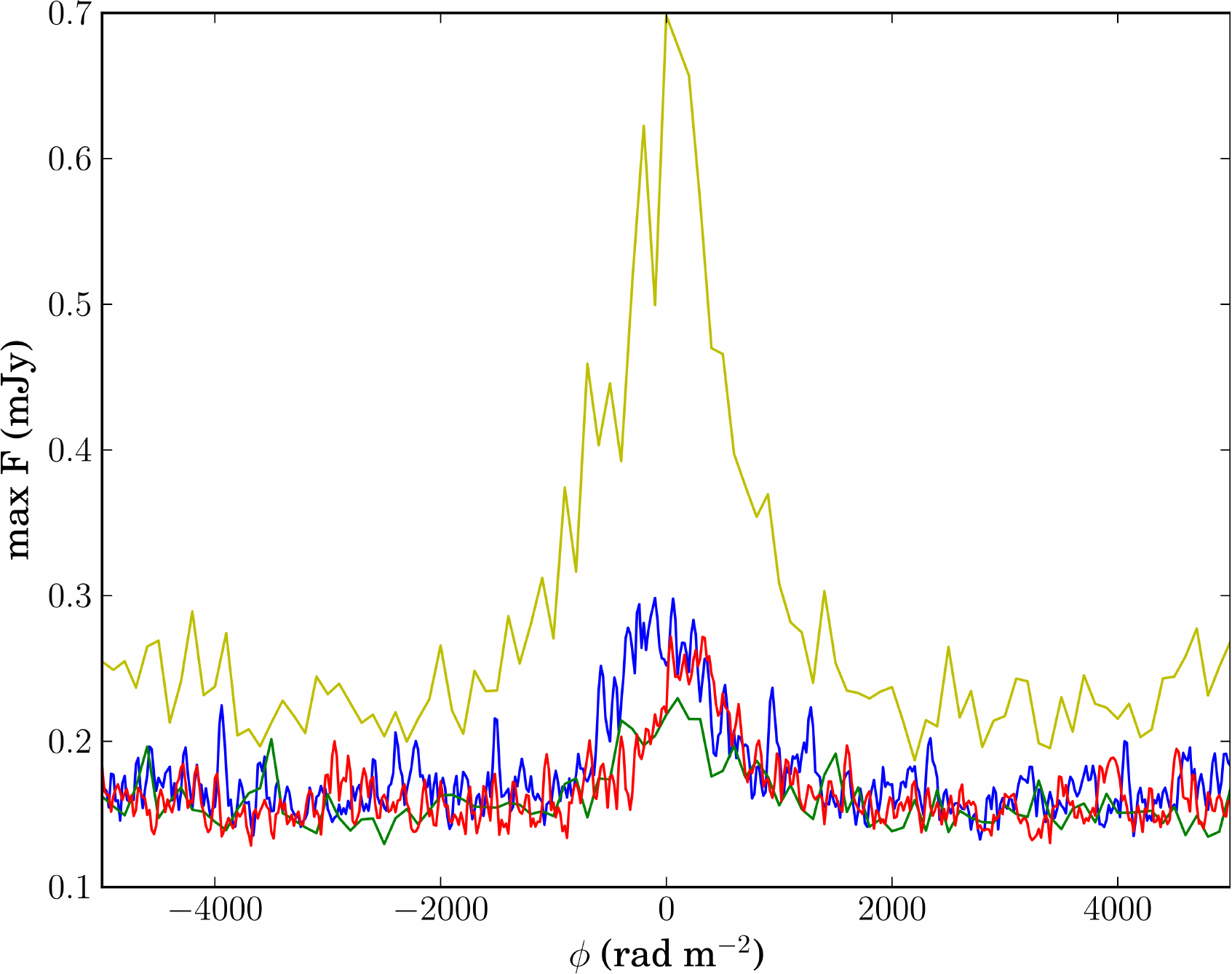}}
   \caption{Each of these functions show the maximum value of all the 500 off-positions at every Faraday depth.  The colours of the functions match the colours of the rectangles shown in Figure \ref{fig:offregions} as to which SNR is nearest.}
   \label{fig:maxoff} 
\end{figure}

\indent	RM-synthesis is done on spectra extracted from each of the 500 off-position boxes.  Since noise dominates Stokes $I$ emission in these regions, RM-synthesis was done without dividing by Stokes $I$.  The method for doing RM-synthesis on noise profiles is the same as outlined for the SNRs.  From these 500 noise Faraday depth spectra we find the maximum value at each Faraday depth.  The result of this analysis is shown in Figure \ref{fig:maxoff}.  The colour of each of these profiles matches the region of the same colour in Figure \ref{fig:offregions}.  From these maximum noise profiles one can see more power for $|\phi| < 2000 \text{ rad m}^{-2}$.    Small-scale structure of polarized emission can be observed by THOR, but polarized emission with angular scales larger than $\sim 6 \arcmin$ at 1 GHz and $\sim 11 \arcmin$ at 2 GHz are filtered out by the VLA in C-array.  Our sensitivity is limited by artifacts around bright sources that change with frequency.  We find that the G43.3$-$0.2 image cube was more affected by noise than the image cubes for the SNRs G46.8$-$0.3, 41.1$-$0.3 and G39.2$-$0.3, thereby resulting in a much higher amplitude in the maximum Faraday depth profile.  For each SNR we exclude any peaks that are below the maximum of its noise profile in Figure \ref{fig:maxoff}.  The values for these detection thresholds are given in Table \ref{tab:det}.  By doing this, any peak greater than the detection threshold has a probability less than 0.2\% of being an effect of noise in the image.  

\subsection{The Effect of Instrumental Polarization}

\begin{figure*}[htb!]
\gridline{\rotatefig{0}{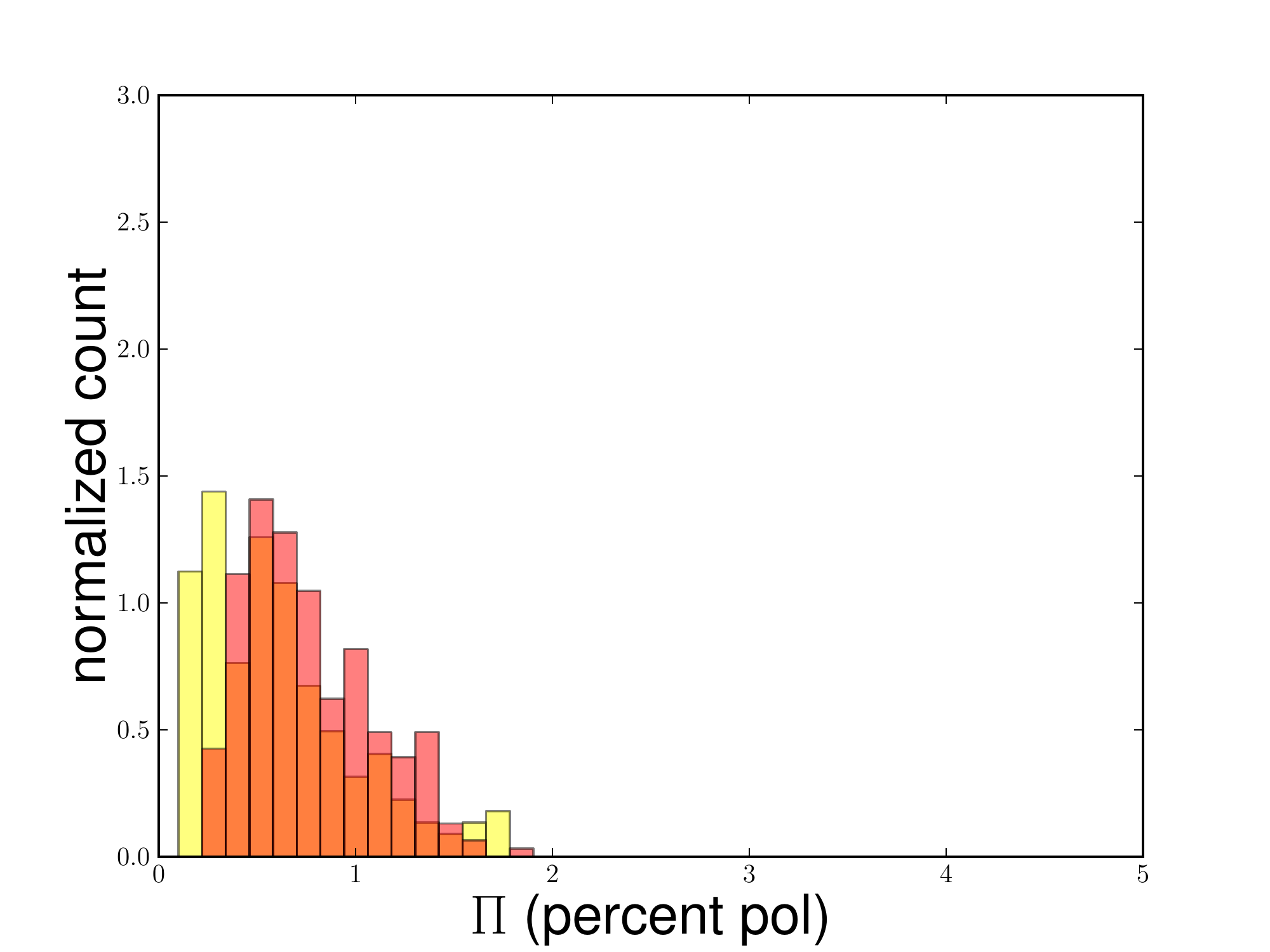}{0.25\textwidth}{(a0)}
          \rotatefig{0}{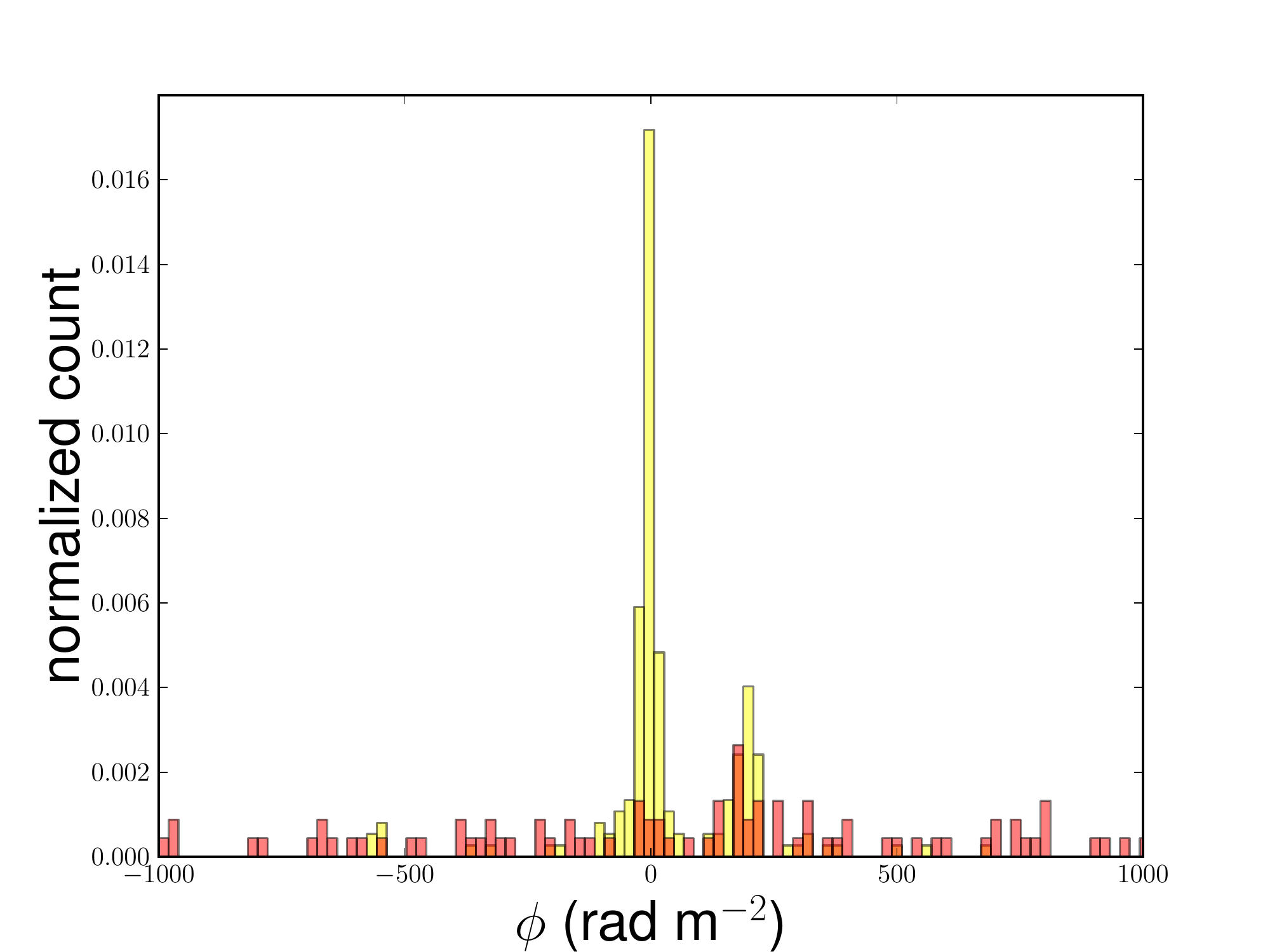}{0.25\textwidth}{(b0)}
          \rotatefig{0}{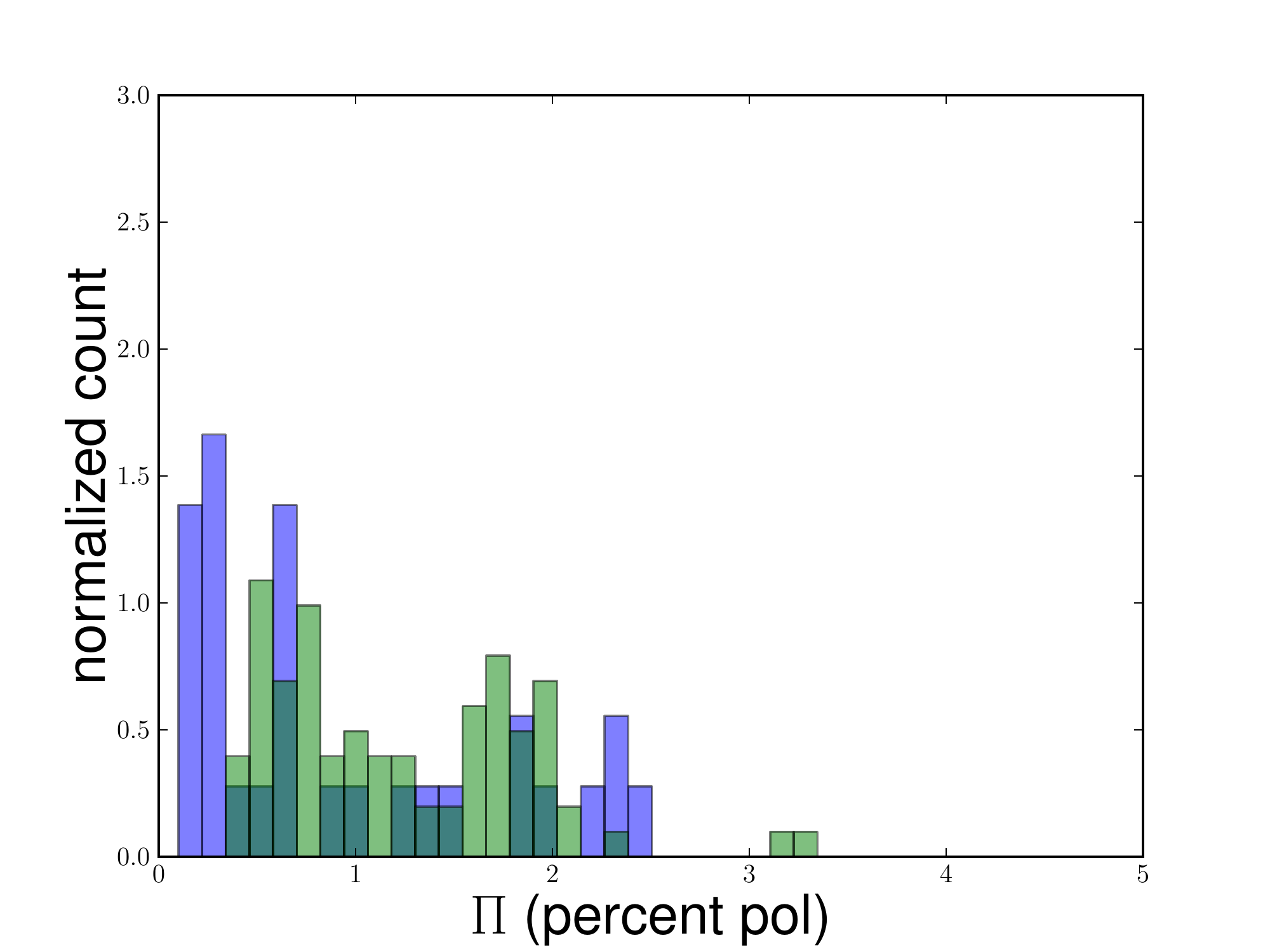}{0.25\textwidth}{(c0)}
          \rotatefig{0}{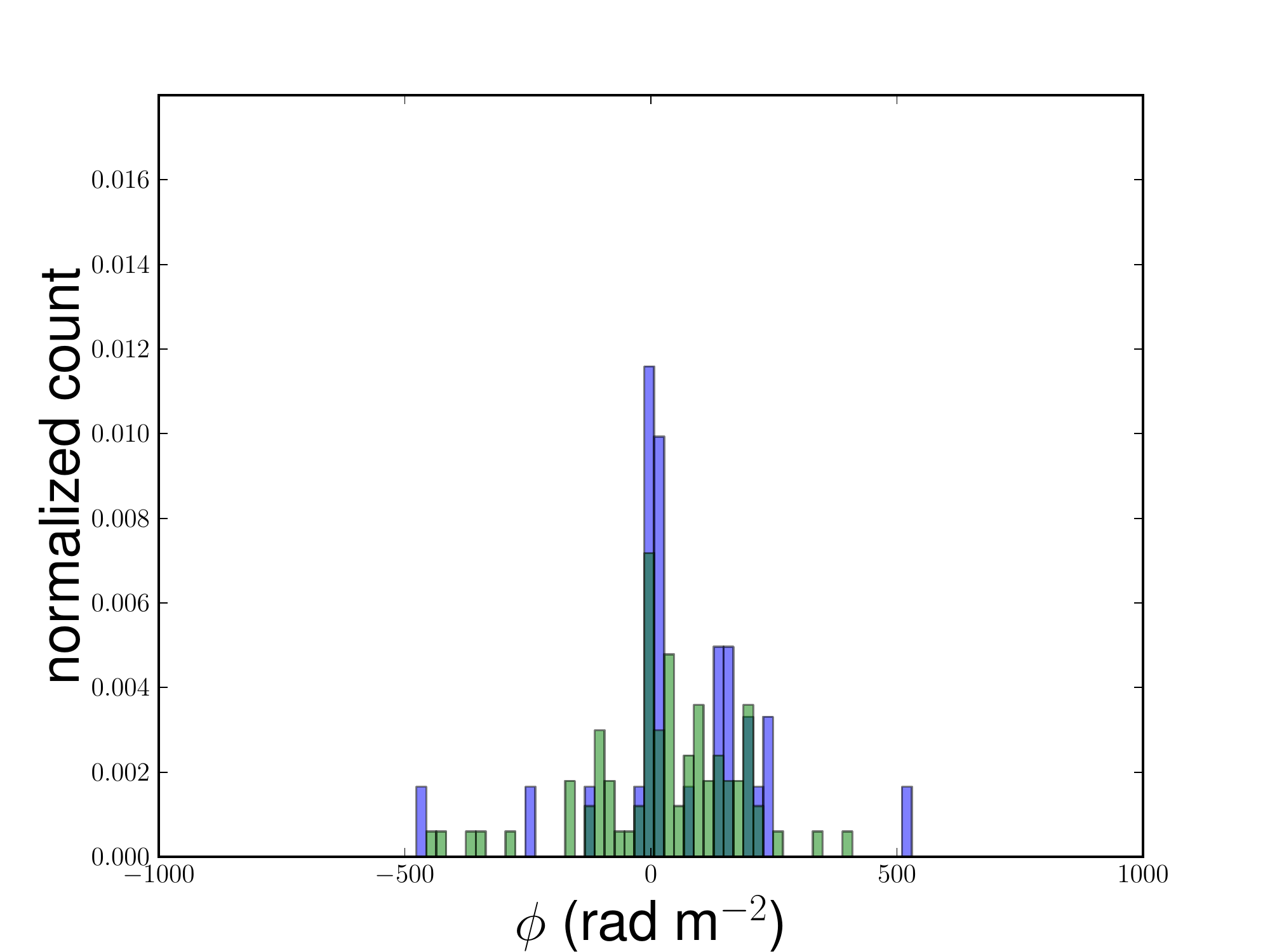}{0.25\textwidth}{(d0)}}
\gridline{\rotatefig{0}{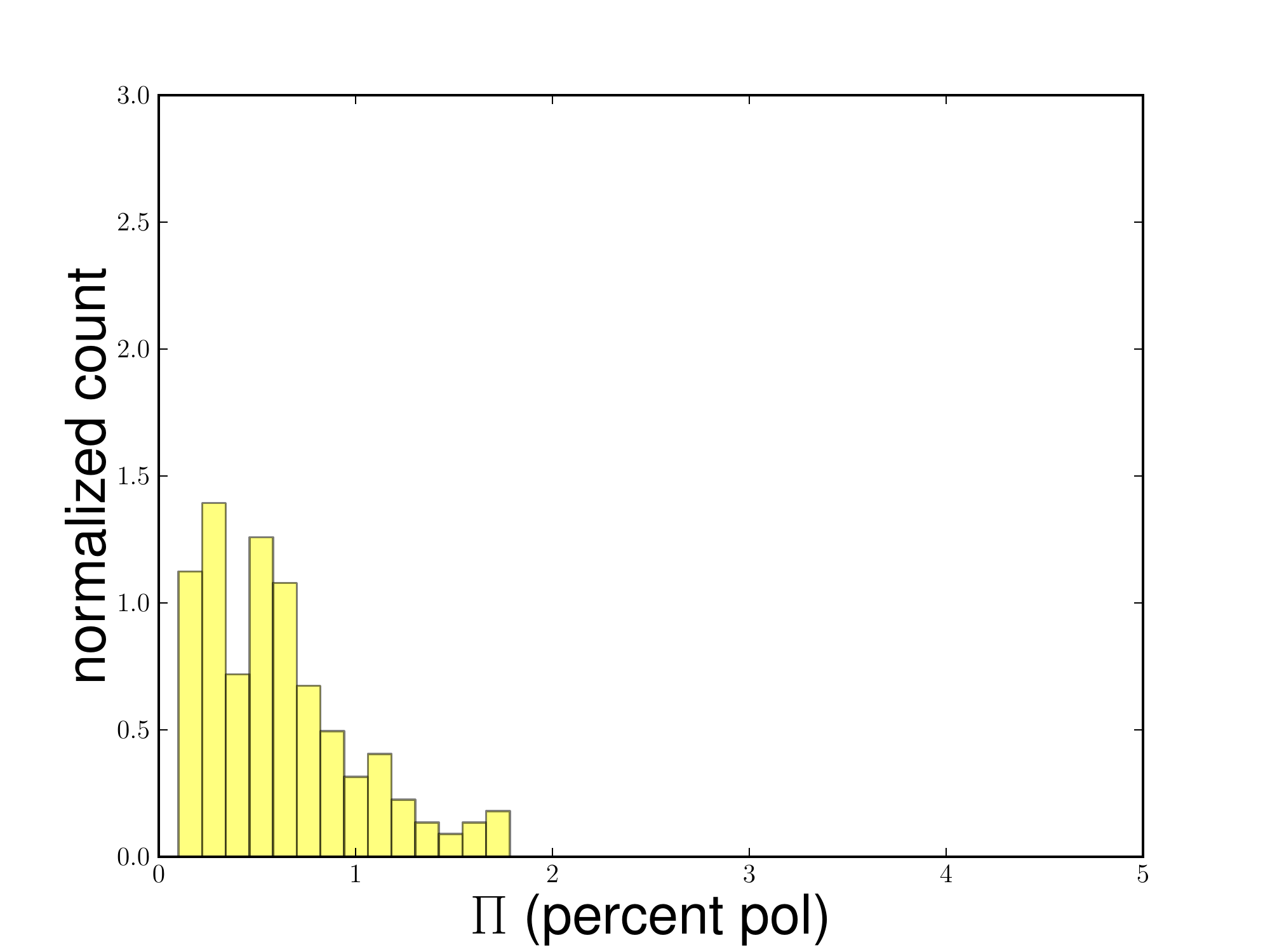}{0.25\textwidth}{(a1)}
          \rotatefig{0}{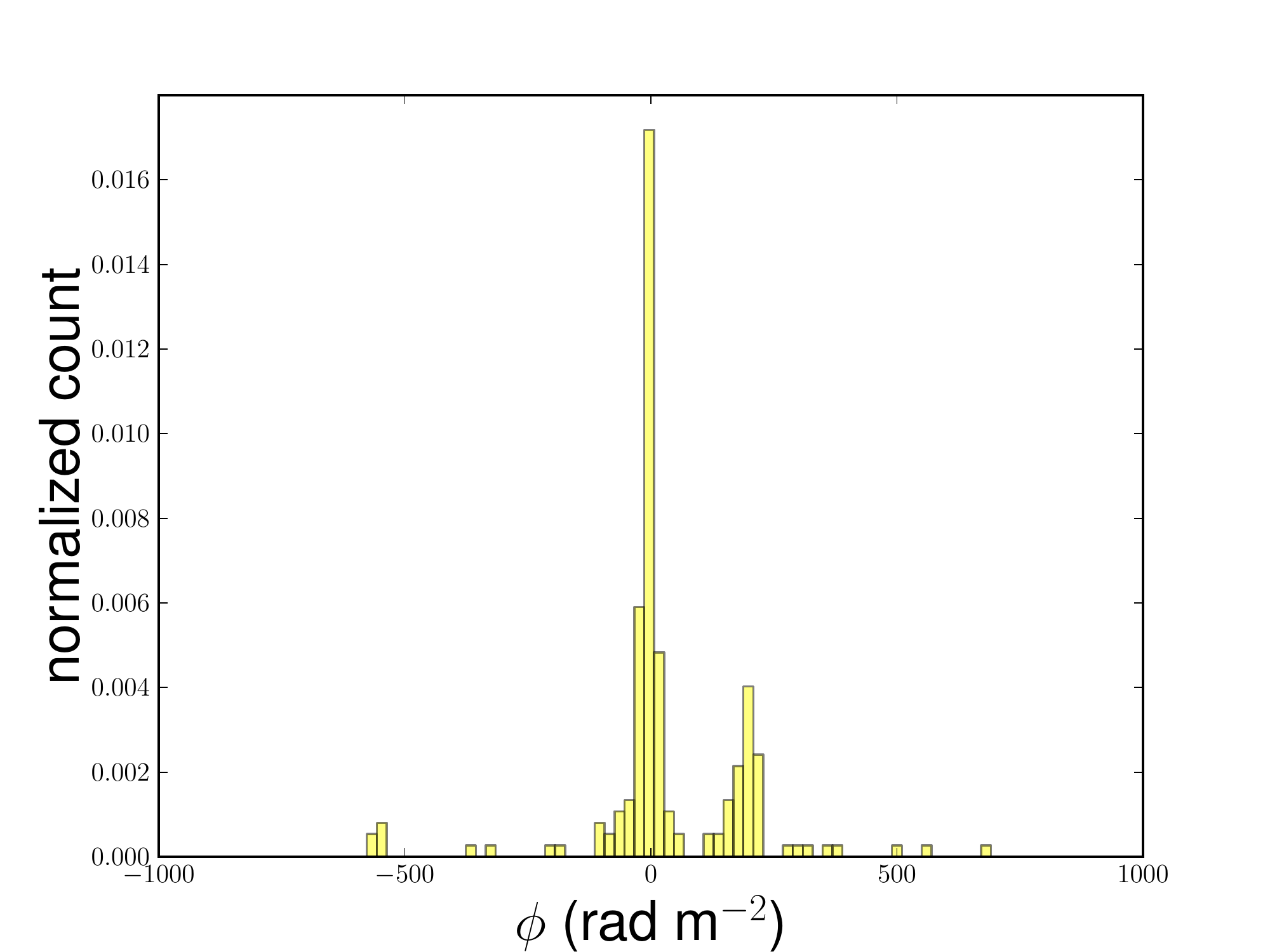}{0.25\textwidth}{(b1)}
          \rotatefig{0}{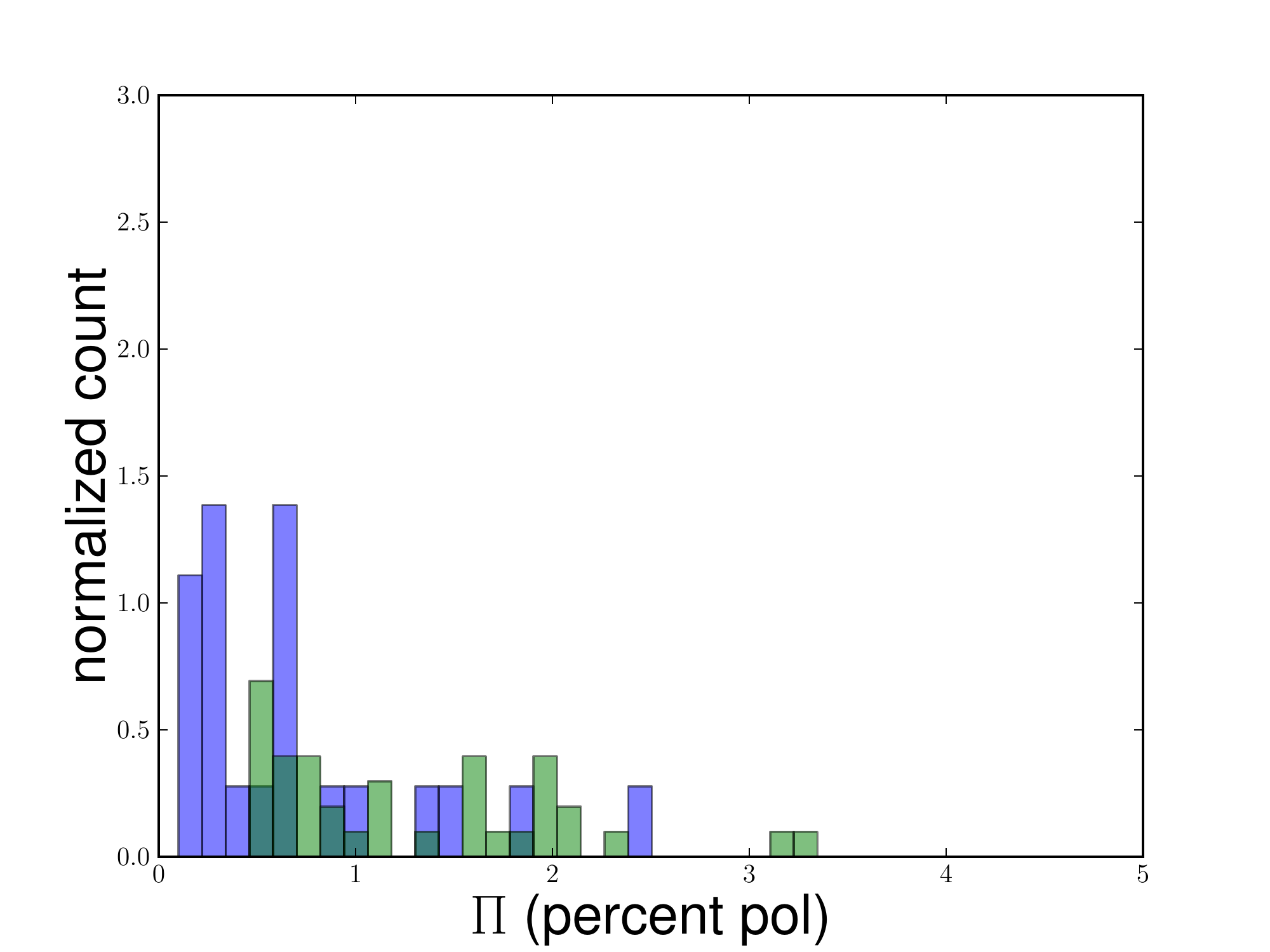}{0.25\textwidth}{(c1)}
          \rotatefig{0}{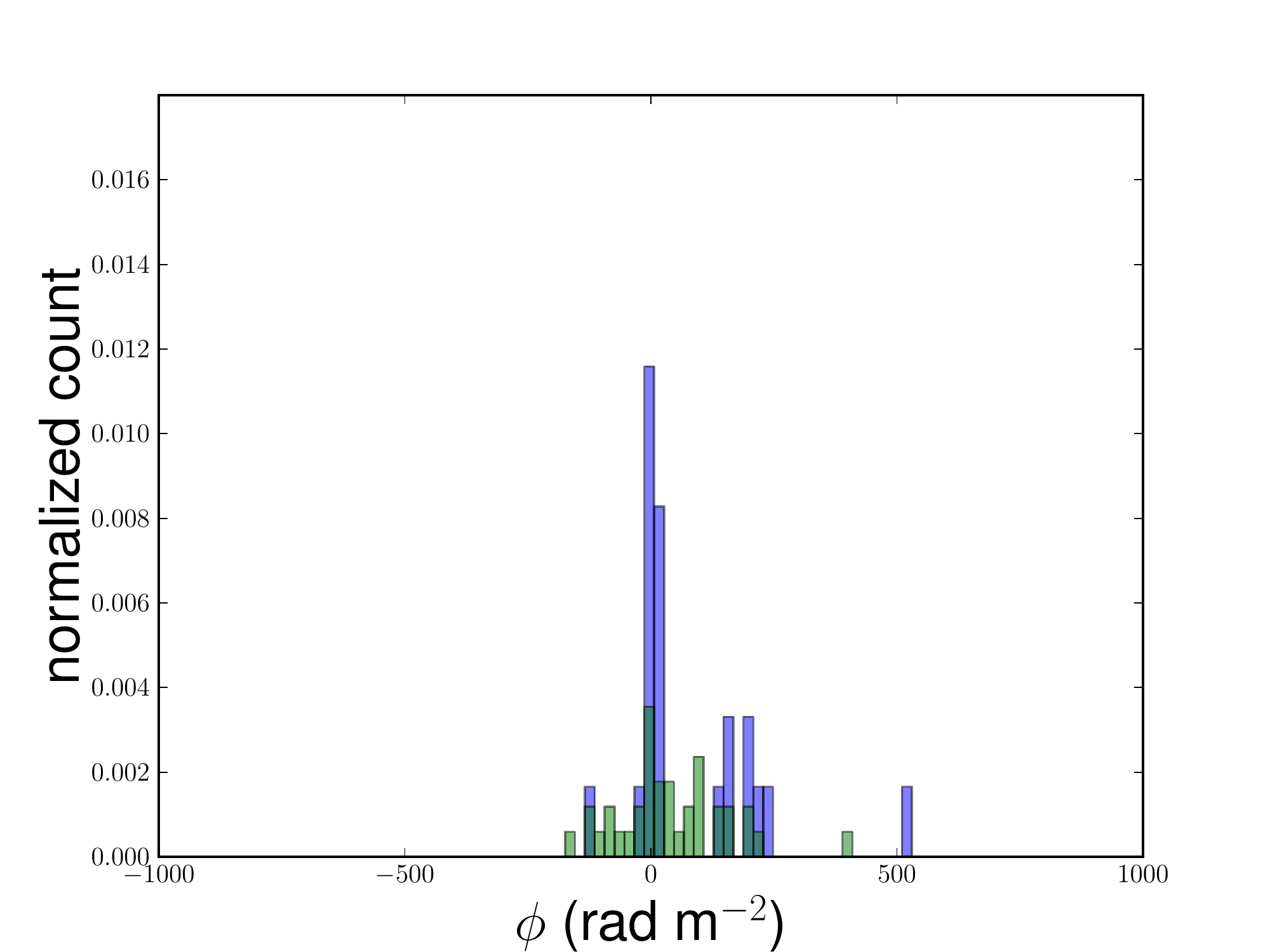}{0.25\textwidth}{(d1)}}
\gridline{\rotatefig{0}{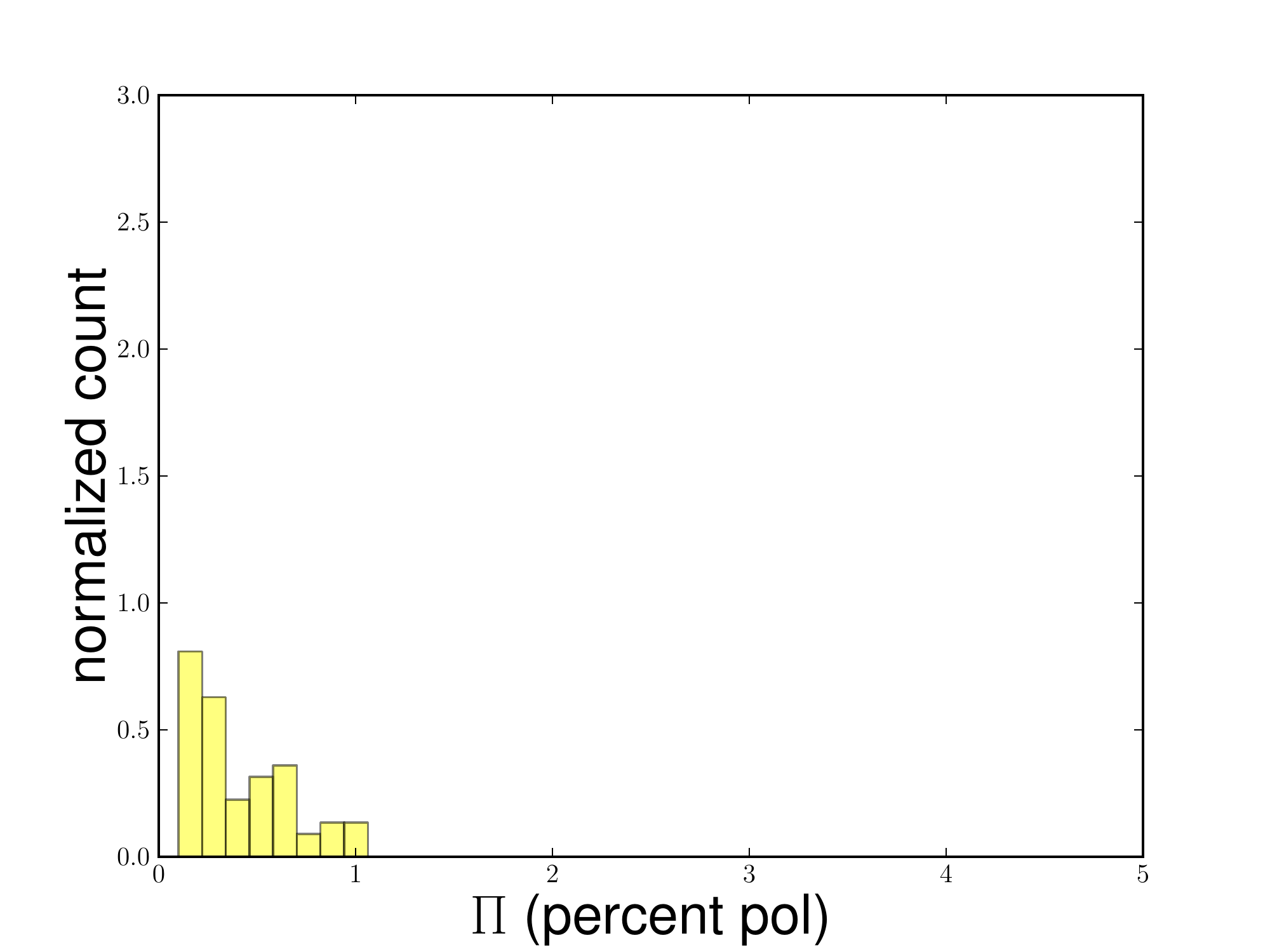}{0.25\textwidth}{(a2)}
          \rotatefig{0}{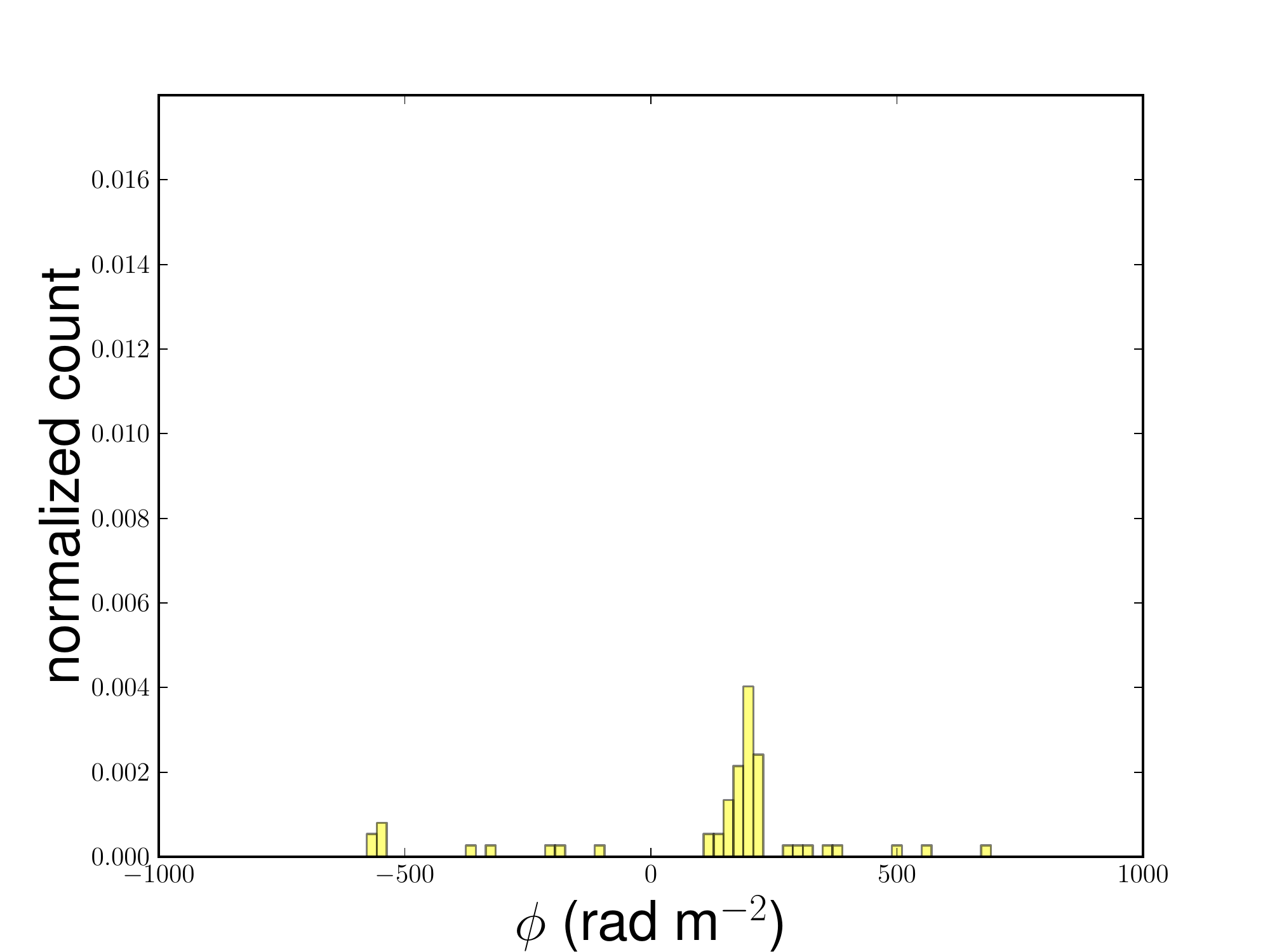}{0.25\textwidth}{(b2)}
          \rotatefig{0}{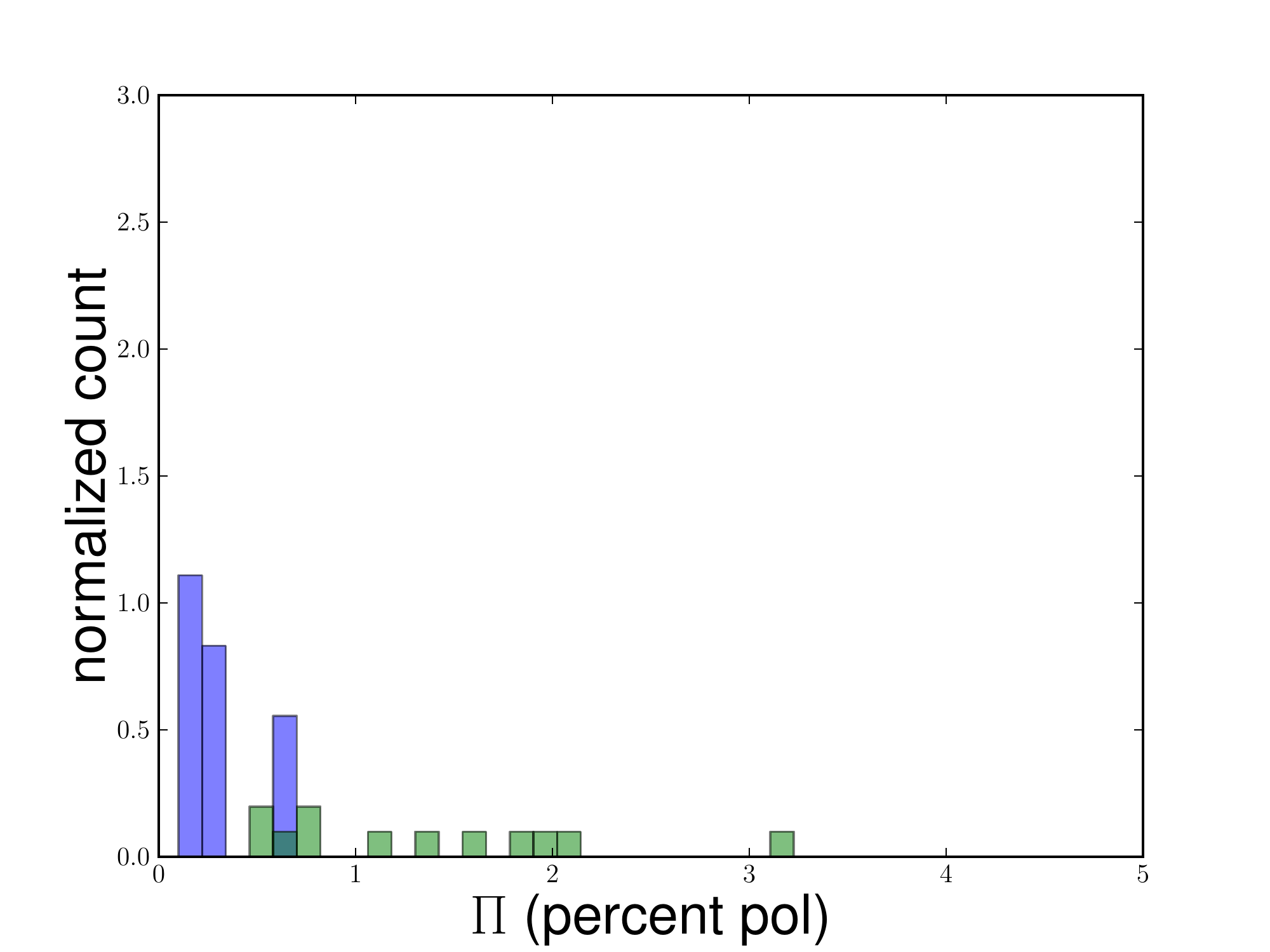}{0.25\textwidth}{(c2)}
          \rotatefig{0}{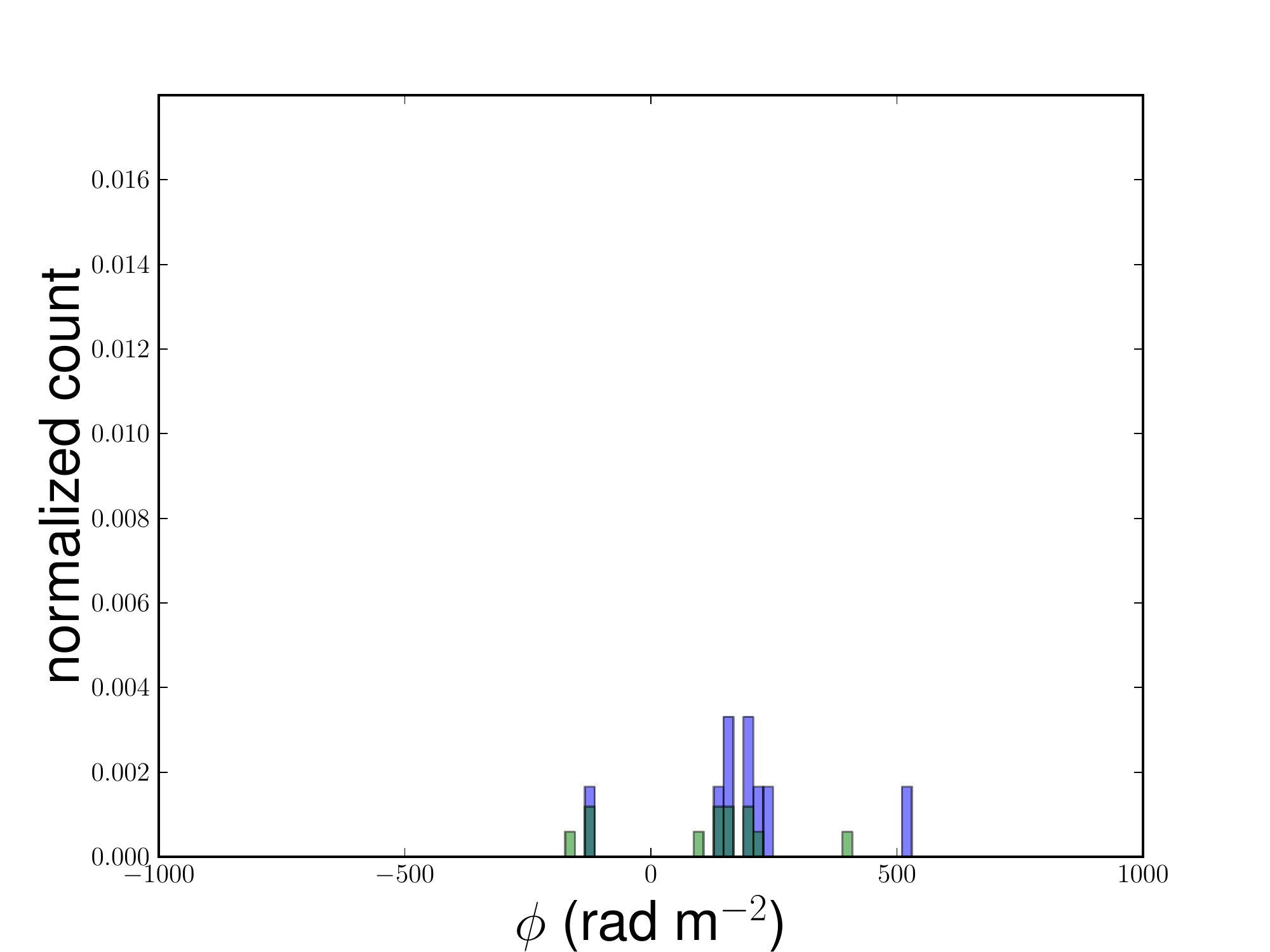}{0.25\textwidth}{(d2)}}
\gridline{\rotatefig{0}{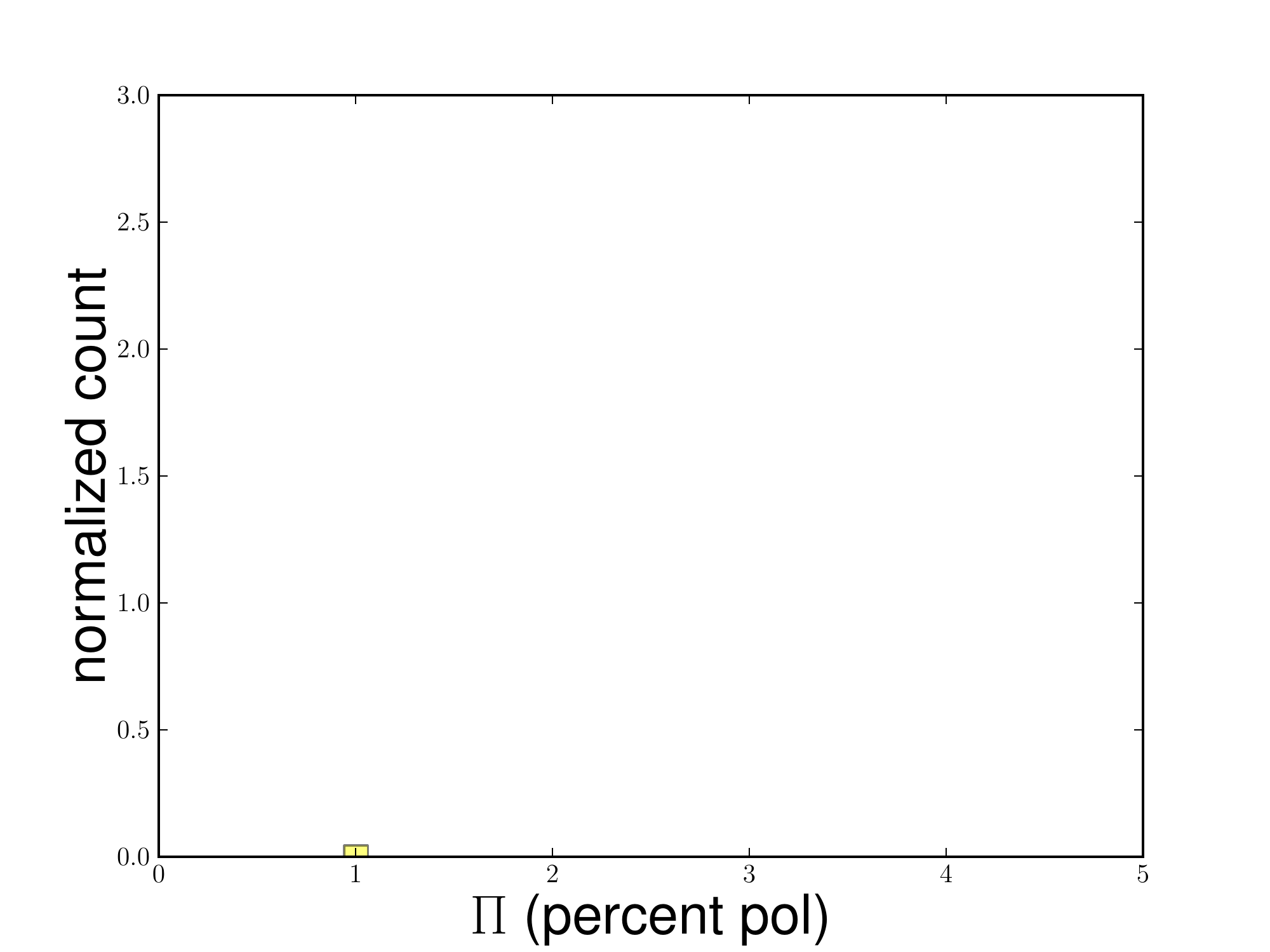}{0.25\textwidth}{(a3)}
          \rotatefig{0}{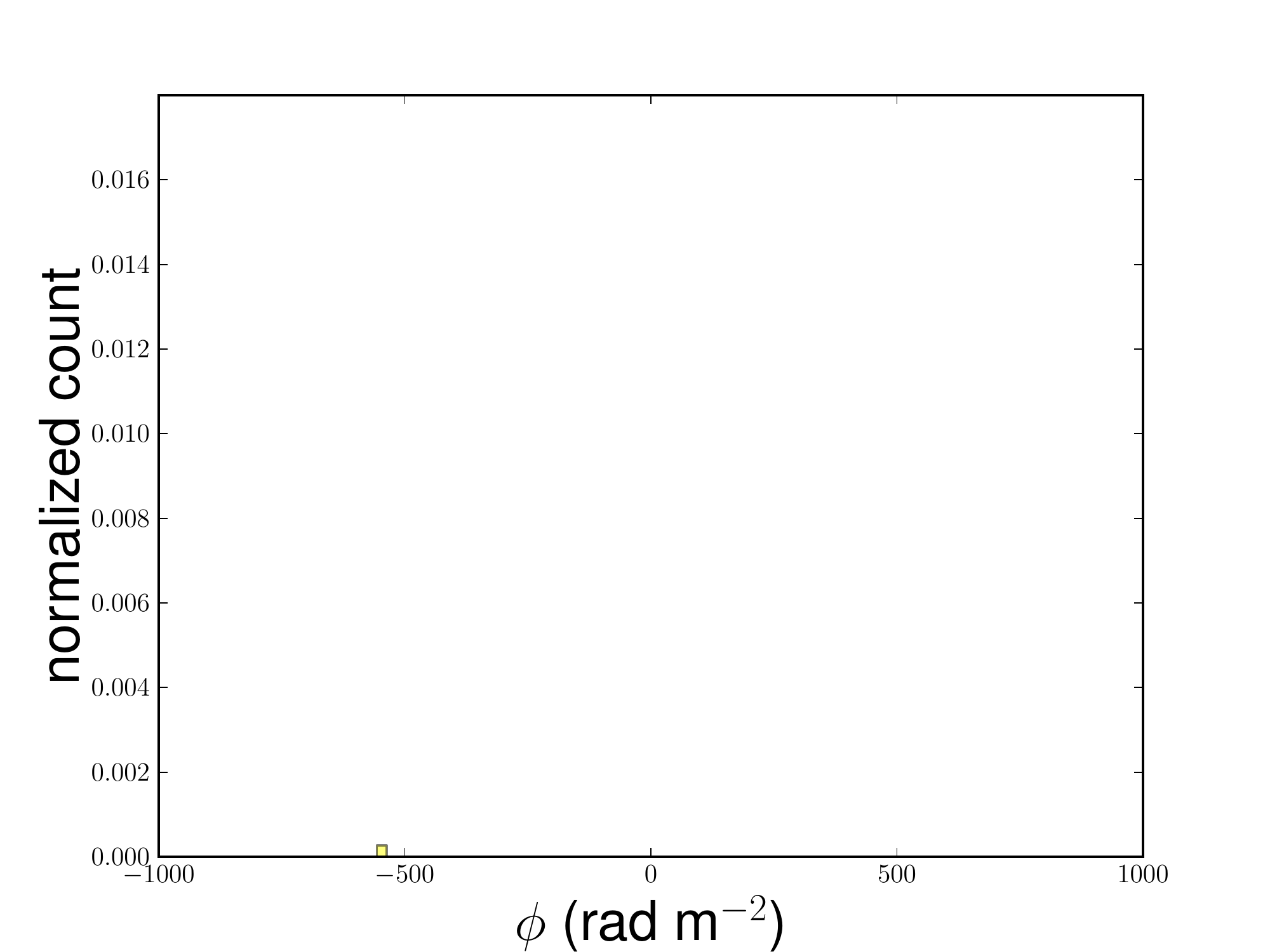}{0.25\textwidth}{(b3)}
          \rotatefig{0}{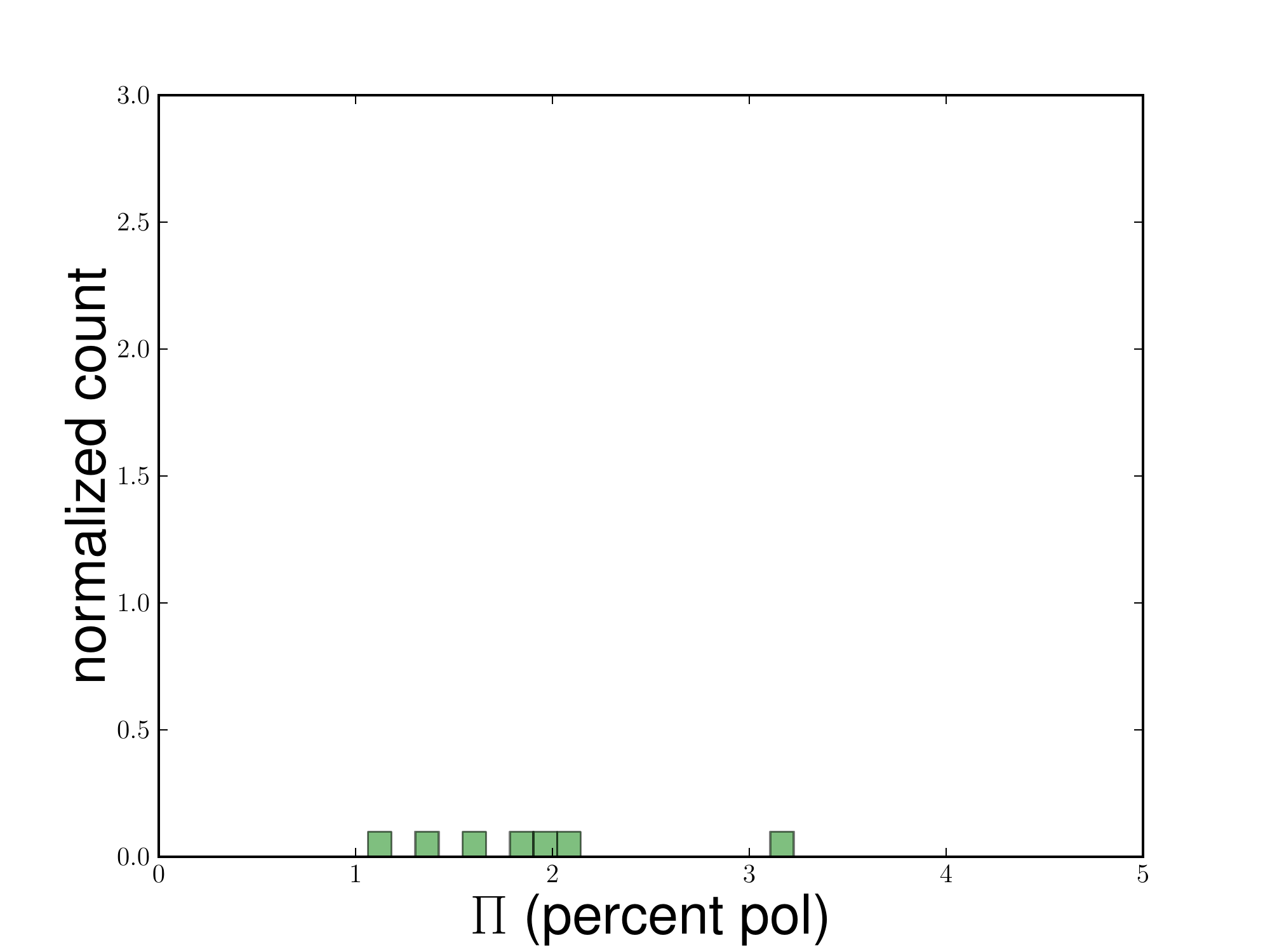}{0.25\textwidth}{(c3)}
          \rotatefig{0}{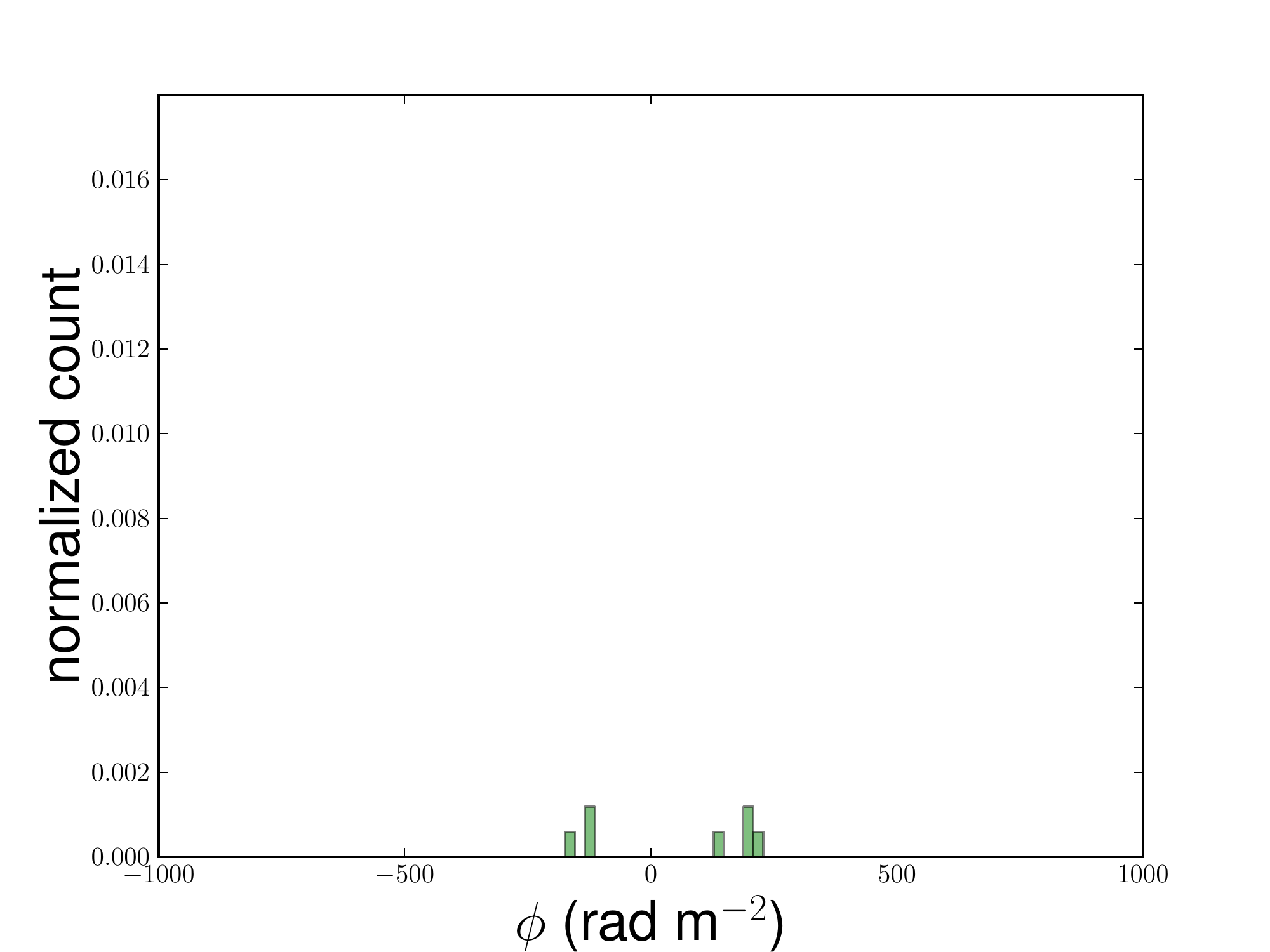}{0.25\textwidth}{(d3)}}
\caption{These figures display the distribution of fractional polarization and Faraday depth for the Faraday profiles extracted from the HII regions shown in Figure \ref{fig:offregions}.  The red and yellow distributions correspond to the HII regions near SNRs G46.8-0.3 and G43.3-0.2, respectively.  The green and blue distributions correspond to the HII regions near SNRs G41.1-0.3 and G39.2-0.3, respectively. Columns (a) and (c) show the distribution of fractional polarization which was found using the same method as for the SNRs (see Figure \ref{fig:G46.8-0.3_PP-Map}). Columns (b) and (d) show the distribution of Faraday depth. }
\label{fig:crits}
\end{figure*}

\indent	The two properties of instrumental polarization are a Faraday depth range in which instrumental polarization is most prevalent as well as a fractional polarization below which instrumental polarization is most observed.  Both of these effects need to be taken into account in our detection criteria. 

\indent	Since Faraday rotation does not occur in the instrument itself, instrumental polarization is most likely to be found at a Faraday depth of 0 rad m$^{-2}$.  Due to this it is important to define a range of Faraday depths in which instrumental polarization is most likely to occur.  \citet{Giessubel2013} define a range around 0 rad m$^{-2}$ that is $\pm$ the FWHM of their RMSF as their Faraday depth rejection range.   

\indent	Since HII regions are typically thought to be unpolarized, we analyzed subregions from HII regions found near the SNRs (pink circles in Figure \ref{fig:offregions}).  For this analysis each subregion has the same size as the SNR subregions. Figure \ref{fig:crits} depicts the distribution of fractional polarization and the highest peak in the Faraday depth spectra for every subregion analyzed in each HII region.  The majority of the Faraday depth peaks are found in a range around 0 rad m$^{-2}$.   Any signal from instrumental polarization is convolved with the RMSF which causes peaks in the Faraday depth spectrum to be broadened to the FWHM of the RMSF, $\delta\phi$.  We take the Faraday depth range of $|\phi| < \delta\phi $ as our rejection range where instrumental polarization is most prevalent. 

\indent	Without applying any detection criteria we find the distributions as shown in row (0) of Figure \ref{fig:crits}.  Columns (a) and (b) are the fractional polarization and Faraday depth distribution of the HII regions found near the SNRs with the highest Galactic longitudes.  Columns (c) and (d) are the same of the HII regions found near the SNRs with the lowest Galactic longitudes.  In row (1) of Figure \ref{fig:crits} we applied the respective detection thresholds (given in Table \ref{tab:det}).  This eliminates noise peaks that can occur at any Faraday depth.

\indent	In row (2) of Figure \ref{fig:crits},  we apply the second detection criterium where peaks found within the Faraday depth range $|\phi| < \delta\phi$ are eliminated.  Comparing rows (1) and (2), subregions with the highest fractional polarization have been rejected by this criterium.  In row (3) of Figure \ref{fig:crits} are peaks left after applying our criterium that any peak below 1\% polarized is rejected.  

\indent 	After applying these detection criteria we eliminate most of the signal that is heavily affected by instrumental polarization.  In row (2) we observe a concentration of peaks near 200 rad m$^{-2}$.  These peaks are a possible outcome of small scale fluctuations in polarization angle of diffuse polarized emission.  Figure \ref{fig:G46_noise} (a) displays an increase in polarized intensity in $\sim 100 < \phi < 700 \text{ rad m}^{-2}$ which is found from subregions where only noise is present.  It is possible that the HII regions act as a Faraday screen for background diffuse polarized emission and the leftover peaks (as seen in row (3) of Figure \ref{fig:crits}) are similar detections.

\indent	The detection criteria we used for each SNR in this study are outlined in Table \ref{tab:det}, along with the requirement that fractional polarization be larger than 1\%.  Our detection criteria allow us to confidently eliminate signal that is heavily affected by either noise or instrumental polarization.  We cannot eliminate all effects instrumental polarization has on real signal, but by applying our detection criteria we eliminate signal that is heavily affected.

\begin{deluxetable*}{cccc}
\tablenum{2}
\tablecaption{Detection criteria\label{tab:det}}
\tablewidth{0pt}
\tablehead{
\colhead{SNR Name} & \colhead{Detection Threshold} & \colhead{$\delta\phi$} & \colhead{$\Pi_0$ Threshold} \\
 & \colhead{(mJy)} & \colhead{(rad m$^{-2}$)} & 
}
\decimalcolnumbers
\startdata
G46.8$-$0.3 & 0.27 & 102.8 & 1\% \\
G43.3$-$0.2 & 0.70 & 102.8 & 1\% \\
G41.1$-$0.3 & 0.23 & 104.4 & 1\% \\
G39.2$-$0.3 & 0.30 & 101.7 & 1\% \\
\enddata
\tablecomments{Detection criteria applied to each subregion to ensure a confident detection of polarized signal.  }
\label{tab:det}
\end{deluxetable*}

\clearpage

\bibliography{refs}{}
\bibliographystyle{aasjournal}

\end{document}